\documentclass[12pt]{iopart}

\usepackage{iopams}
\usepackage{graphicx}

\begin{document}

\title[Circadian pattern and burstiness]{Circadian pattern and burstiness in mobile phone communication}

\author{Hang-Hyun Jo$^1$, M\'arton Karsai$^1$, J\'anos Kert\'esz$^{1,2}$ and Kimmo Kaski$^1$}
\address{$^1$ BECS, Aalto University School of Science, P.O. Box 12200, FI-00076}
\address{$^2$ Institute of Physics and BME-HAS Cond. Mat. Group, BME, Budapest, Budafoki \'ut 8., H-1111}
\ead{joh1@aalto.fi}

\begin{abstract}
The temporal communication patterns of human individuals are known to be inhomogeneous or bursty, which is reflected as the heavy tail behavior in the inter-event time distribution. As the cause of such bursty behavior two main mechanisms have been suggested: a) Inhomogeneities due to the circadian and weekly activity patterns and b) inhomogeneities rooted in human task execution behavior. Here we investigate the roles of these mechanisms by developing and then applying systematic de-seasoning methods to remove the circadian and weekly patterns from the time-series of mobile phone communication events of individuals. We find that the heavy tails in the inter-event time distributions remain robustly with respect to this procedure, which clearly indicates that the human task execution based mechanism is a possible cause for the remaining burstiness in temporal mobile phone communication patterns. 
\end{abstract}

\submitto{\NJP}
\maketitle

\section{Introduction}

Recently modern information-communication-technology (ICT) has opened us access to large amounts of stored digital data on human communication, which in turn has enabled us to have unprecedented insights into the patterns of human behavior and social interaction. For example we can now study the structure and dynamics of large-scale human communication networks~\cite{Onnela2007a,Onnela2007b,Krings2009,Palla2007} and the laws of mobility~\cite{Gonzalez2008,Song2010a,Song2010b}, as well as the motifs of individual behavior~\cite{Barabasi2005,Malmgren2008,Malmgren2009a,Malmgren2009b,Anteneodo2010,Miritello2011,Karsai2011}. One of the robust findings of these studies is that human activity over a variety of communication channels is inhomogeneous, such that high activity bursts of rapidly occurring events are separated by long periods of inactivity~\cite{Johansen2001,Eckmann2004,Harder2006,Goncalves2008,Zhou2008,Radicchi2009}. This feature is usually characterized by the distribution of inter-event times $\tau$, defined as time intervals between, e.g., consecutive e-mails sent by a single user. This distribution has been found to have a heavy tail and show a power-law decay as $P(\tau)\sim \tau^{-1}$~\cite{Barabasi2005}.

In human behavior obvious causes of inhomogeneity are the circadian and other longer cycles of our lives as results of natural and societal factors. Malmgren~\textit{et al.}~\cite{Malmgren2008,Malmgren2009a} suggested that an approximate power-law scaling found in the inter-event time distribution of human correspondence activity is a consequence of circadian and weekly cycles affecting us all, such that the large inter-event times are attributed to nighttime and weekend inactivity. As an explanation they proposed a cascading inhomogeneous Poisson process, which is a combination of two Poisson processes with different time scales. One of them is characterized by the time-dependent event rate representing the circadian and weekly activity patterns, while the other corresponds to the cascading bursty behavior with a shorter time scale. Their model was able to reproduce an apparent power-law behavior in the inter-event time distribution of email and postal mail correspondence. In addition they calculated the Fano and Allan factors to indicate the existence of some correlations for the email data as well as for their model of inhomogeneous Poisson process, with quite good comparison~\cite{Anteneodo2010}. 

However, the question remains whether in addition to the circadian and weekly cycle driven inhomogeneities there are also other correlations due to human task execution that contribute to the inhomogeneities observed in communication patterns, as suggested, e.g., by the queuing models ~\cite{Barabasi2005,Vazquez2006}. There is evidence for this by Goh and Barab\'asi~\cite{Goh2008}, who introduced a measure that indicates the communication patterns to have correlations. Recently, Wu~\textit{et al.} have studied the modified version of the queuing process proposed in~\cite{Barabasi2005} by introducing a Poisson process as the initiator of localized bursty activity~\cite{Wu2010}. This was aimed at explaining the observation that the inter-event time distributions in Short Message (SM) correspondence follow a bimodal combination of power-law and Poisson distributions. The power-law (Poisson) behavior was found dominant for $\tau<\tau_0$ ($\tau>\tau_0$). Since the event rates extracted from the empirical data have the time scales larger than $\tau_0$ (also measured empirically), a bimodal distribution was successfully obtained. However, in their work the effects of circadian and weekly activity patterns were not considered, thus needing to be investigated in detail.

As the circadian and weekly cycles affect human communication patterns in quite obvious ways, taking place mostly during the daytime and differently during the weekends, our aim in this paper is to remove or de-season from the data the temporal inhomogeneities driven by these cycles. Then the study of the remaining de-seasoned data would enable us to get insight to the existence of other human activity driven correlations. This is important for two reasons. First, communication patterns tell about the nature of human behavior. Second, in devising models of human communication behavior the different origins of inhomogeneities should be properly taken into account; is it enough to describe the communication pattern by an inhomogeneous Poissonian process or do we need a model to reflect correlations in other human activities, such as those due to task execution?

In this paper, we provide a systematic method to de-season the circadian and weekly patterns from the mobile phone communication data. Firstly, we extract the circadian and weekly patterns from the time-stamped communication records and secondly, these patterns are removed by rescaling the timings of the communication events, i.e. phone calls and SMs. The rescaling is performed such that the time is dilated or contracted at times of high or low event activity, respectively. Finally, we obtain the inter-event time distributions by using the rescaled timings and comparing them with the original distributions to check how the heavy tail and burstiness behavior are affected. As the main results we find that the de-seasoned data still shows heavy tail inter-event time distributions with power-law scalings thus indicating that human task execution is a possible cause of remaining burstiness in mobile phone communication.

This paper is organized as follows. In Section~\ref{sect:analysis}, we introduce the methods for de-seasoning the circadian and weekly patterns systematically in various ways. By applying these methods the values of burstiness of inter-event time distributions are obtained and subsequently discussed. Finally, we summarize the results in Section~\ref{sect:summary}.

\section{De-seasoning analysis}\label{sect:analysis}

We investigate the effect of circadian and weekly cycles on the heavy-tailed inter-event time distribution and burstiness in human activity by using the mobile phone call (MPC) dataset from a European operator (national market share $\sim 20\%$) with time-stamped records over a period of $119$ days starting from January 2, 2007. The data of January 1, 2007 are not considered due to its rather unusual pattern of human communication. We have only retained links with bidirectional interaction, yielding $N=5.2\times 10^6$ users, $L=10.6\times 10^6$ links, and $C=322\times 10^6$ events (calls). For the analysis of Short Message (SM) dataset, see Appendix.

We perform the de-seasoning analysis by defining first the observable. For an individual service user $i$, $n_i(t)$ denotes the number of events at time $t$, where $t$ ranges from $0$ seconds, i.e. the start of January 2, 2007 at midnight, to $T_f\approx 1.03\times 10^7$ seconds (119 days). The total number of events $s_i\equiv \sum_{t=0}^{T_f}n_i(t)$ is called the strength of user $i$. In general, for a set of users $\Lambda$, the number of events at time $t$ is denoted by $n_\Lambda(t)\equiv \sum_{i\in\Lambda}n_i(t)$. $\Lambda$ can represent one user, a set of users, or the whole population. When the period of cycle $T$ is given, the event rate $\rho_{\Lambda,T}(t)$ with $0\leq t<T$ is defined as
\begin{equation}
    \rho_{\Lambda,T}(t)=\frac{T}{s_\Lambda} \sum_{k=0}^{\lfloor T_f/T\rfloor} n_\Lambda(t+kT),\ s_\Lambda=\sum_{t=0}^{T_f} n_\Lambda(t).
    \label{eq:eventRate}
\end{equation}
For convenience, we redefine the periodic event rate with period $T$ as $\rho_{\Lambda,T}(t)=\rho_{\Lambda,T}(t+kT)$ with any non-negative integer $k$ for $0\leq t<\infty$. By means of the event rate, we define the rescaled time $t^*(t)$ as following~\cite{Anteneodo2010}
\begin{equation}
    t^*(t)=\sum_{0\leq t'<t} \rho_{\Lambda,T}(t').
    \label{eq:rescaledTime}
\end{equation}
This rescaling corresponds to the transformation of the time variable by $\rho^*(t^*)dt^*=\rho(t)dt$ with $\rho^*(t^*)=1$. Here $\rho^*(t^*)=1$ means that there exists no cyclic pattern in the frame of rescaled time. The time is dilated (contracted) at the moment of high (low) activity.

In order to check whether the rescaling affects the inter-event time distributions and whether still some burstiness exists we reformulate the inter-event time distributions by using rescaled event times and compare them with the original distributions. The definition of the rescaled inter-event time from the rescaled time is straightforward. Considering two consecutive events of a user $i\in\Lambda$ occurring at times $t_j$ and $t_{j+1}$, the original inter-event time is $\tau \equiv t_{j+1}-t_j$, then the corresponding rescaled inter-event time is defined as follows
\begin{equation}
    \tau^* \equiv t^*(t_{j+1})-t^*(t_j)=\sum_{t_j\leq t'<t_{j+1}}\rho_{\Lambda,T}(t').
    \label{eq:rescaledTau}
\end{equation}
To find out how much the de-seasoning affects burstiness, we measure the burstiness of events, as proposed in~\cite{Goh2008}, where the burstiness parameter $B$ is defined as
\begin{equation}
B\equiv\frac{\sigma_\tau-m_\tau}{\sigma_\tau+m_\tau}.
\end{equation}
Here $\sigma_\tau$ and $m_\tau$ are the standard deviation and the mean of the inter-event time distribution $P(\tau)$, respectively. The value of $B$ is bounded within the range of $[-1,1]$ such that $B=1$ for the most bursty behavior, $B=0$ for neutral or homogeneous Poisson behavior, and $B=-1$ for completely regular behavior. The burstiness of the original inter-event time distribution, denoted by $B_0$, is to be compared with that of $B_T$ of the rescaled inter-event time distribution for given period $T$. With the de-seasoning the burstiness is expected to decrease and here we are most interested in by what amount the burstiness decreases when using $T=1$ day or $7$ days, i.e. removing the circadian or weekly patterns. 

\begin{figure*}[!t]
\begin{tabular}{cc}
\includegraphics[width=.45\columnwidth]{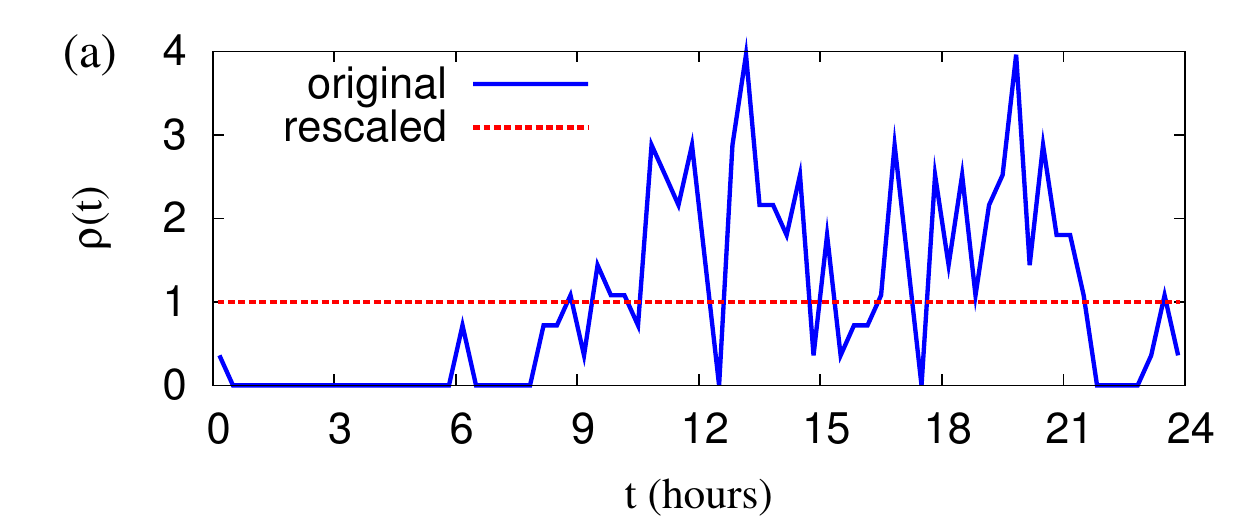}&
\includegraphics[width=.45\columnwidth]{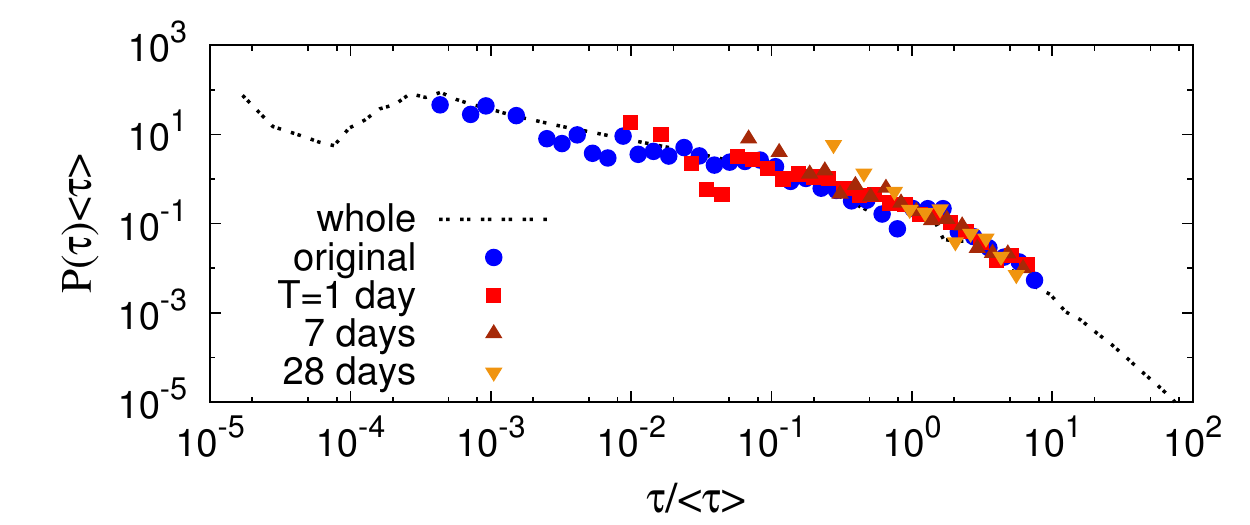}\\
\includegraphics[width=.45\columnwidth]{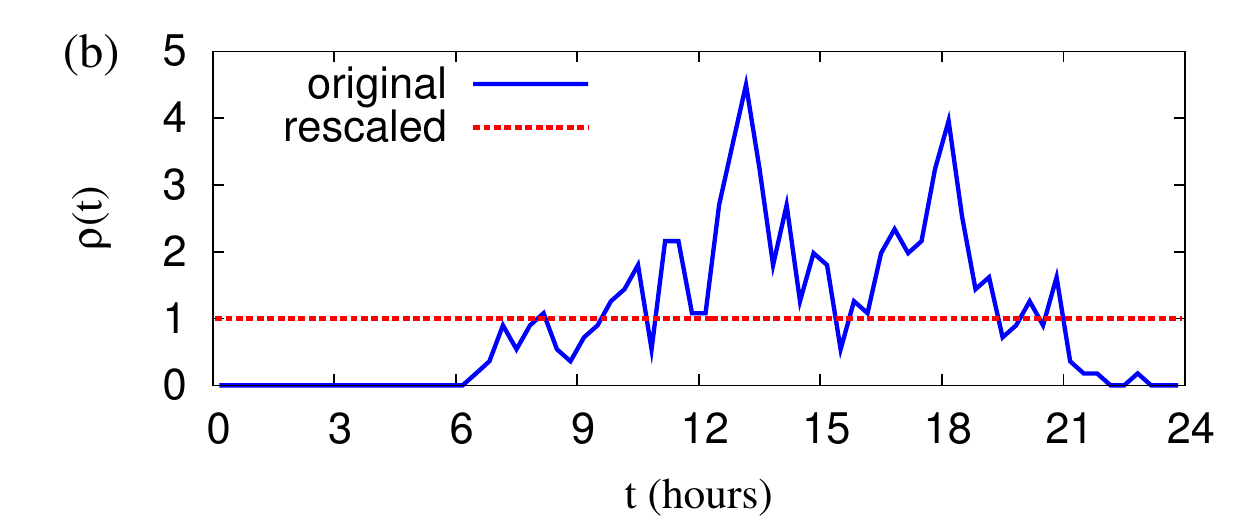}&
\includegraphics[width=.45\columnwidth]{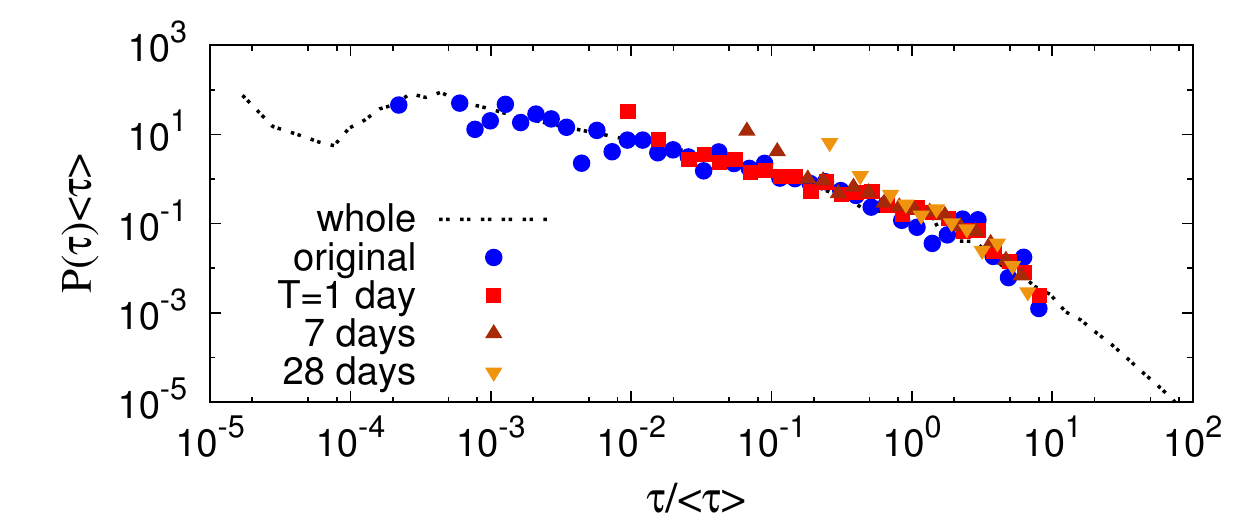}\\
\includegraphics[width=.45\columnwidth]{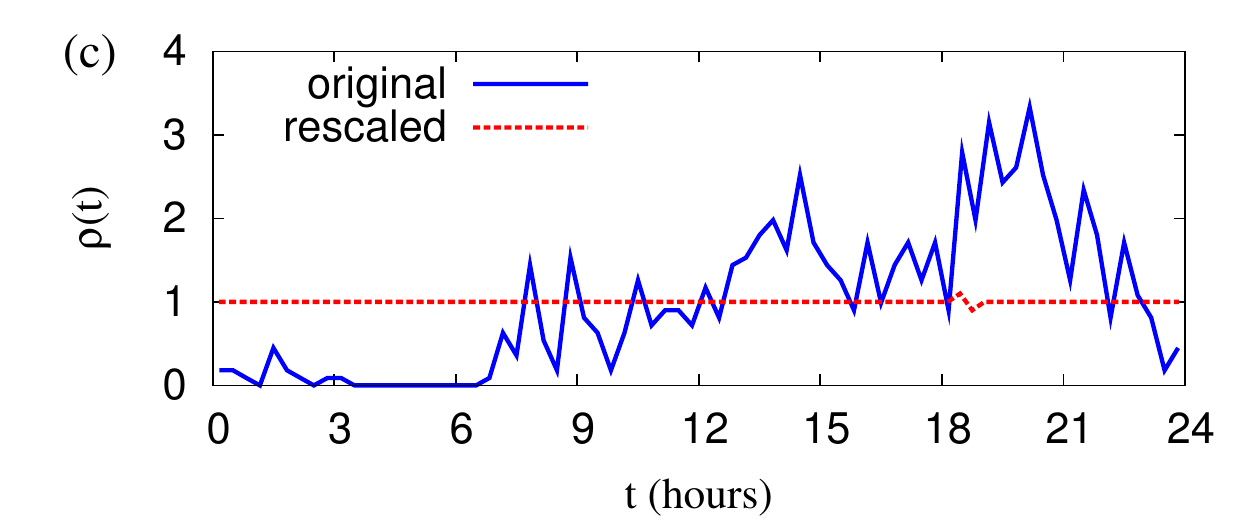}&
\includegraphics[width=.45\columnwidth]{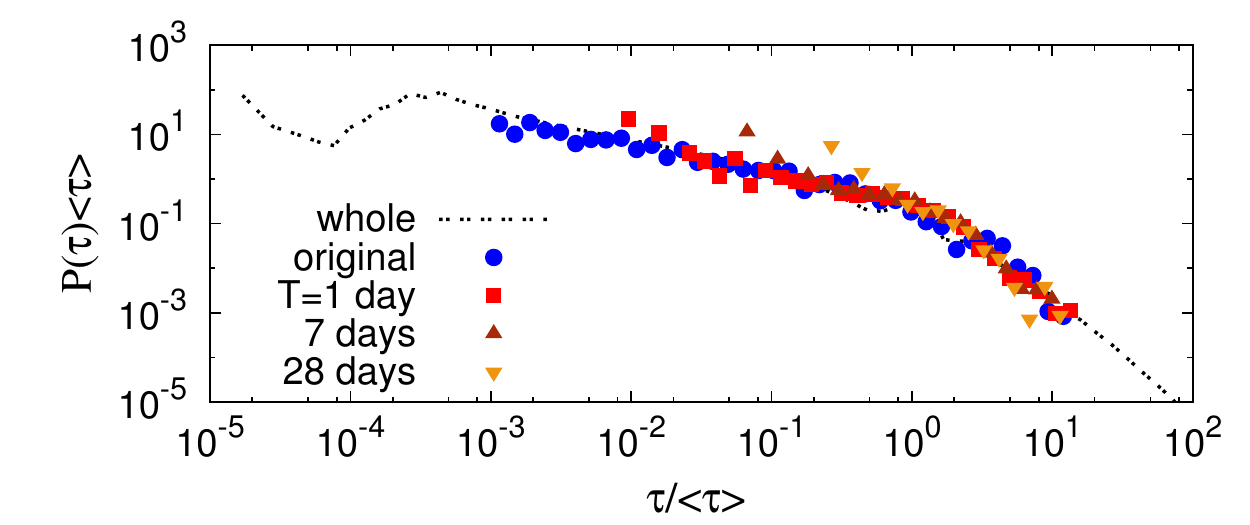}\\
\includegraphics[width=.45\columnwidth]{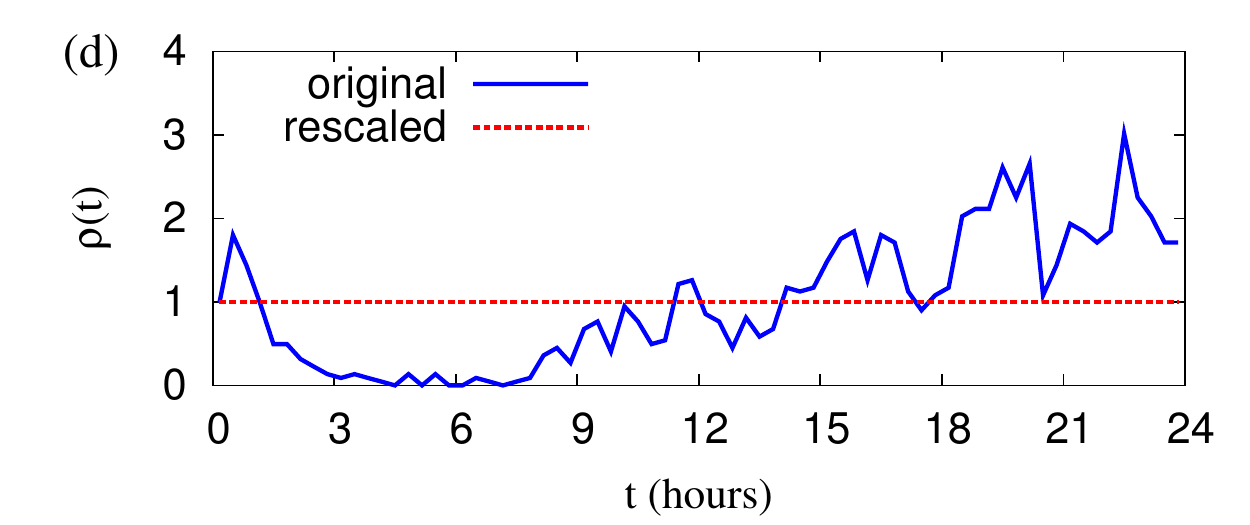}&
\includegraphics[width=.45\columnwidth]{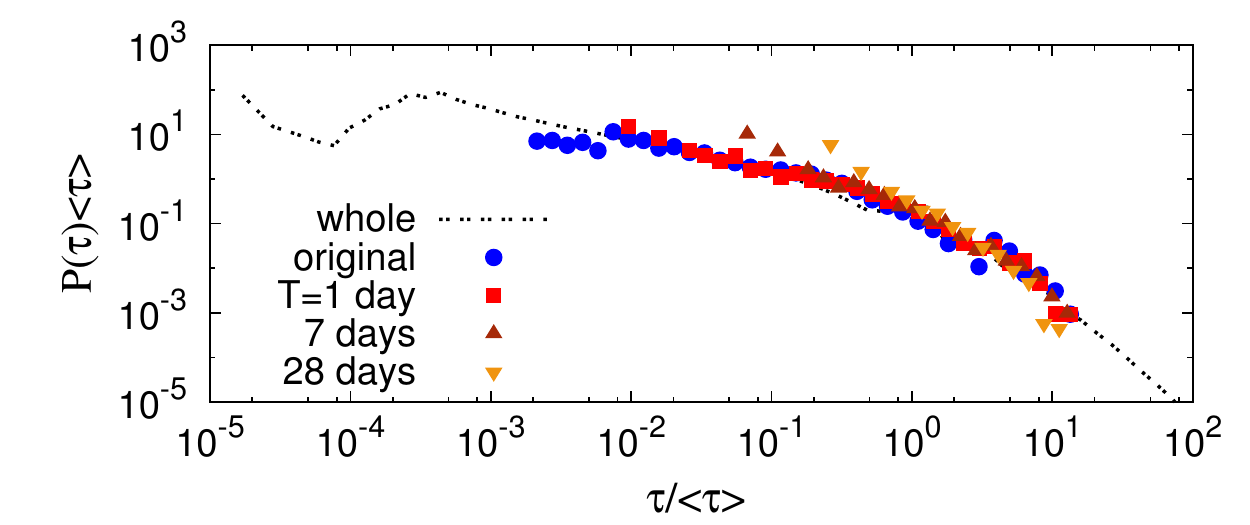}\\
\includegraphics[width=.45\columnwidth]{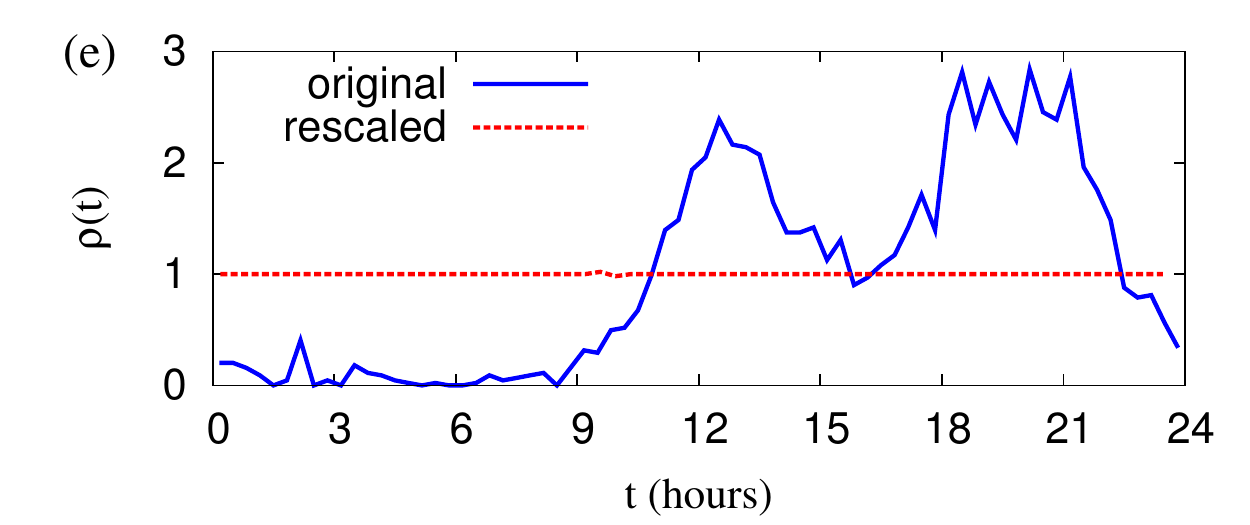}&
\includegraphics[width=.45\columnwidth]{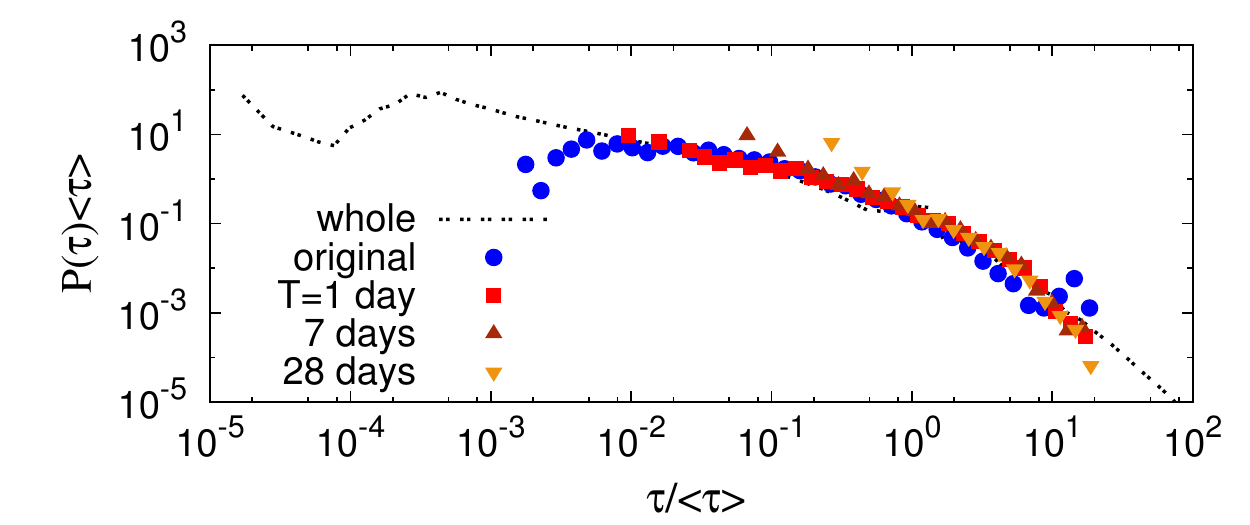}
\end{tabular}
\caption{De-seasoning MPC patterns of individual users: the original and the rescaled event rates with period of $T=1$ day (left) and the original and the rescaled inter-event time distributions with various periods of $T$ (right). Individual users with the strength $s_i=200$ (a), 400 (b), 800 (c), 1600 (d), and 3197 (e) are analyzed. The original inter-event time distribution of the whole population is also plotted as a dashed curve for comparison.}
\label{fig:call_individual}\end{figure*}

\subsection{Individual de-seasoning}

First, we perform the de-seasoning analysis for individual users with various values of $T$. For some sample individuals in the MPC dataset, we obtain the original and the rescaled event rates. The event rates in case of $T=1$ day are depicted in the left column of Fig.~\ref{fig:call_individual}. The strengths of individual users are 200, 400, 800, 1600, and 3197. In most cases we find the characteristic circadian pattern, i.e. inactive nighttime and active daytime with one peak in the afternoon and another peak in the evening. The rescaled event rate successfully shows the expected de-seasoning effect, i.e. $\rho^*(t^*)=1$, except for weak fluctuations.

The rescaled inter-event time distributions up to $T=28$ days are compared with the original distributions in the right column of Fig.~\ref{fig:call_individual}. Note that the possible minimum value of rescaled inter-event time is $T/s_i$. We find that the rescaled inter-event time distributions still show the heavy tails. For the user with strength $200$, the burstiness decreases from the original value of $B_0\approx 0.202$ to value $B_7\approx 0.174$ (weekly pattern removed), then dropping  further to value $B_{28}\approx 0.104$ (i.e. monthly pattern removed). For the most active user with strength $s_i=3197$, the burstiness decays faster as $T$ increases: $B_0\approx 0.469$, $B_7\approx 0.254$, and $B_{28}\approx 0.219$. However, the values of $B$ are overall larger than those of the less active user. The results imply that de-seasoning the circadian and weekly patterns does not considerably affect the temporal burstiness patterns of individuals. Finally, in a limiting case of $T=T_f$, since $n_i(t)$ has the value of either 0 or 1, all $\tau^*$ are the same as $T/s_i$ in Eqs.~(\ref{eq:eventRate}) and (\ref{eq:rescaledTau}), leading to $B_{T_f}=-1$.

\begin{figure*}[!t]
\begin{tabular}{cc}
\includegraphics[width=.45\columnwidth]{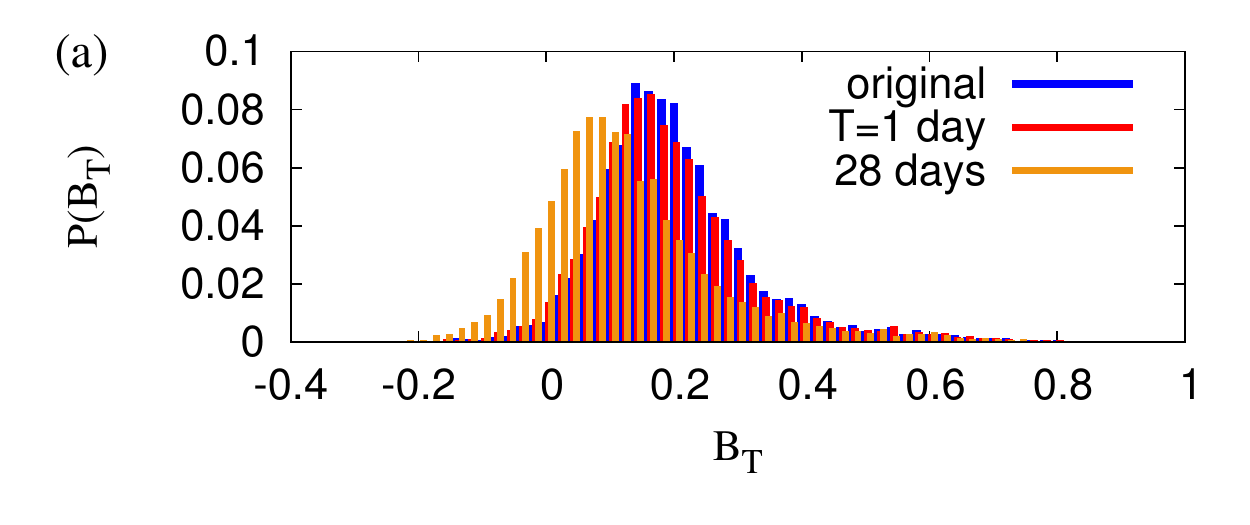}&
\includegraphics[width=.45\columnwidth]{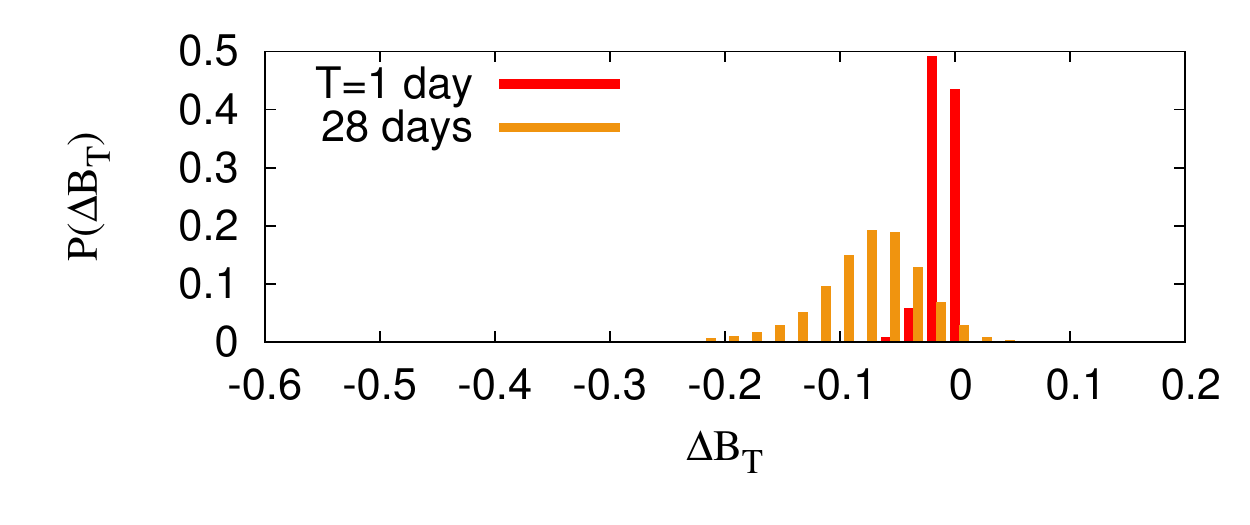}\\
\includegraphics[width=.45\columnwidth]{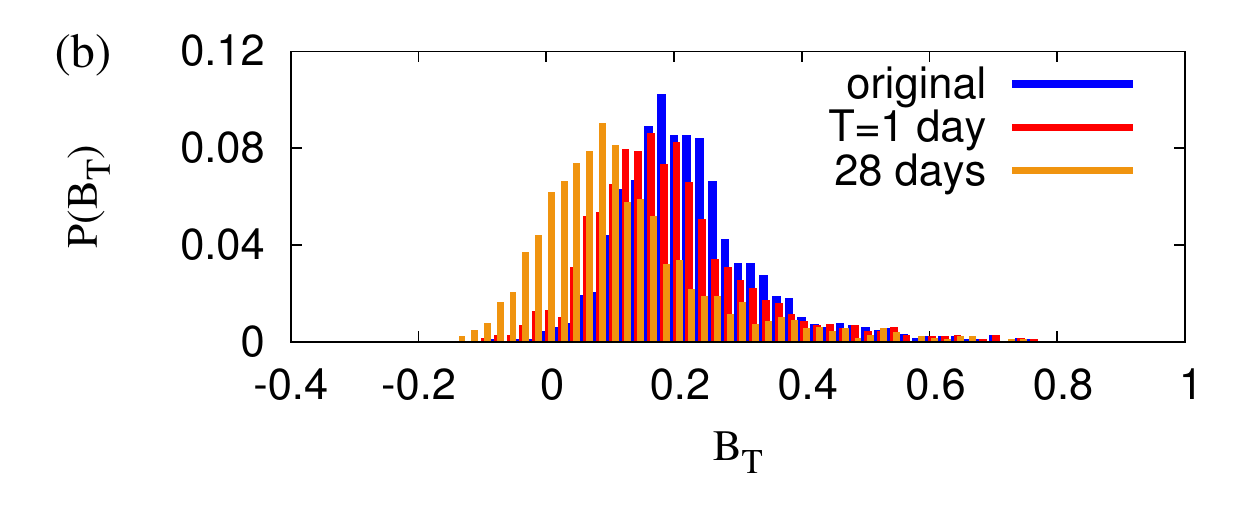}&
\includegraphics[width=.45\columnwidth]{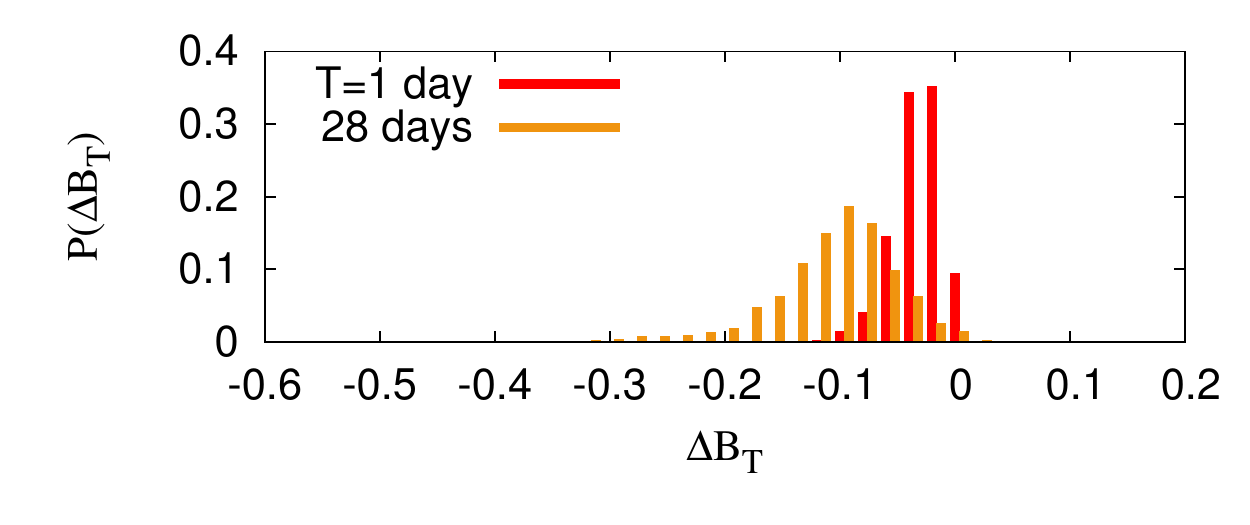}\\
\includegraphics[width=.45\columnwidth]{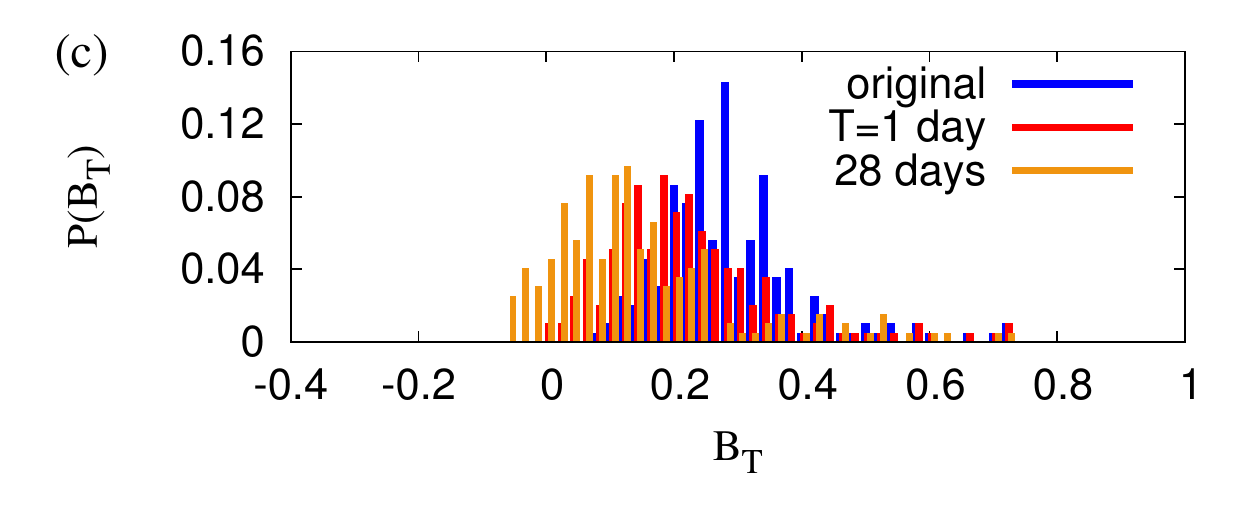}&
\includegraphics[width=.45\columnwidth]{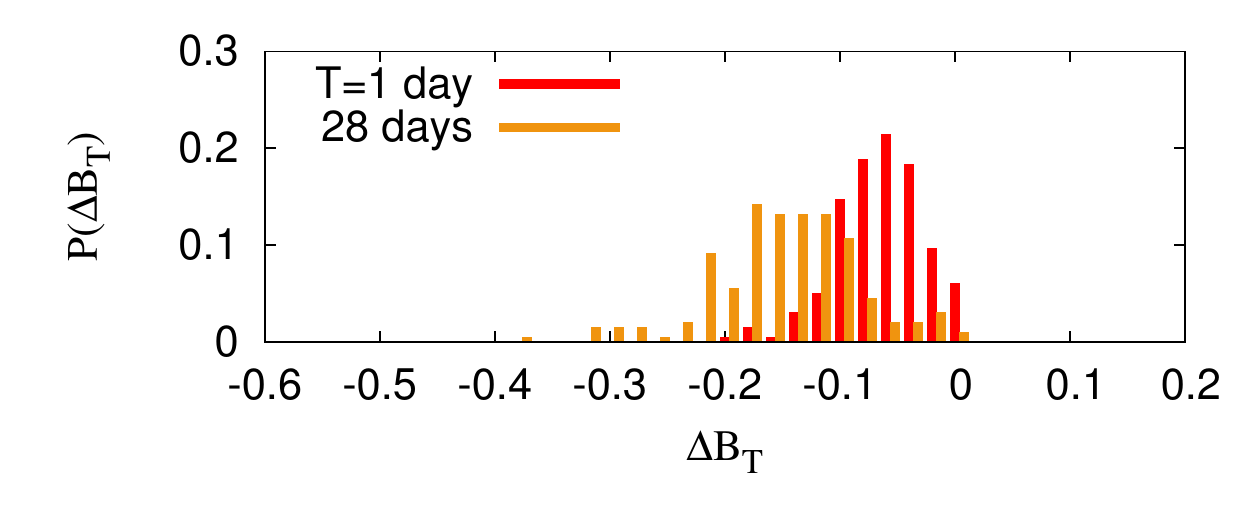}\\
\includegraphics[width=.45\columnwidth]{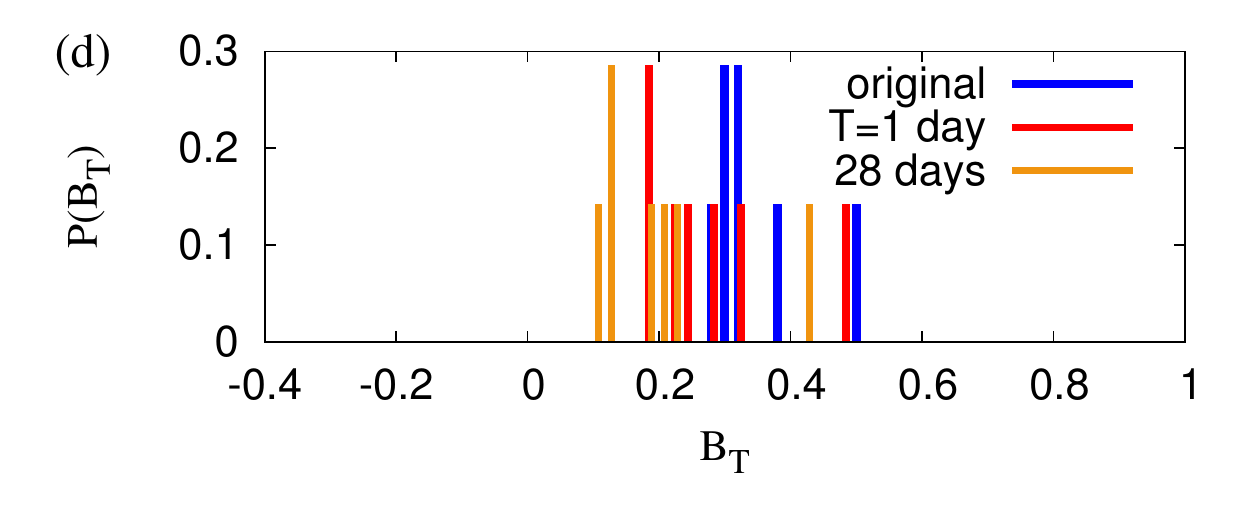}&
\includegraphics[width=.45\columnwidth]{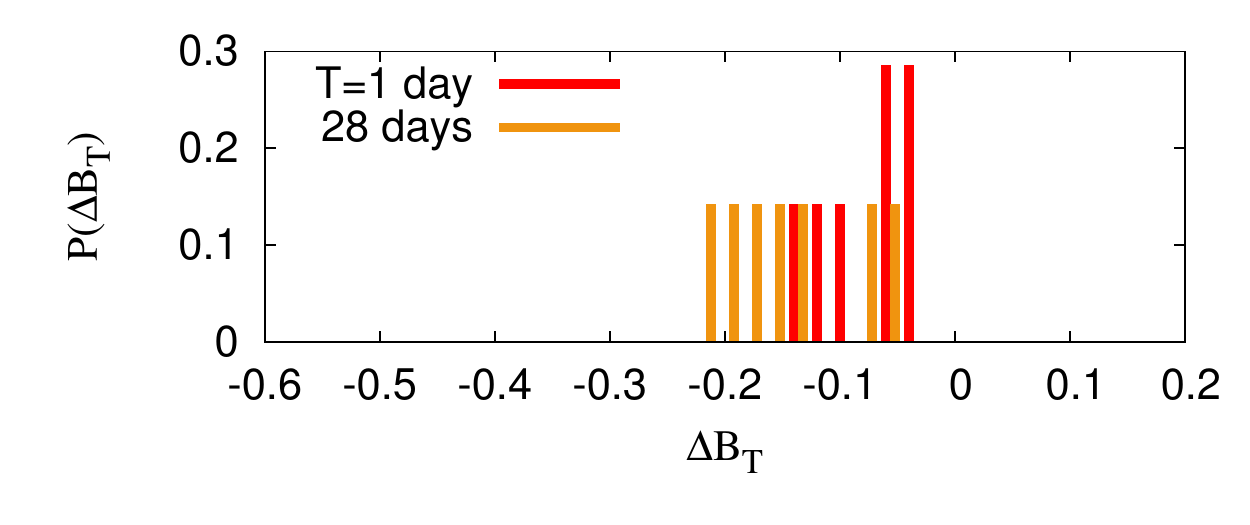}
\end{tabular}
\caption{Distributions $P(B_T)$ of the original and rescaled burstiness of invididual users with the same strength (left) and distributions $P(\Delta B_T)$ of the difference in burstiness, defined as $\Delta B_T=B_T-B_0$ (right). The individual users with the strengths $s_i=200$ (a), 400 (b), 800 (c), and 1600 (d) are analyzed. The numbers of users are correspondingly 6397, 1746, 196, and 7.}
\label{fig:call_strength_sep}\end{figure*}

\begin{figure*}[!t]
\begin{tabular}{ccc}
\includegraphics[width=.3\columnwidth]{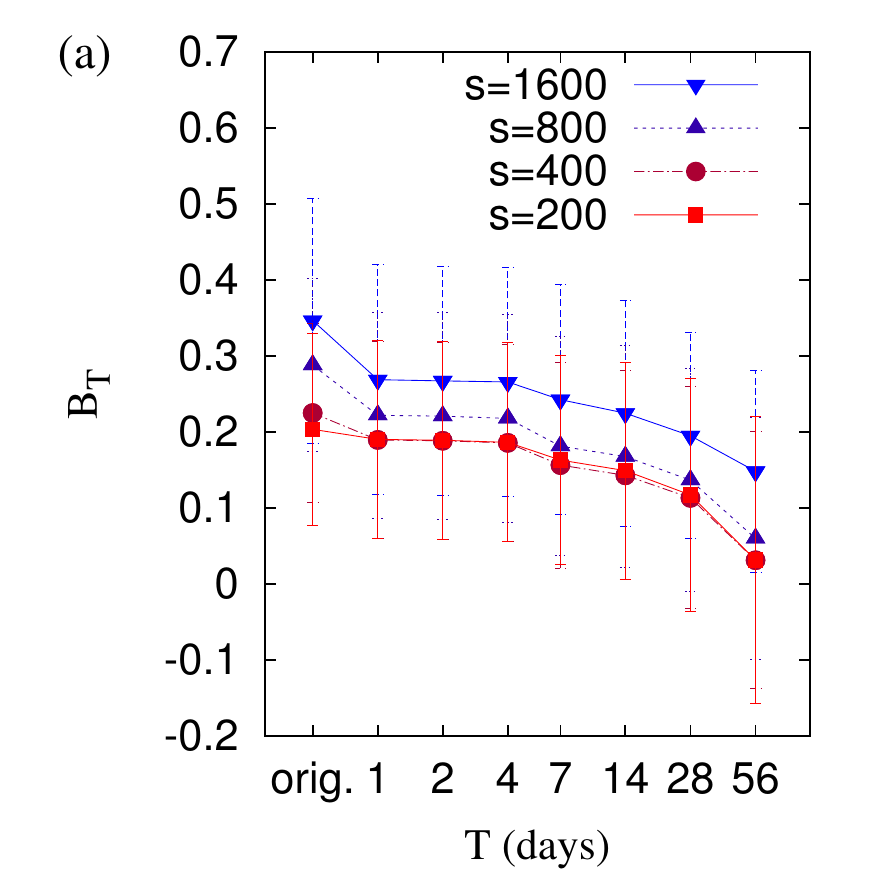}&
\includegraphics[width=.3\columnwidth]{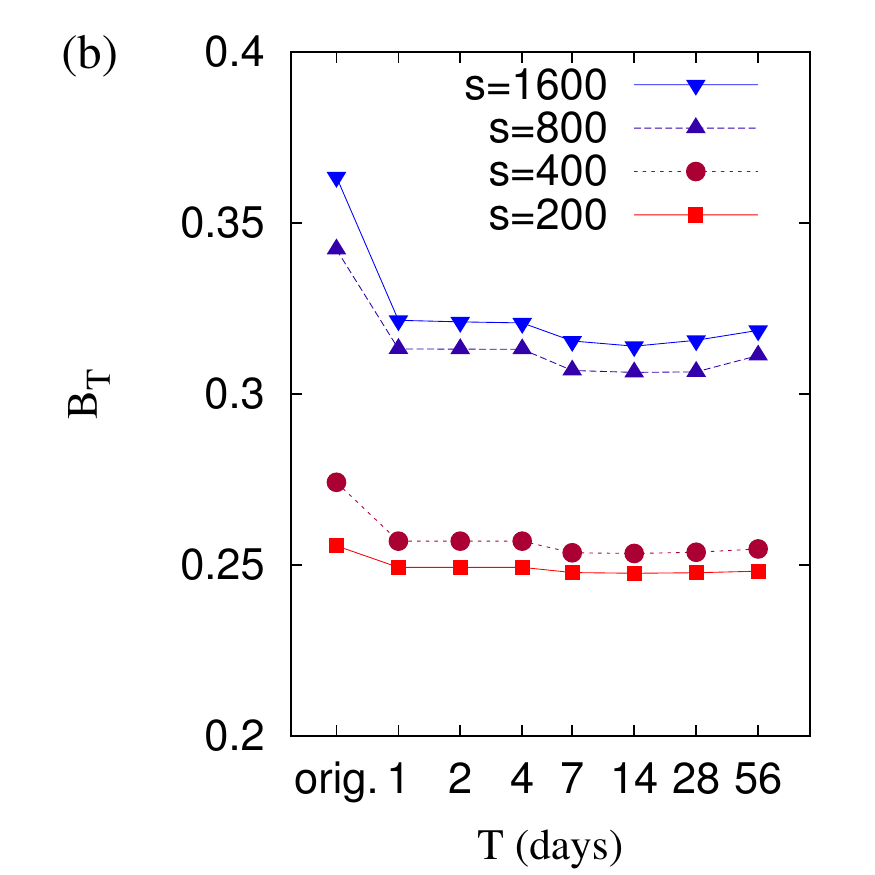}&
\includegraphics[width=.3\columnwidth]{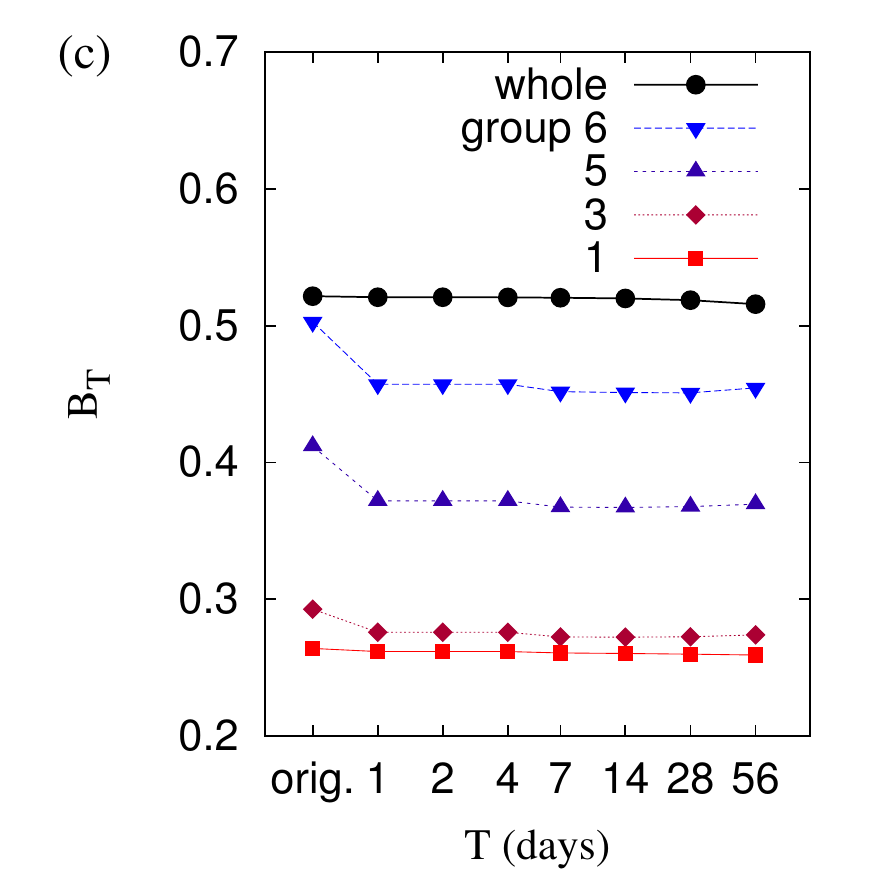}
\end{tabular}
\caption{Burstiness $B_T$ as a function of period of $T$: (a) the average and the standard deviation of $B_T$ obtained from the burstiness distribution in Fig.~\ref{fig:call_strength_sep}, (b) burstiness from groups with the same strength, and (c) burstiness from groups with broad ranges of strength.}
\label{fig:call_burstiness}\end{figure*}

Next, we obtain the distributions $P(B_T)$ of original and rescaled values of burstiness of individual users with the same strength and the distributions $P(\Delta B_T)$ of the difference in burstiness, defined as $\Delta B_T\equiv B_T-B_0$, see Fig.~\ref{fig:call_strength_sep}. The averages and the standard deviations of burstiness distributions for different periods of $T$ are plotted in Fig.~\ref{fig:call_burstiness}(a). The more active users have the larger values of burstiness, while the values of burstiness of the more active users decays faster (slower) than those of the less active users before (after) $T=7$ days. The overall behavior of the distributions shows that de-seasoning the circadian and weekly patterns does not destroy the bursty behavior of most individual users irrespective of their strengths. In addition, we find some exceptional users whose original values of burstiness are negative, indicating for more regular behavior than the Poisson process, and we also find a few individual users whose values of burstiness have grown as a result of de-seasoning, i.e. $\Delta B_T>0$.

\begin{figure*}[!ht]
\begin{tabular}{cc}
\includegraphics[width=.45\columnwidth]{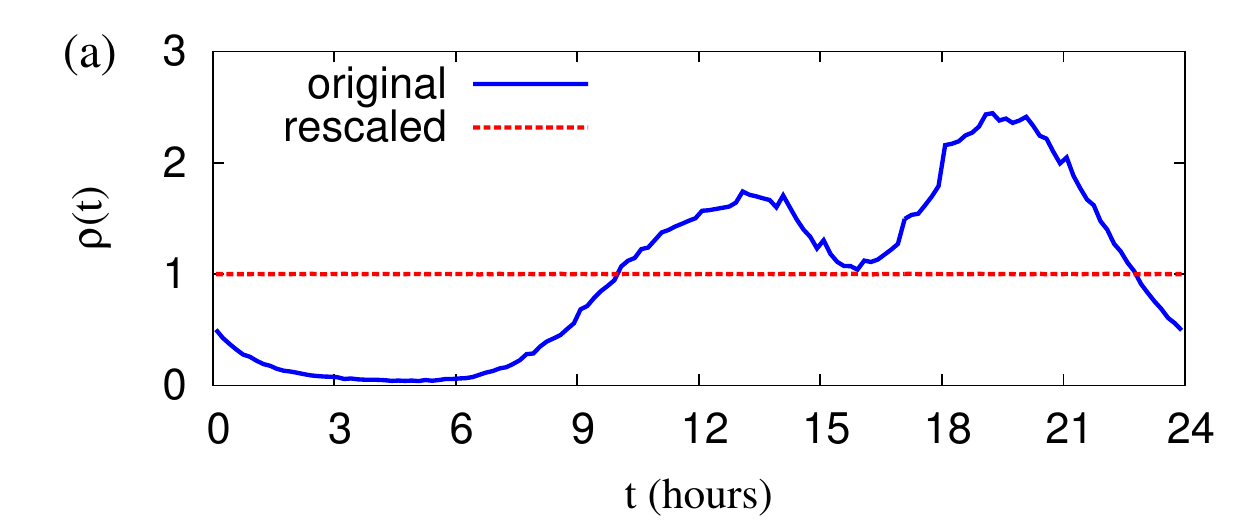}&
\includegraphics[width=.45\columnwidth]{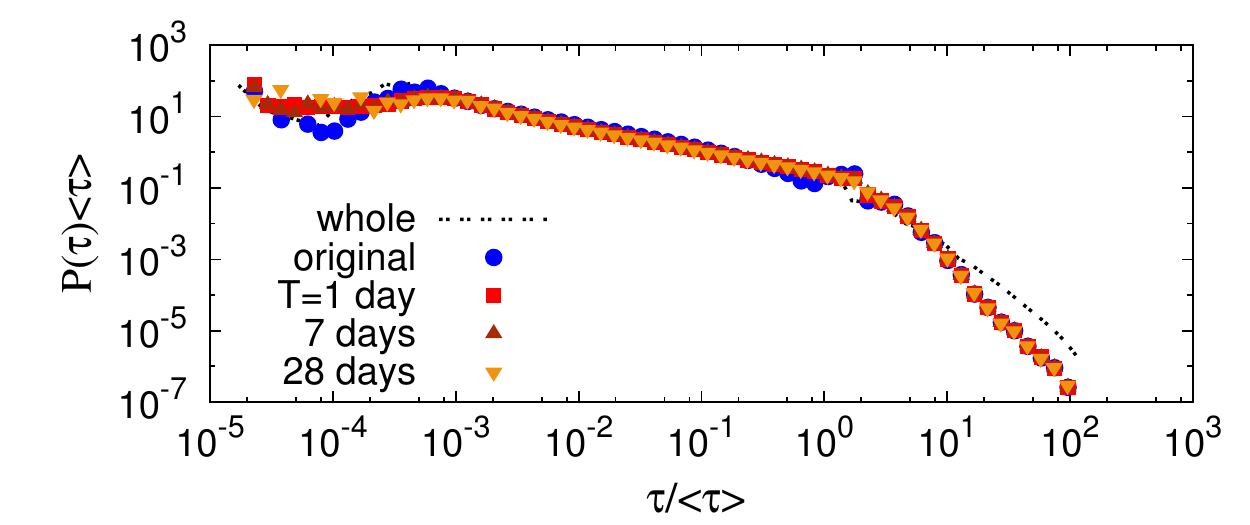}\\
\includegraphics[width=.45\columnwidth]{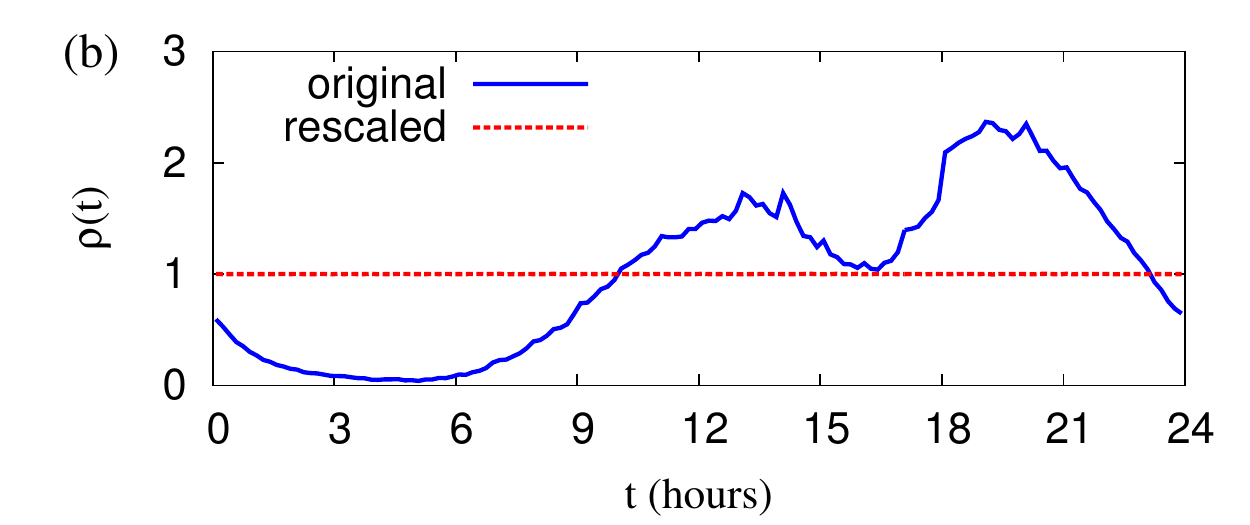}&
\includegraphics[width=.45\columnwidth]{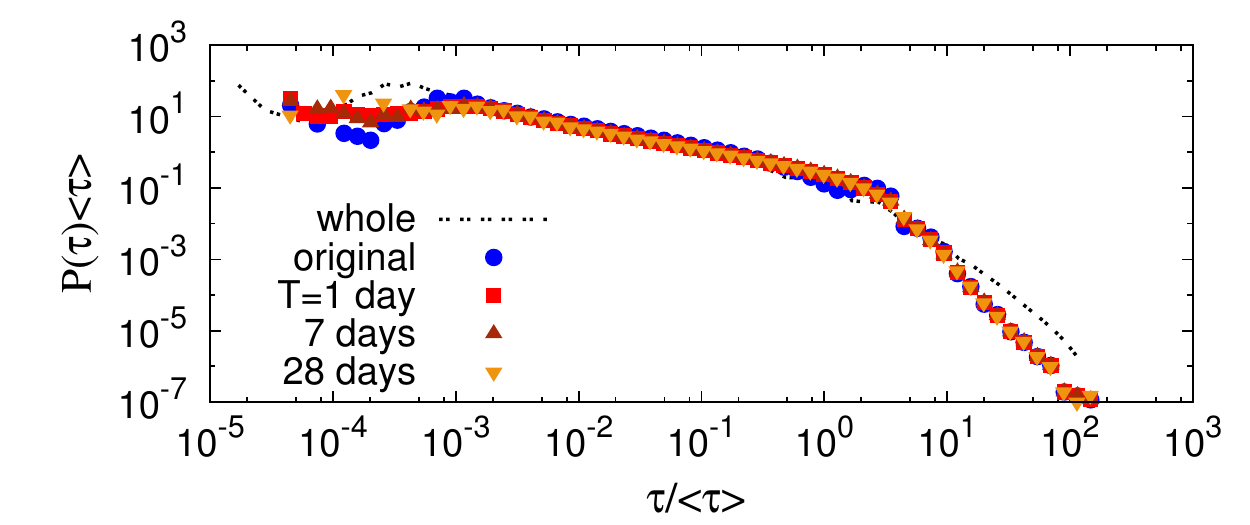}\\
\includegraphics[width=.45\columnwidth]{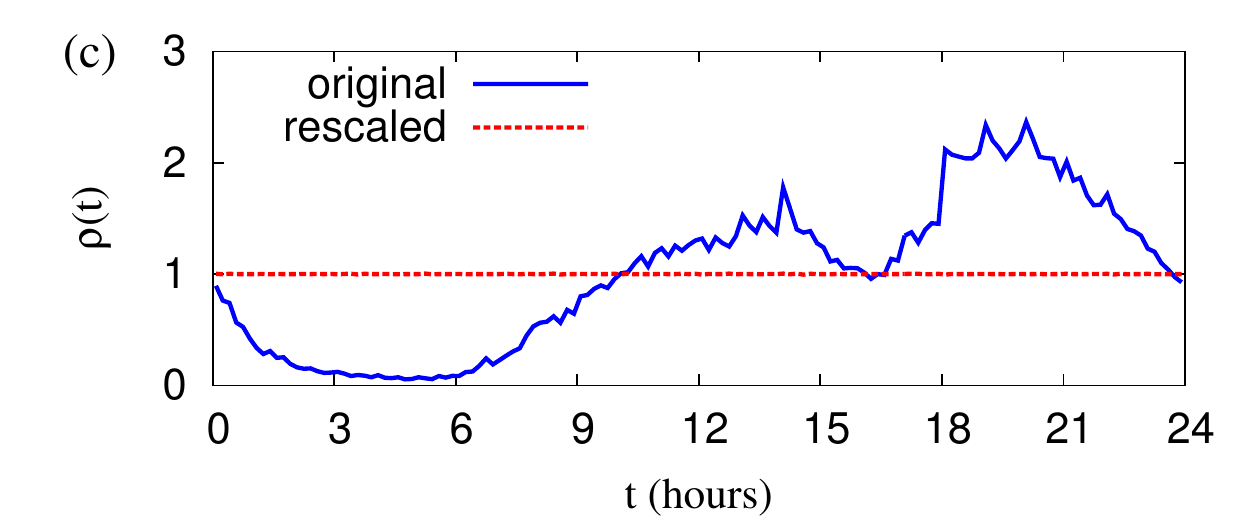}&
\includegraphics[width=.45\columnwidth]{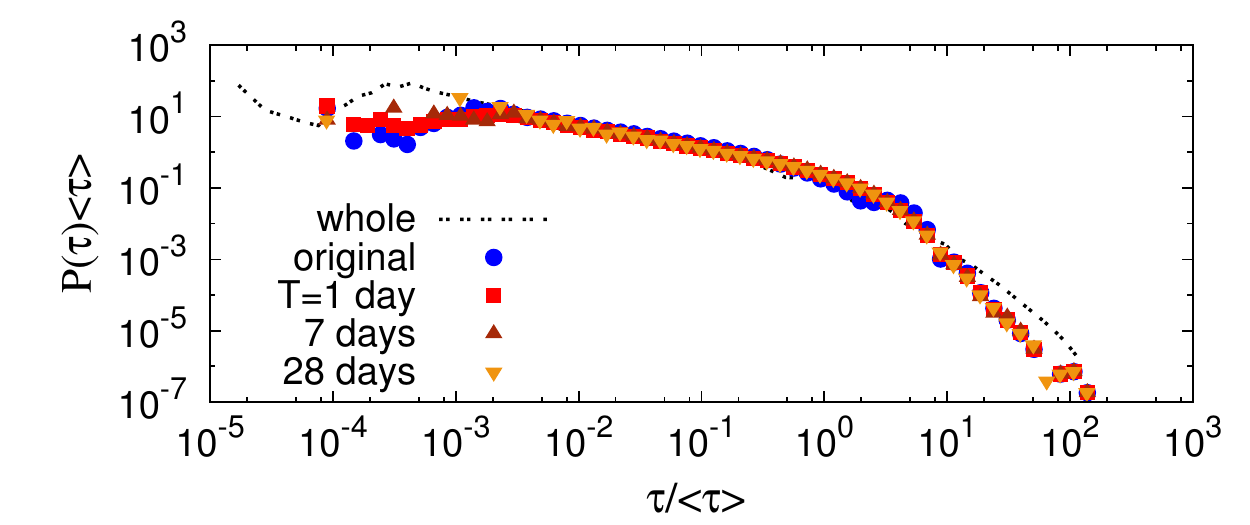}\\
\includegraphics[width=.45\columnwidth]{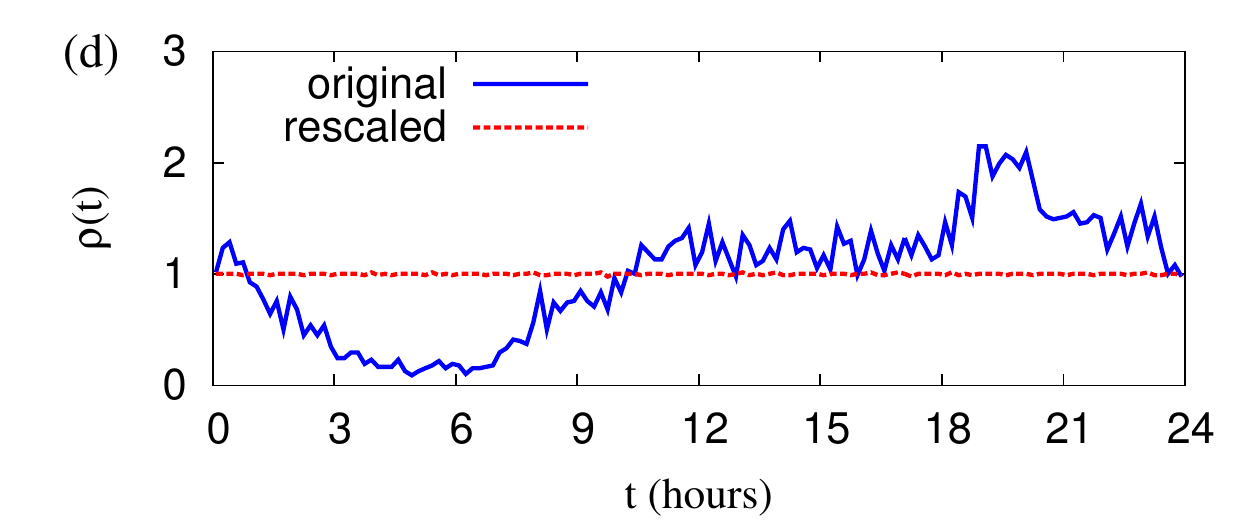}&
\includegraphics[width=.45\columnwidth]{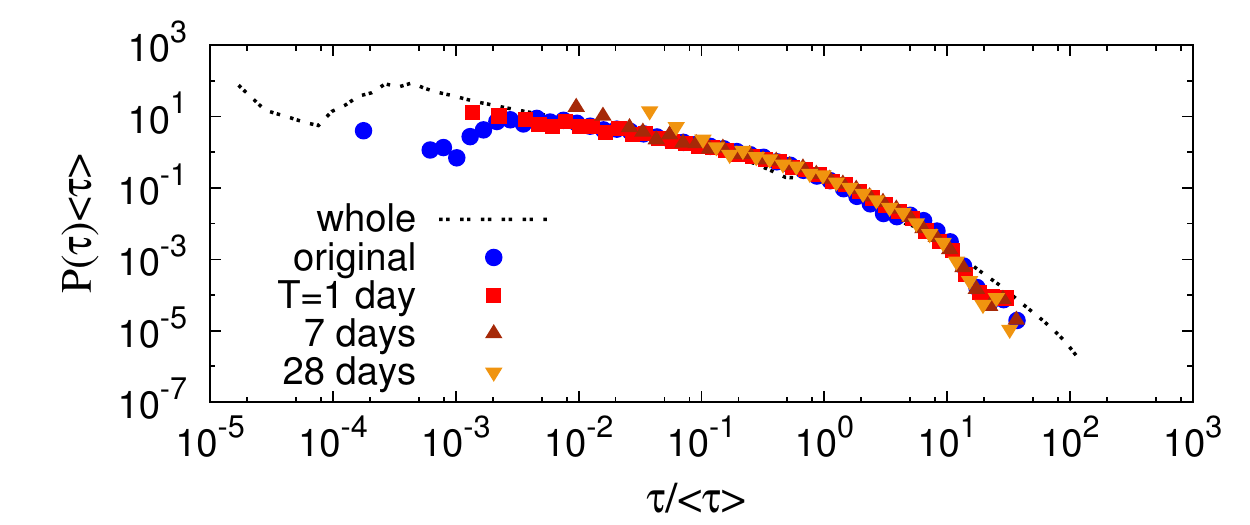}
\end{tabular}
\caption{De-seasoning of MPC patterns for groups with the same strengths: $s=200$ (a), 400 (b), 800 (c), and 1600 (d). The original and rescaled distributions of burstiness are plotted in Fig.~\ref{fig:call_burstiness}(b).}
\label{fig:call_strength_merge}\end{figure*}

\subsection{De-seasoning the groups of individuals with the same strength}

Here we analyze the group of individual users with the same strength, i.e. $\Lambda_s\equiv \{i|s_i=s\}$. The averaged event rate of a group is measured by merging individual event rates, precisely by obtaining $n_{\Lambda_s}(t)= \sum_{i\in\Lambda_s} n_i(t)$. Figure~\ref{fig:call_strength_merge} shows the original and the rescaled event rates with $T=1$ day (left) and the original and the rescaled inter-event time distributions with various periods of $T$ (right) for groups with strengths $s=200$, 400, 800, and 1600. The values of burstiness decrease only slightly as $T$ increases, but are smaller than those of the original burstiness, as shown in Fig.~\ref{fig:call_burstiness}(b). 

The burstiness of groups of individuals with the same strength is larger than the average values of individual burstiness from $P(B)$ of the same strength. For example, $B_0\approx 0.256$ for the group of strength $s=200$ turns out to be larger than $\int P(B_0)dB_0\approx 0.204$. Regarding to this difference, we would like to note that the de-seasoning of individual event times by means of the averaged event rate may cause systematic errors due to the different circadian and weekly patterns between the individual and the group. For resolving this issue, the various data clustering methods, such as self-organizing maps, can be used to classify users' activity patterns beyond their strengths and then perform the de-seasoning separately for the different groups.

\subsection{De-seasoning the groups of individuals with broad ranges of strength}

\begin{figure*}[!ht]
\begin{tabular}{cc}
\includegraphics[width=.45\columnwidth]{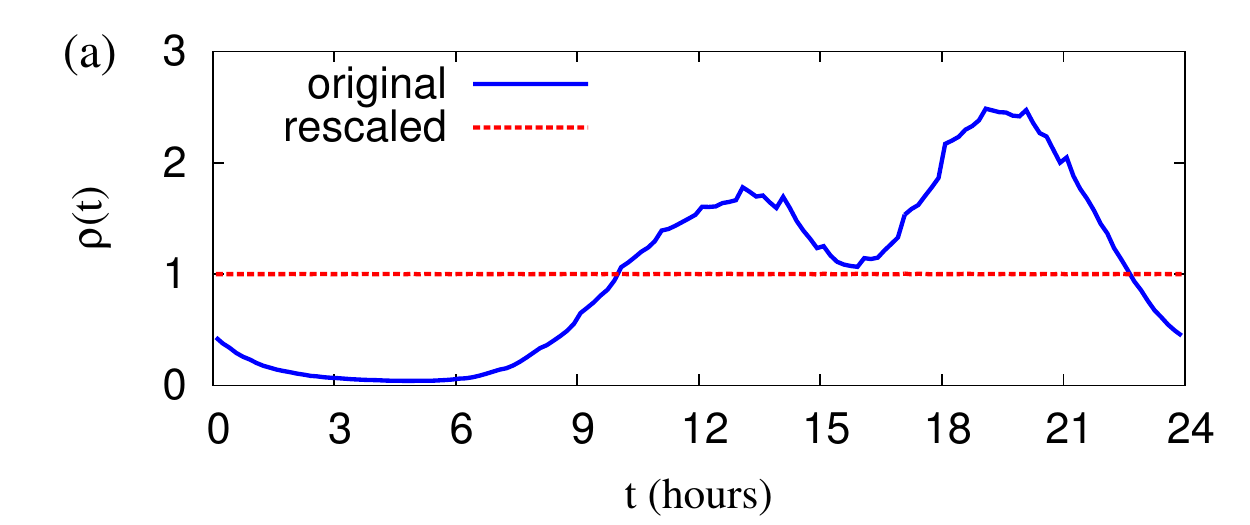}&
\includegraphics[width=.45\columnwidth]{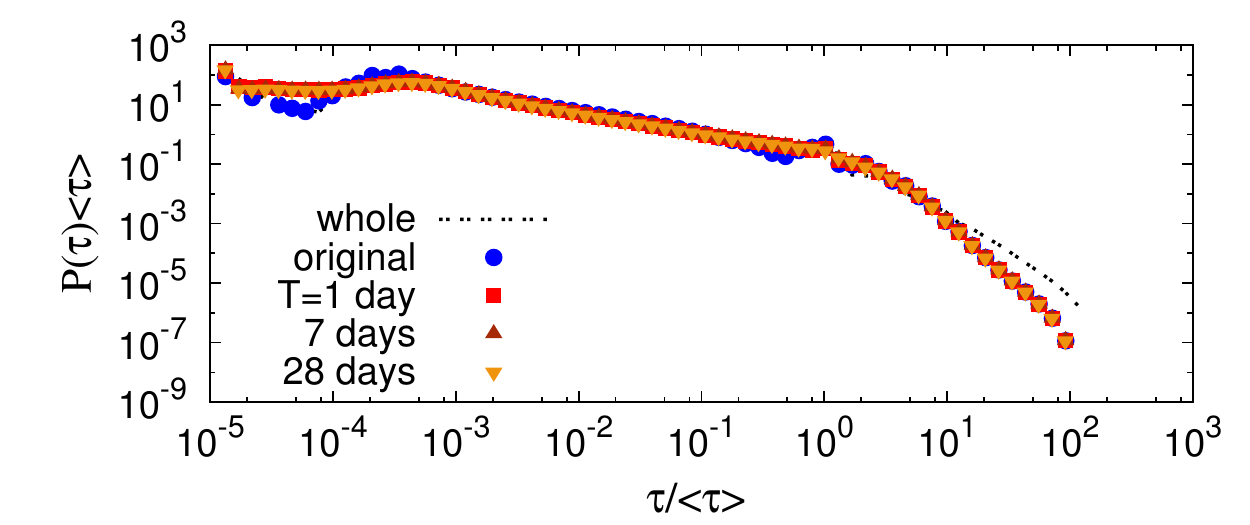}\\
\includegraphics[width=.45\columnwidth]{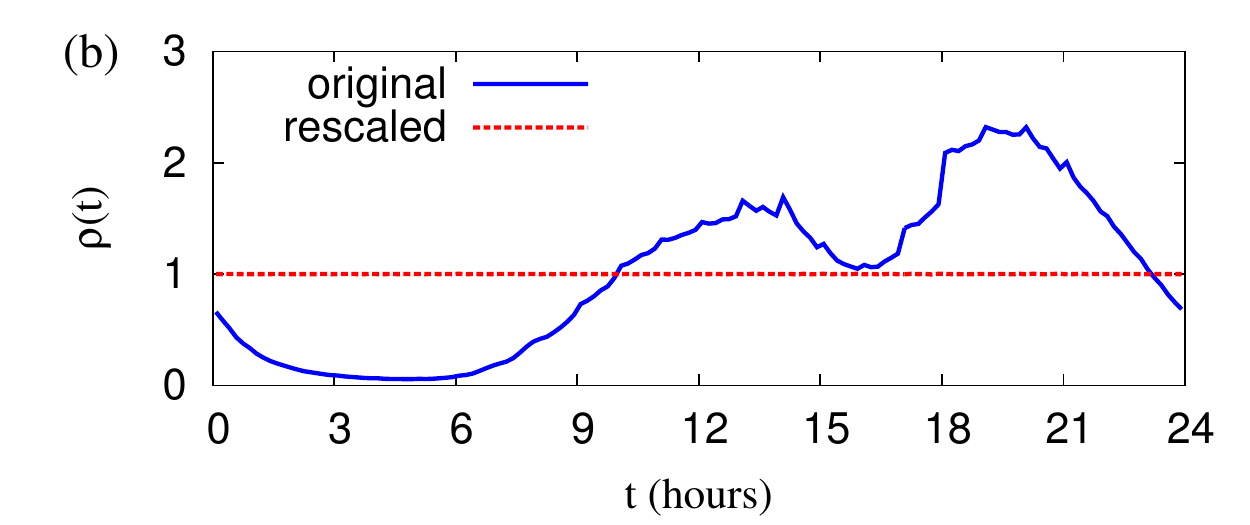}&
\includegraphics[width=.45\columnwidth]{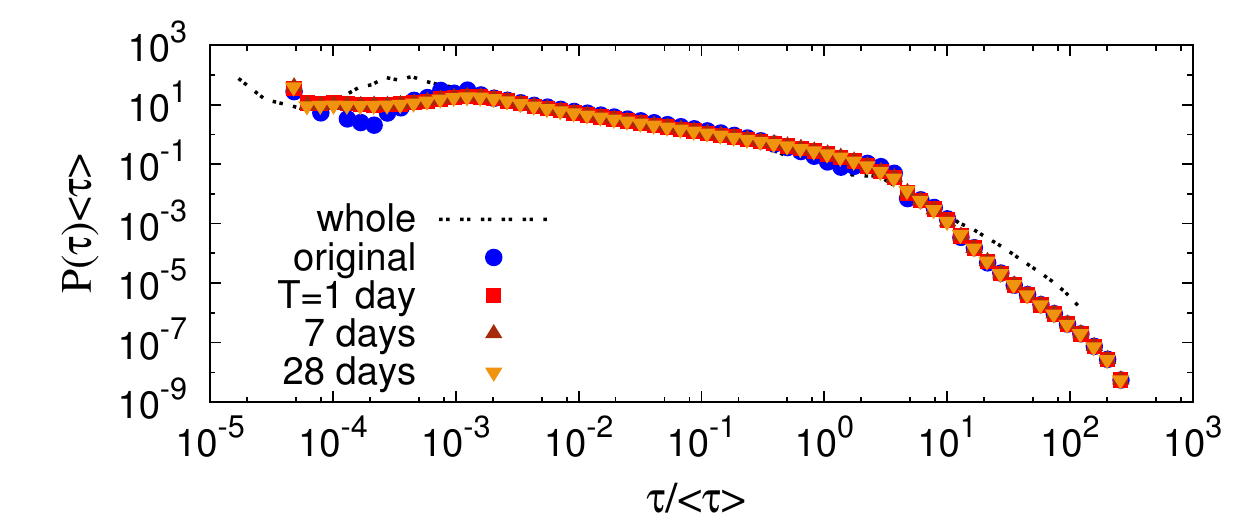}\\
\includegraphics[width=.45\columnwidth]{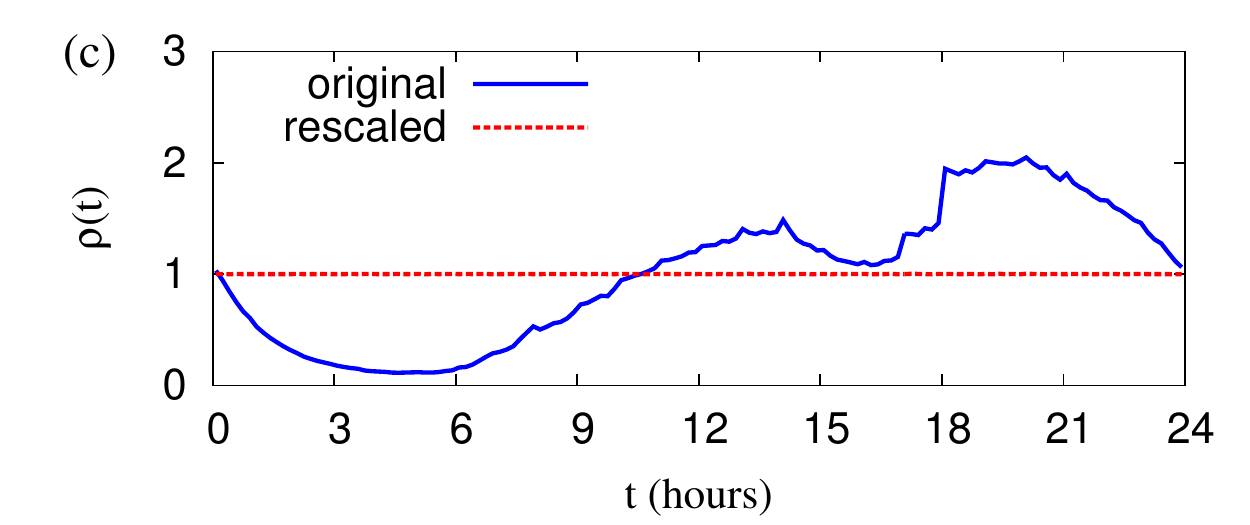}&
\includegraphics[width=.45\columnwidth]{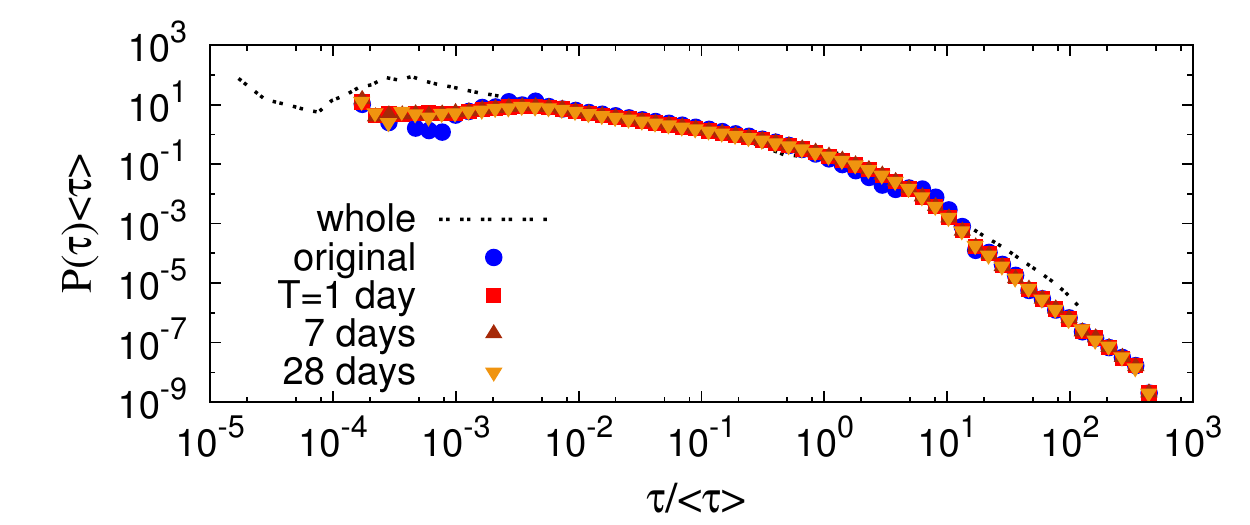}\\
\includegraphics[width=.45\columnwidth]{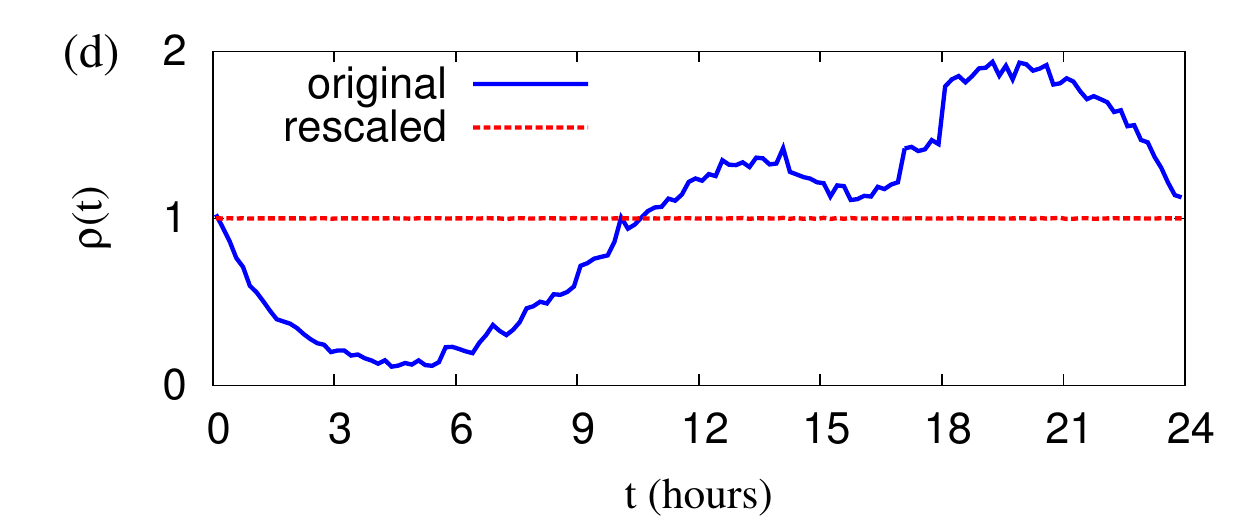}&
\includegraphics[width=.45\columnwidth]{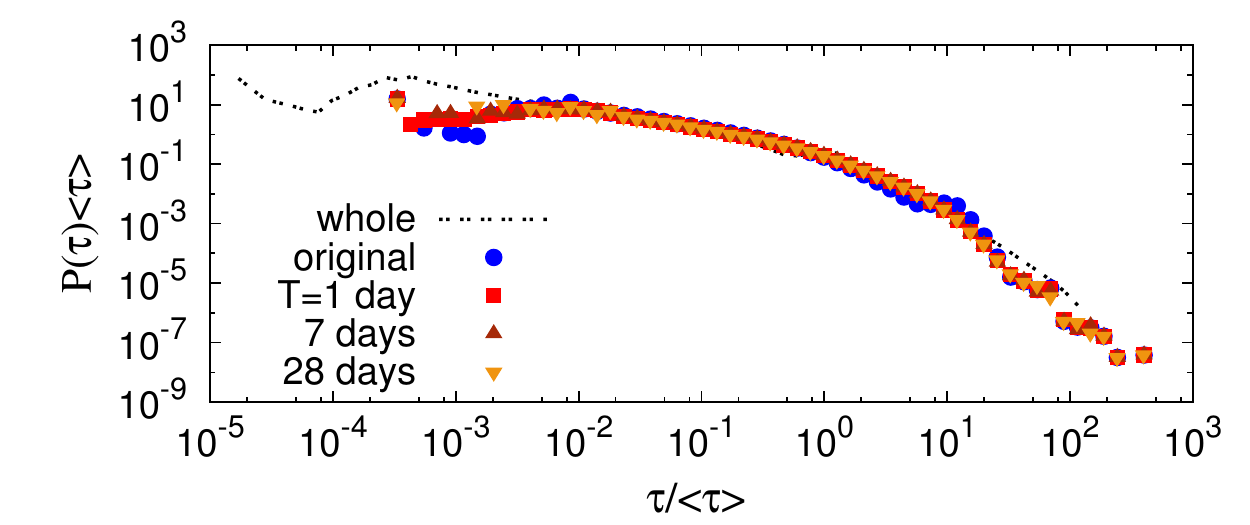}\\
\includegraphics[width=.45\columnwidth]{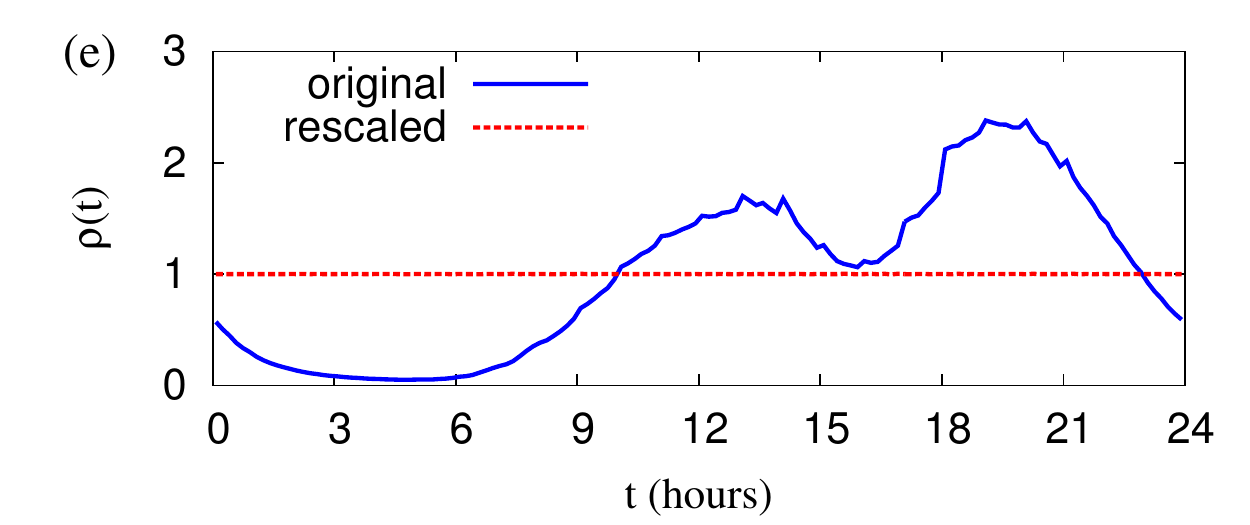}&
\includegraphics[width=.45\columnwidth]{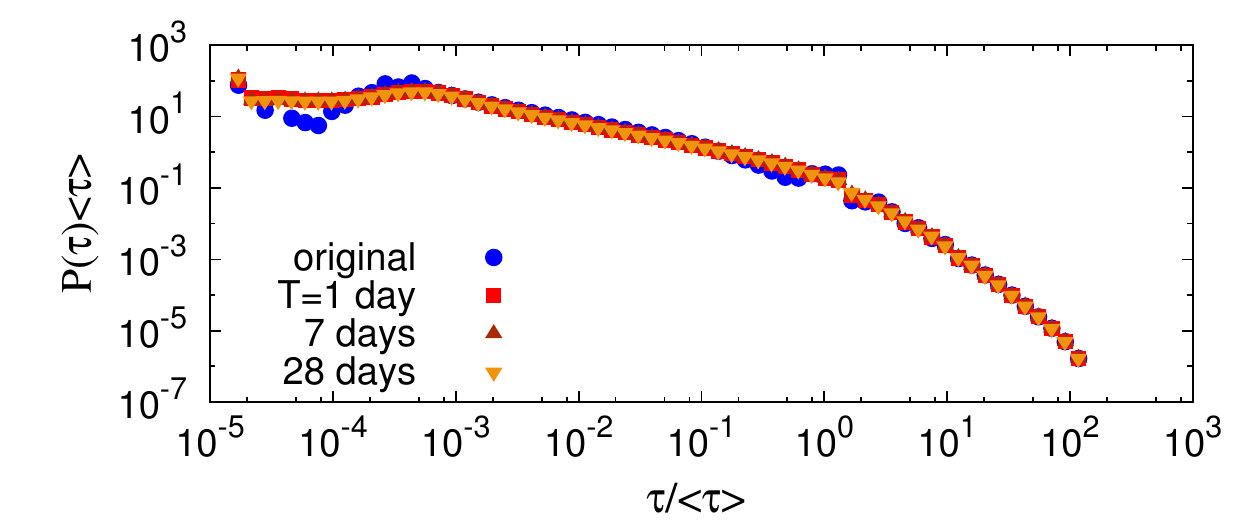}
\end{tabular}
\caption{De-seasoning of the MPC patterns for groups of individuals with broad ranges of strength: groups 1 (a), 3 (b), 5 (c), 6 (d), and the whole population (e). For the details of groups, see Table~\ref{table:call_group}. The original and rescaled distributions of burstiness are plotted in Fig.~\ref{fig:call_burstiness}(c).}
\label{fig:call_group}\end{figure*}

For the larger scale analysis, we consider the strength dependent grouping of users, i.e. groups of individual users with a broad range of strengths, denoted by $\Lambda_{m_1,m_2}\equiv \{i| m_1\leq s_i<m_2\}$, as similarly done in~\cite{Zhou2008,Radicchi2009}. The values of $m$s are determined in terms of the ratio to the maximum strength $s_{\rm max}=7911$, see Table~\ref{table:call_group} for the details of the groups. We determine the averaged event rates of the groups and some of them are shown in the left column of Fig.~\ref{fig:call_group}. By means of the event rates, we perform the de-seasoning to get the rescaled inter-event time distributions, see the right column of Fig.~\ref{fig:call_group}. It is found that the values of burstiness initially decrease slightly and then stay constant at relatively large values as $T$ increases, shown in Fig.~\ref{fig:call_burstiness}(c). These results again confirm our conclusions that de-seasoning the circadian and weekly patterns does not wipe out the bursty behavior of human communication patterns.

\begin{table}[!h]
    \caption{Strength dependent grouping of individuals in the MPC dataset. For each group, the range of strength, also in terms of the ratio to the maximum strength $s_{\rm max}=7911$, the number of users, and its fraction to the whole population except for the user with $s_{\rm max}$ are summarized.}
\label{table:call_group}
\begin{indented}
\item[]\begin{tabular}{rrr}
\hline
group index & strength range ($\%$) & the number of users ($\%$)\\ 
\hline
0 & 0-79 (0-1) & 2821103 (54.4)\\
1 & 79-158 (1-2) & 1010923 (19.5)\\
2 & 158-316 (2-4) & 843412 (16.3)\\
3 & 316-632 (4-8) & 418460 (8.1)\\
4 & 632-1265 (8-16) & 89718 (1.7)\\
5 & 1265-2531 (16-32) & 5857 (0.1)\\
6 & 2531-5063 (32-64) & 173 (0.003)\\
7 & 5063-7594 (64-96) & 5 (0.0001)\\
whole & 0-7594 (0-96) & 5189651 (100)\\
\hline
\end{tabular}
\end{indented}
\end{table}

\begin{figure*}[!h]
\begin{tabular}{cc}
\includegraphics[width=.45\columnwidth]{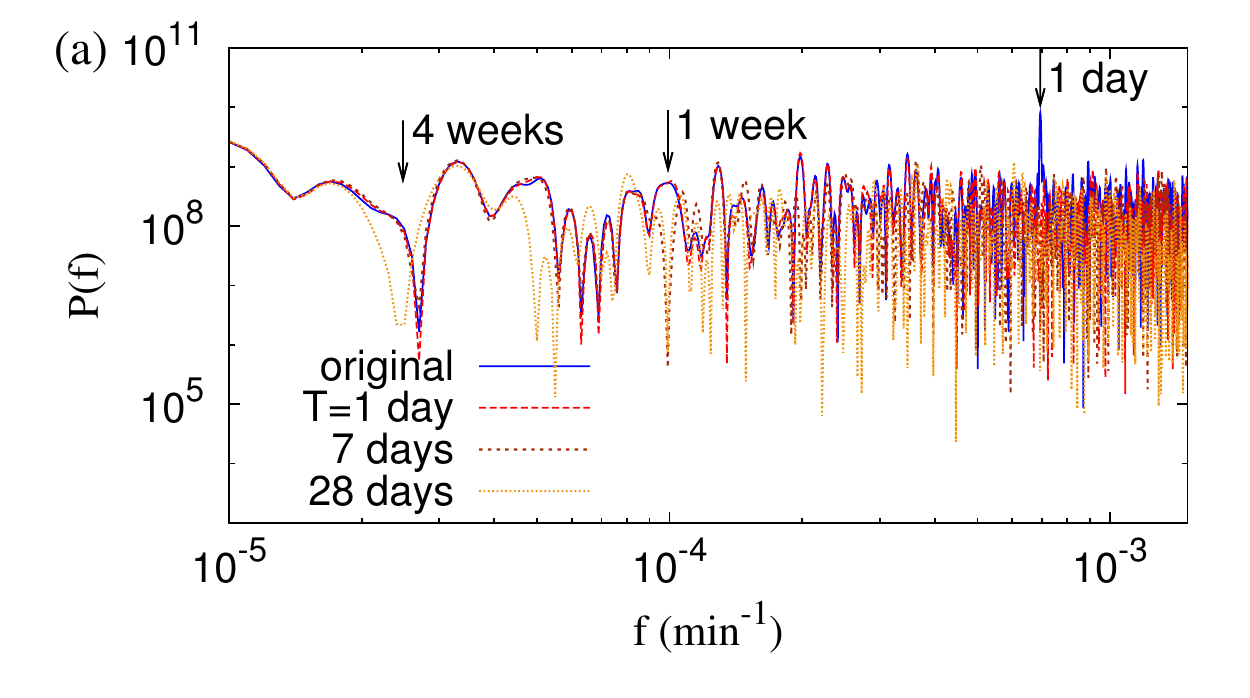}&
\includegraphics[width=.45\columnwidth]{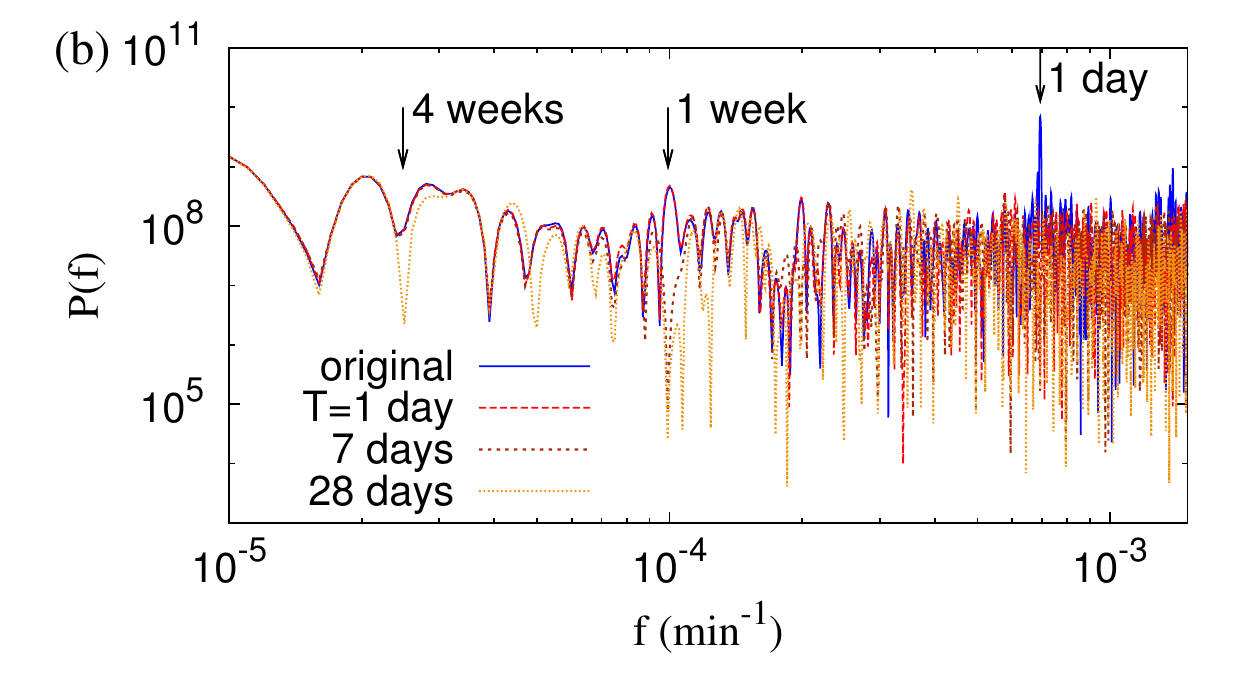}\\
\includegraphics[width=.45\columnwidth]{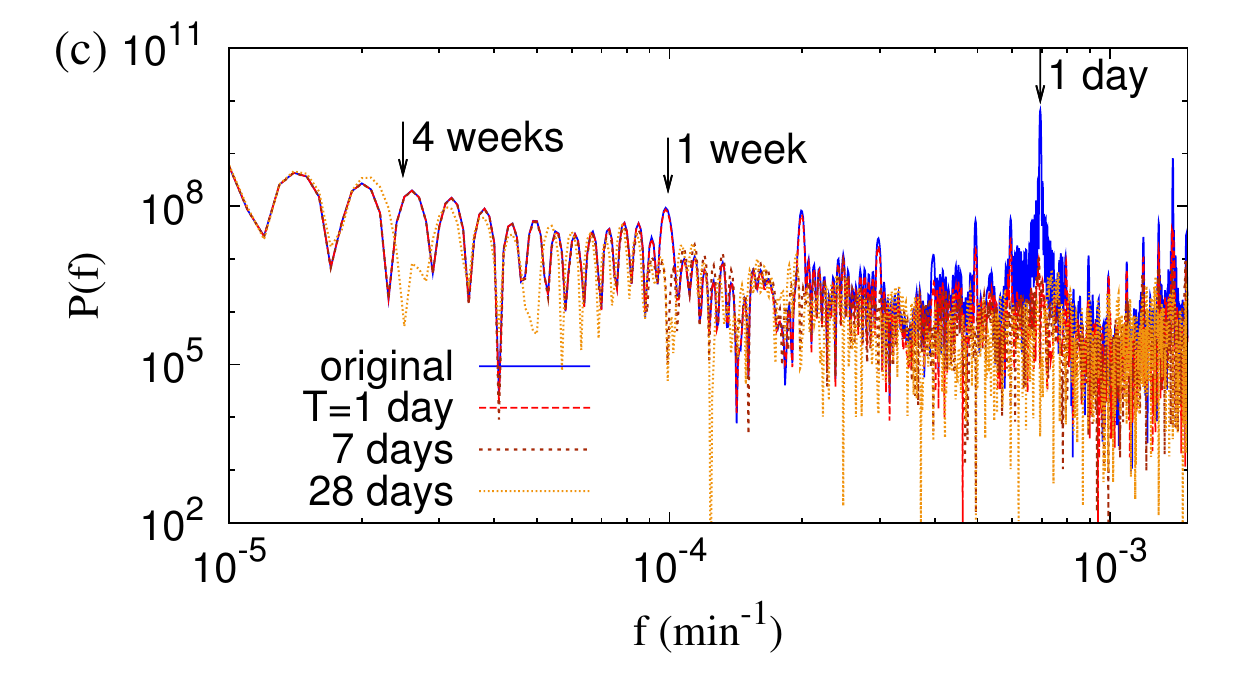}&
\includegraphics[width=.45\columnwidth]{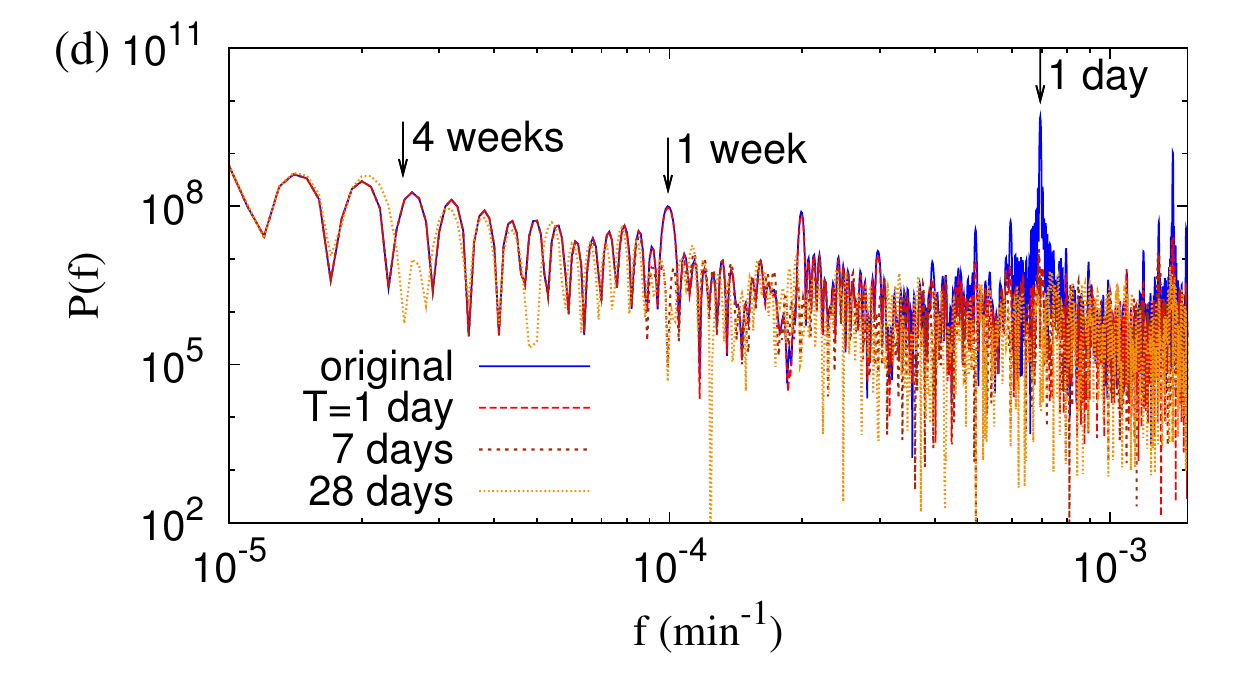}\\
\includegraphics[width=.45\columnwidth]{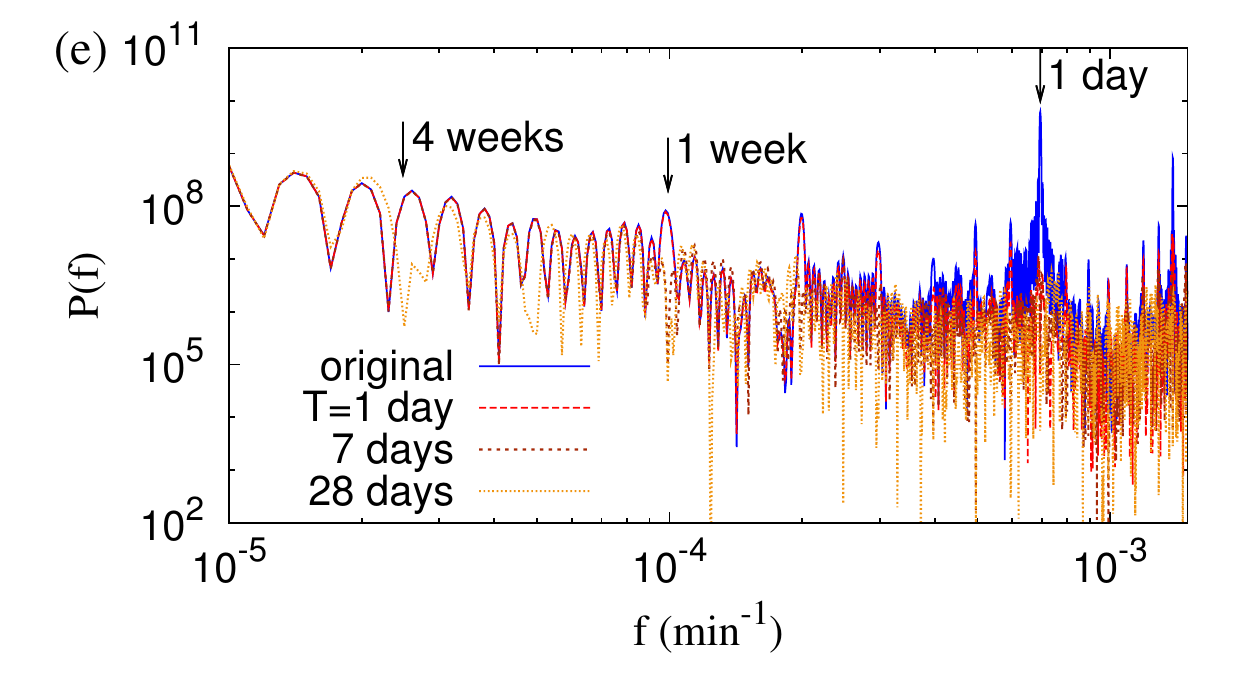}
\end{tabular}
\caption{Power spectra, $P(f)$, of the original and the rescaled event rates with various periods of $T$ for individual users with strengths 200 (a) and 800 (b), for groups with the same strengths of 200 (c) and 800 (d), and for the whole population (e). The circadian and weekly peaks in the original power spectra are successfully removed by the de-seasoning in various ways.}
\label{fig:call_powerSpec}\end{figure*}

\subsection{Power spectra analysis}

In order to see clearly the effect of de-seasoning on the event rates, we compare the power spectra of the rescaled event rates to the original event rates. The power spectrum of the event rate is defined as
\begin{equation}
    P(f)=\left|\sum_{t=0}^{T_f}\rho_{\Lambda,T}(t)e^{2\pi ift}\right|^2,
\end{equation}
where $f$ denotes the frequency. In Fig.~\ref{fig:call_powerSpec} we show the results of this comparison, where it is evident that with our de-seasoning methods, the circadian and weekly peaks of the original power spectrum are successfully removed and do not show in the rescaled spectra. In all the cases, ranging from the individual de-seasoning to the whole population de-seasoning, when $T=1$ day, the circadian peak at $1/f=1$ day is removed while the others, i.e. the weekly and monthly peaks remain. For $T=7$ days, the weekly peak at $1/f=7$ days is removed, and so on. This behavior can be understood because the cyclic patterns longer than $T$ will not be affected by de-seasoning with the period $T$.

\section{Summary}\label{sect:summary}

The heavy tails and burstiness of inter-event time distributions in human communication activity are affected by circadian and weekly cycles as well as by correlations rooted in human task execution. To investigate the existence of correlations rooted in human task execution we devised a systematic method to de-season circadian and weekly cycles appearing in human activity and successfully demonstrated their removal. Here the circadian and weekly patterns extracted from the mobile phone call and Short Messages records are used to rescale the timings of events, i.e. the time is dilated or contracted during high or low call or SM activity of individual service users, respectively.

We have found that after de-seasoning circadian and weekly cycles driven inhomogeneities, the heavy tails and burstiness of inter-event time distributions still remain. Hence our results imply that the heavy tails and burstiness are not only the consequence of circadian and weekly cycles but also due to correlations in human task execution. In addition, we calculated the Fano and Allan factors~\cite{Anteneodo2010} for the de-seasoned mobile phone communication data, which yielded further evidence of the remaining burstiness in both mobile phone call and Short Messages activities of human communication. Although beyond the scope of the current study, as the next step one needs to focus on describing the mechanisms of the remaining burstiness in human communication patterns, served well by building models and analyzing their results.    

%\appendix
\section*{Appendix}\label{appendix}

We also investigated the effect of circadian and weekly cycles on the heavy-tailed inter-event time distribution by using Short Message (SM) dataset from the same European operator mentioned in the main text. We have only retained links with bidirectional interaction, yielding $N=4.2\times 10^6$ users, $L=8.5\times 10^6$ links, and $C=114\times 10^6$ events (SMs). We have merged some consecutive SMs sent by one user to another within $10$ seconds into one SM event because one longer message can be divided into many short messages due to the length limit of a single SM ($160$ characters)~\cite{Kovanen2009}.

\begin{figure*}[!h]
\begin{tabular}{cc}
\includegraphics[width=.45\columnwidth]{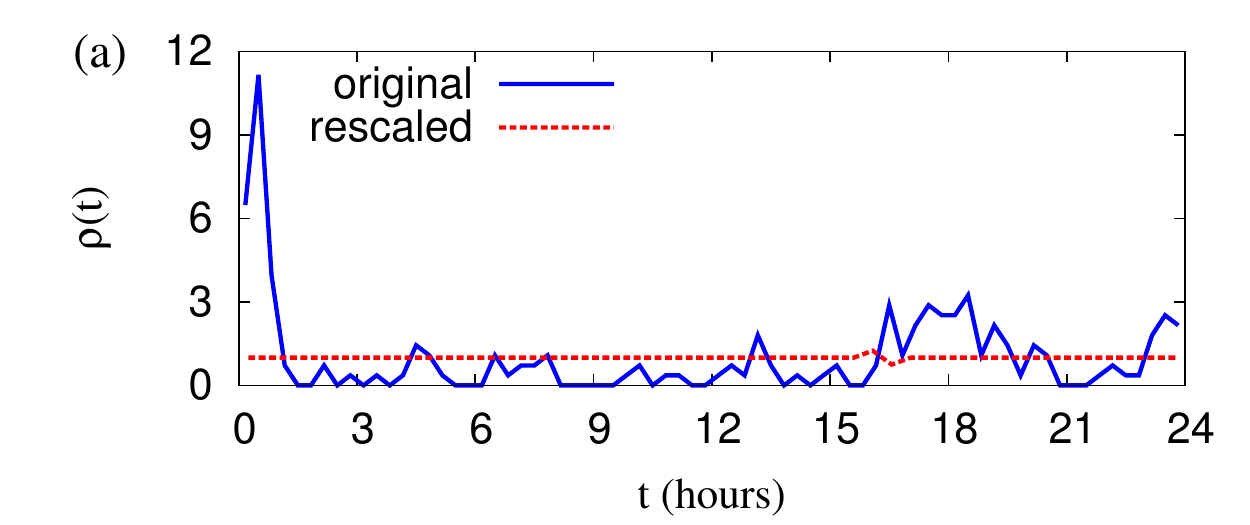}&
\includegraphics[width=.45\columnwidth]{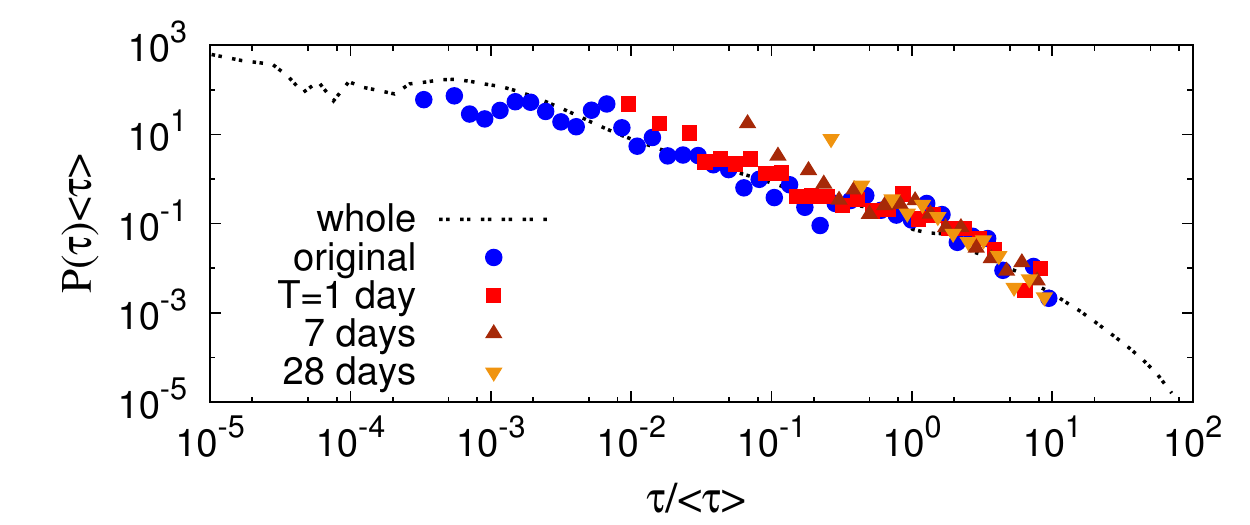}\\
\includegraphics[width=.45\columnwidth]{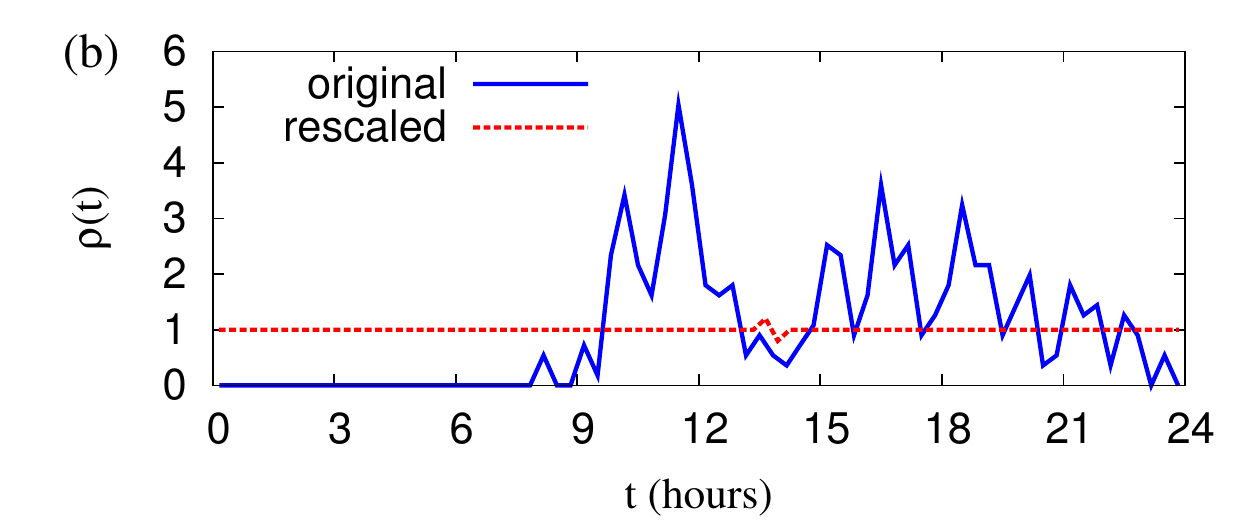}&
\includegraphics[width=.45\columnwidth]{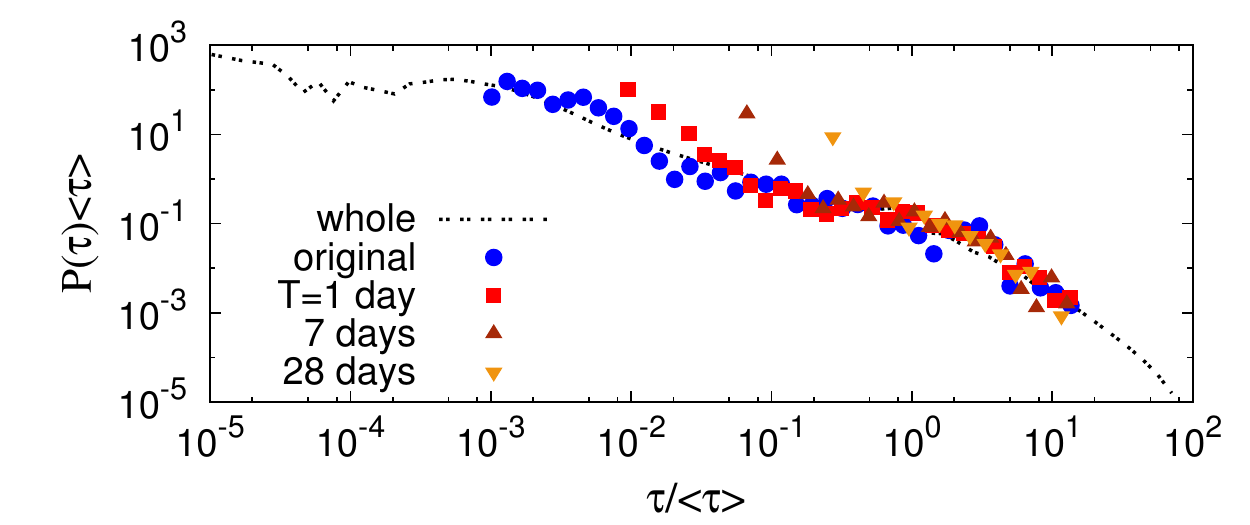}\\
\includegraphics[width=.45\columnwidth]{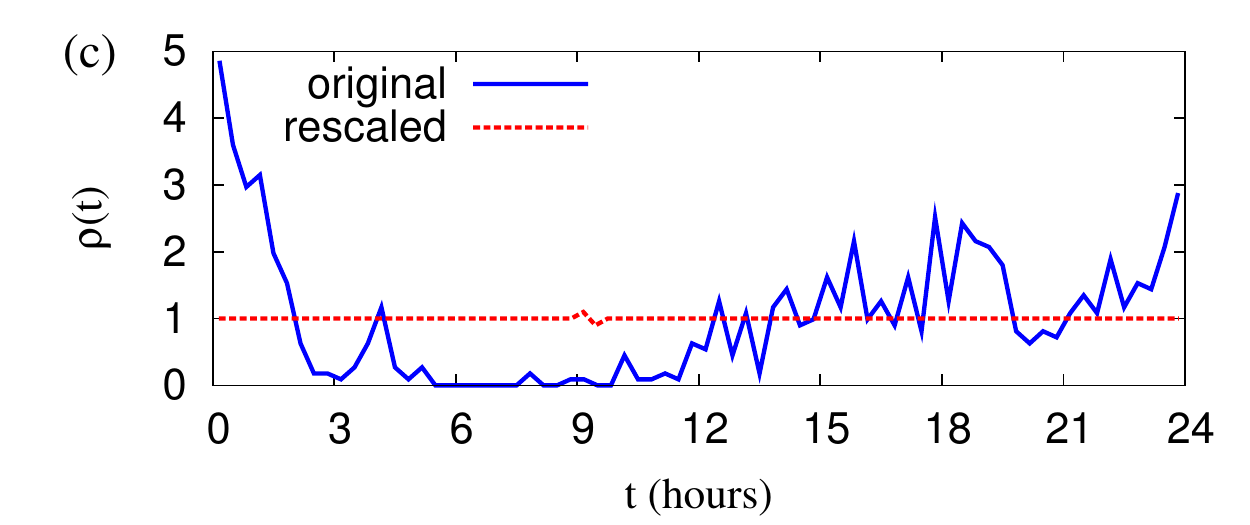}&
\includegraphics[width=.45\columnwidth]{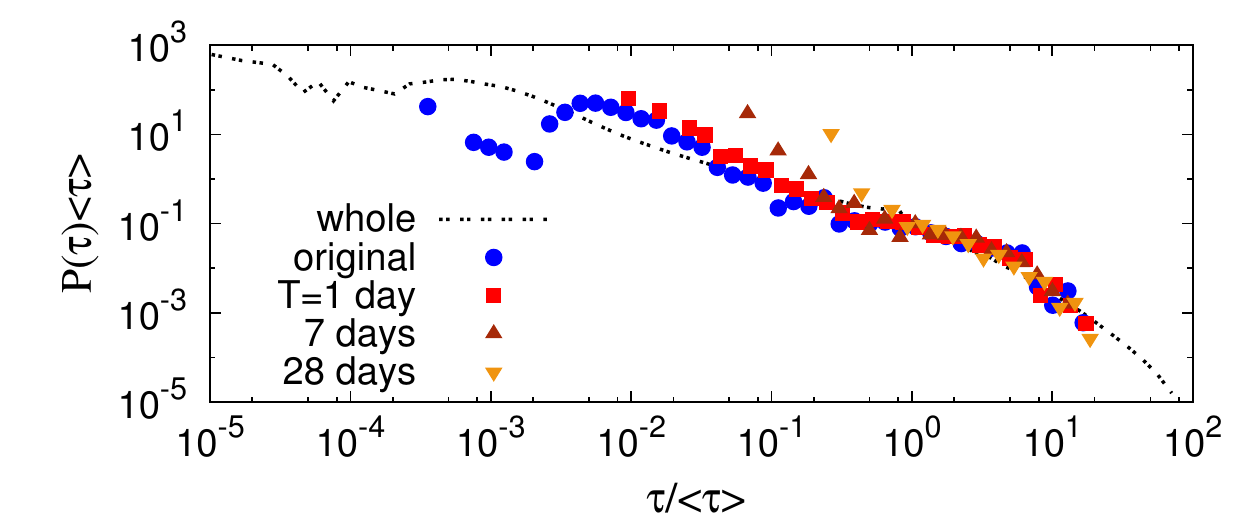}\\
\includegraphics[width=.45\columnwidth]{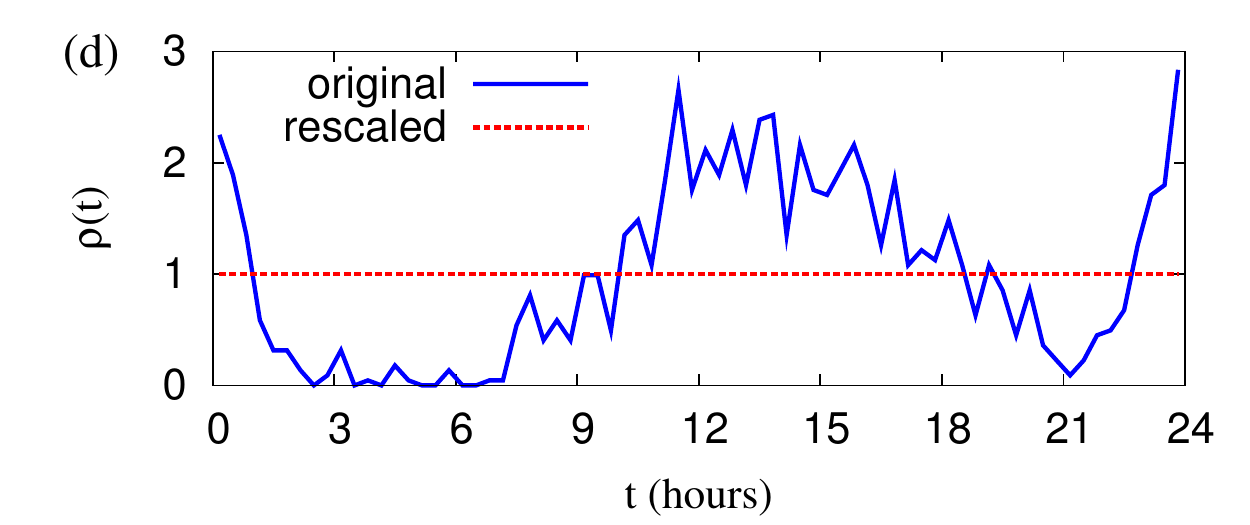}&
\includegraphics[width=.45\columnwidth]{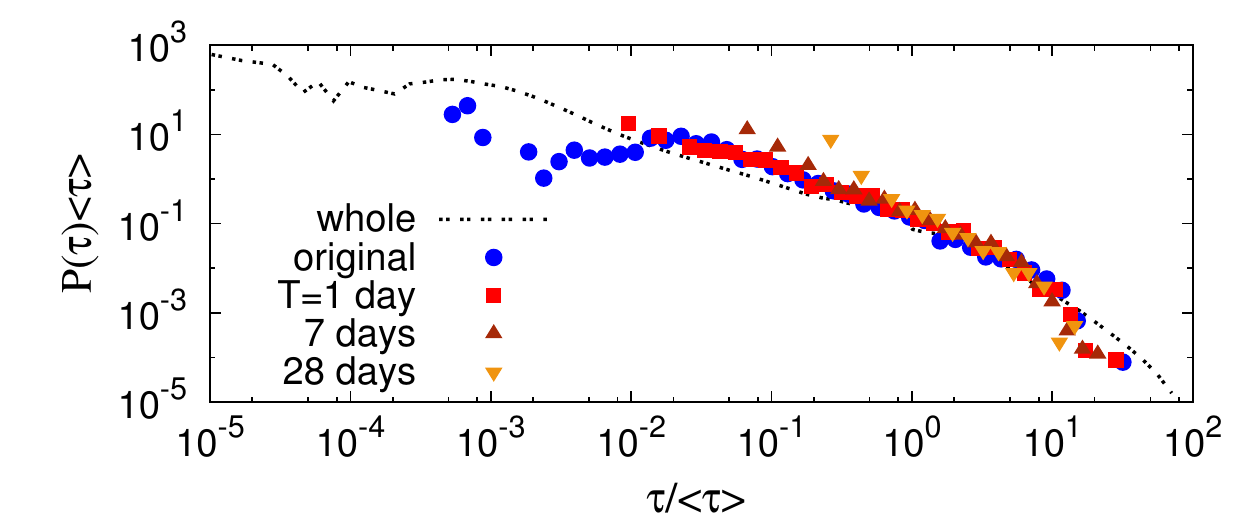}\\
\includegraphics[width=.45\columnwidth]{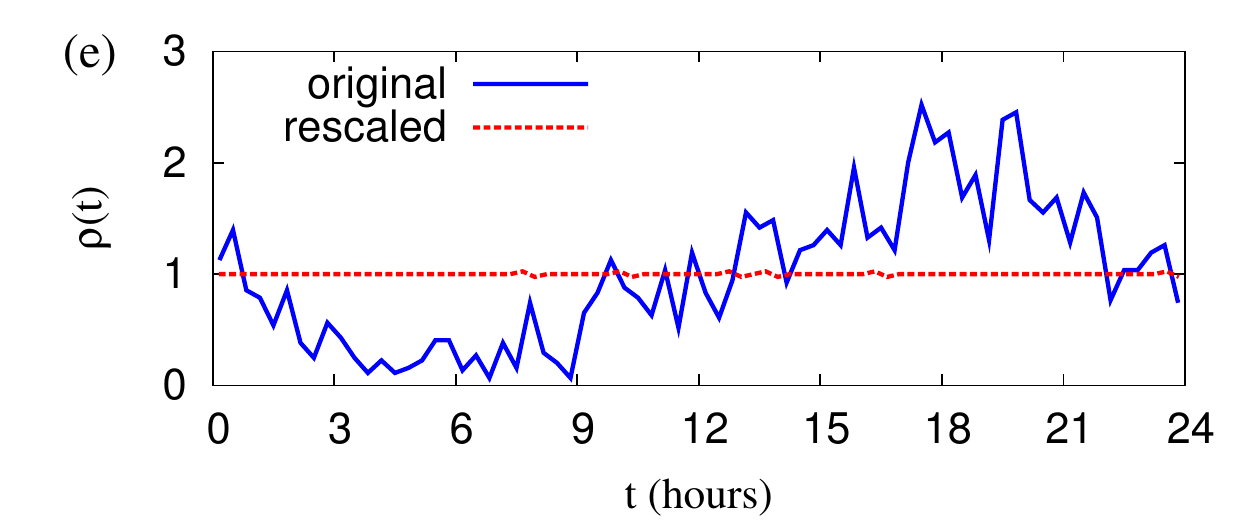}&
\includegraphics[width=.45\columnwidth]{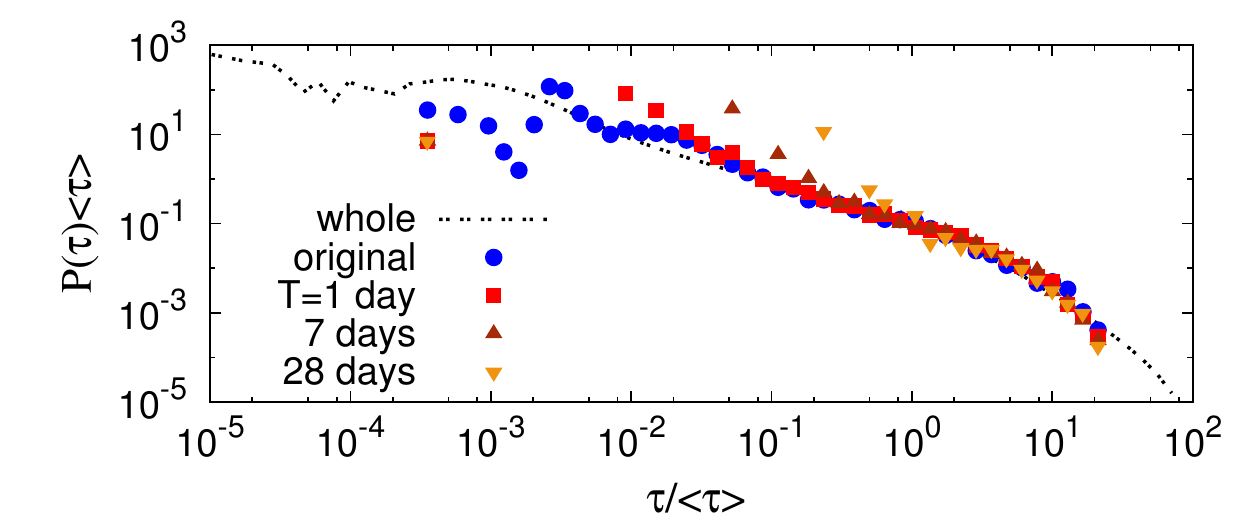}
\end{tabular}
\caption{De-seasoning SM patterns of individual users: the original and the rescaled event rates with period of $T=1$ day (left) and the original and the rescaled inter-event time distributions with various periods of $T$ (right). Individual users with the strength $s_i=200$ (a), 400 (b), 800 (c), 1600 (d), and 3200 (e) are analyzed. The original inter-event time distribution of the whole population is also plotted as a dashed curve for comparison.}
\label{fig:SMS_individual}\end{figure*}

\begin{figure*}[!h]
\begin{tabular}{cc}
\includegraphics[width=.45\columnwidth]{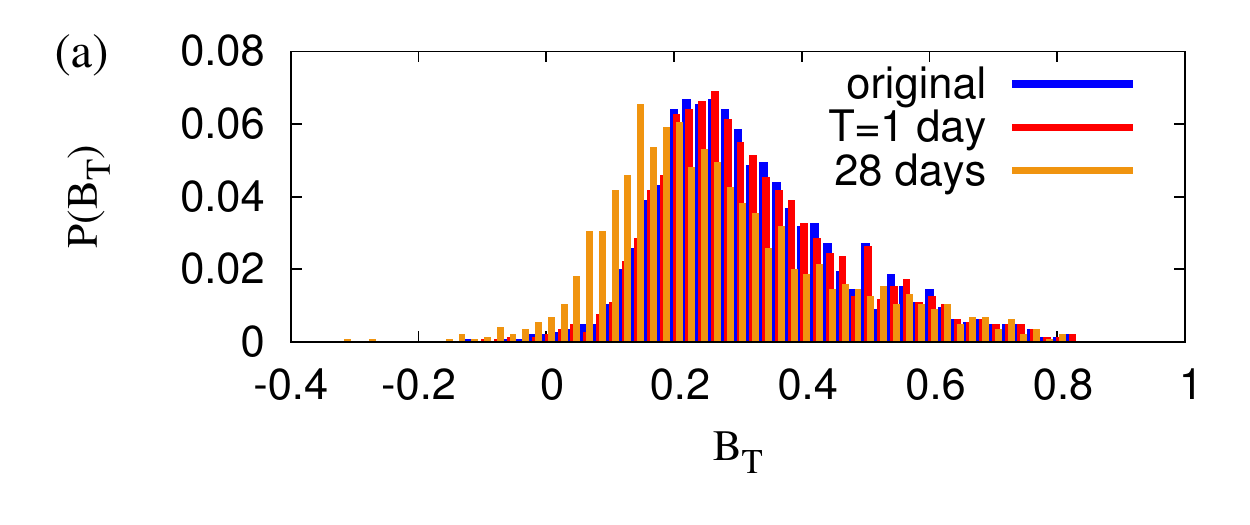}&
\includegraphics[width=.45\columnwidth]{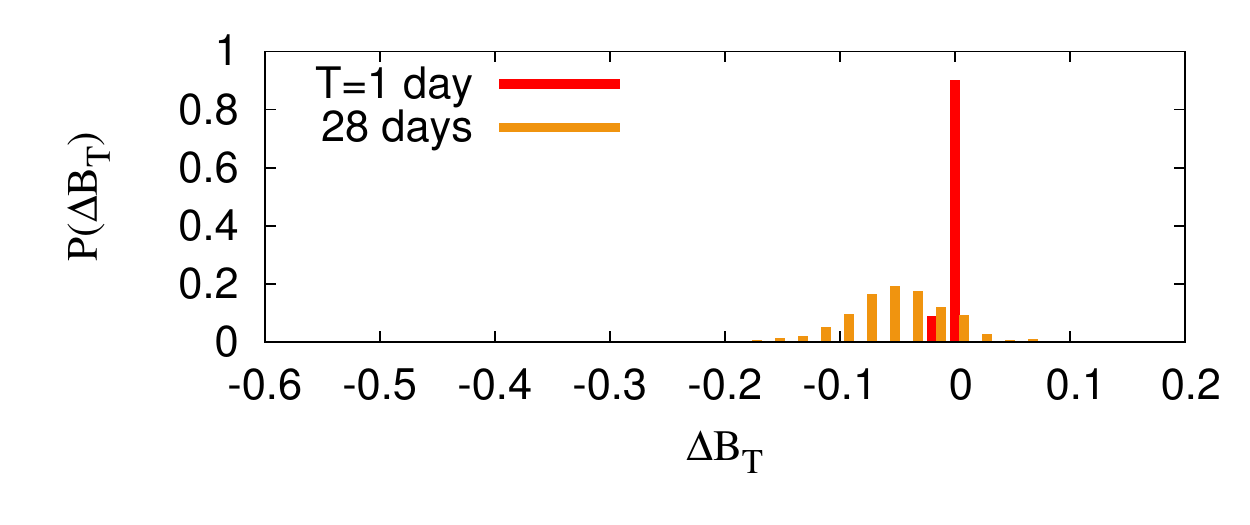}\\
\includegraphics[width=.45\columnwidth]{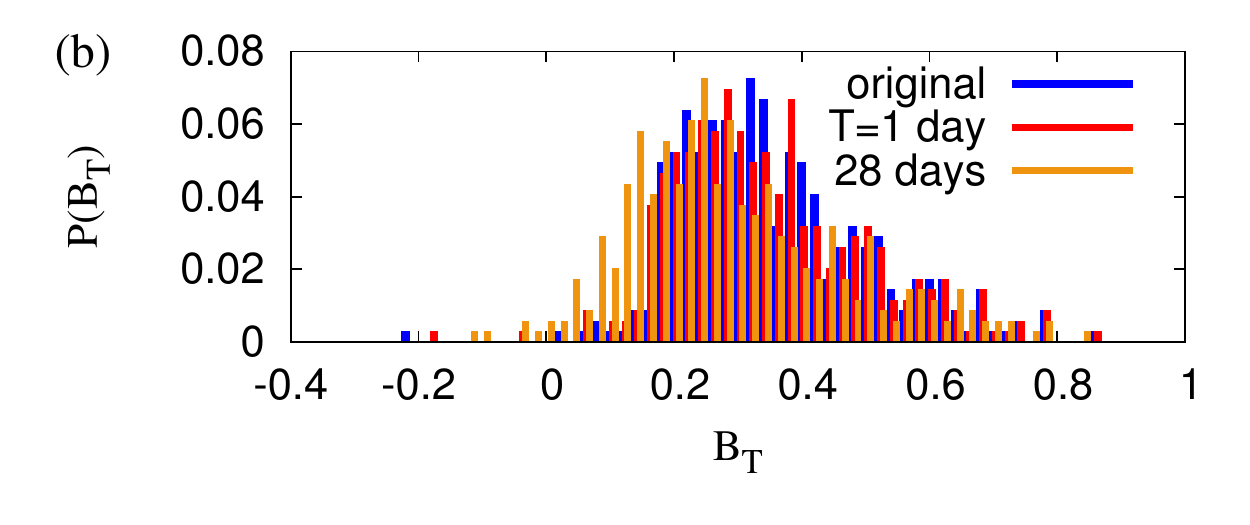}&
\includegraphics[width=.45\columnwidth]{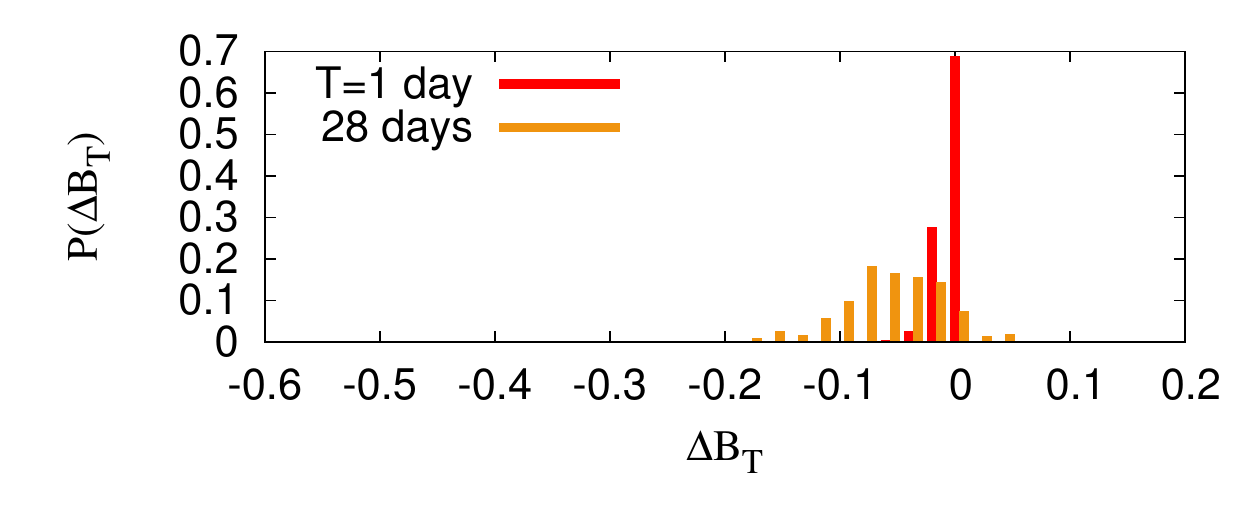}\\
\includegraphics[width=.45\columnwidth]{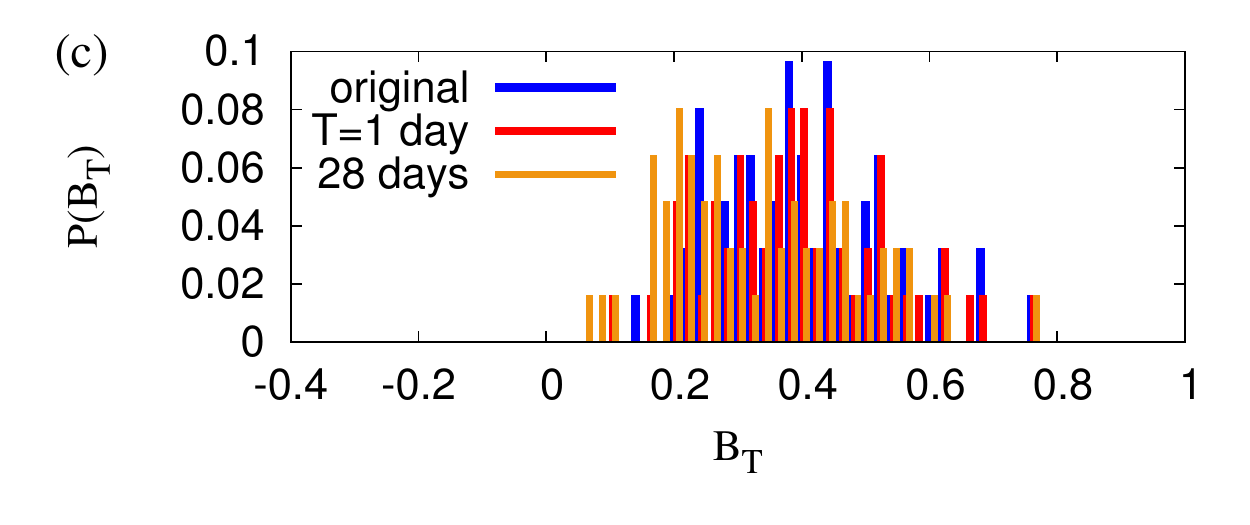}&
\includegraphics[width=.45\columnwidth]{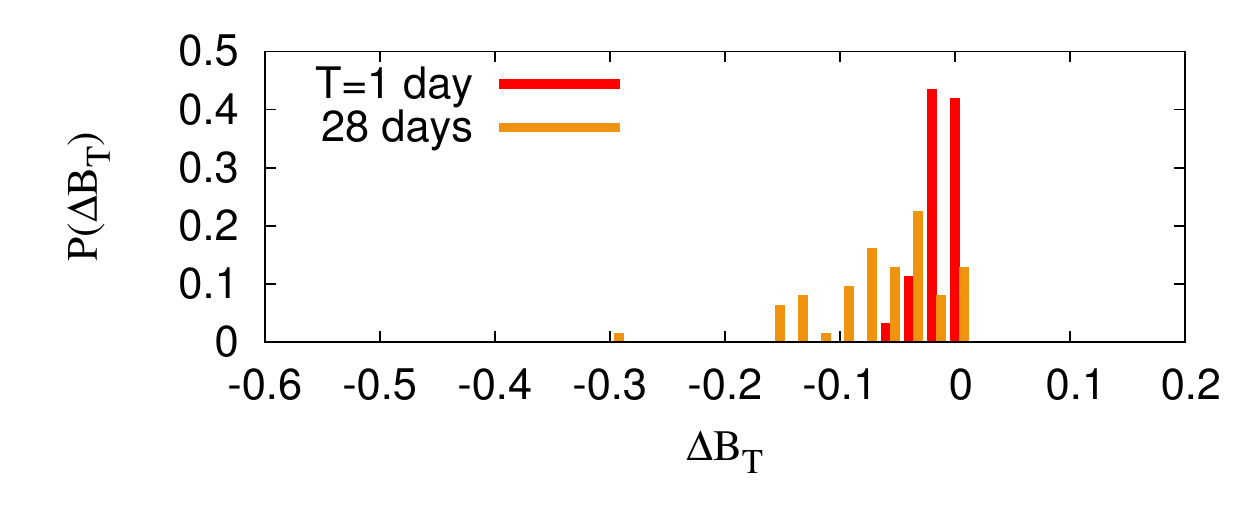}\\
\includegraphics[width=.45\columnwidth]{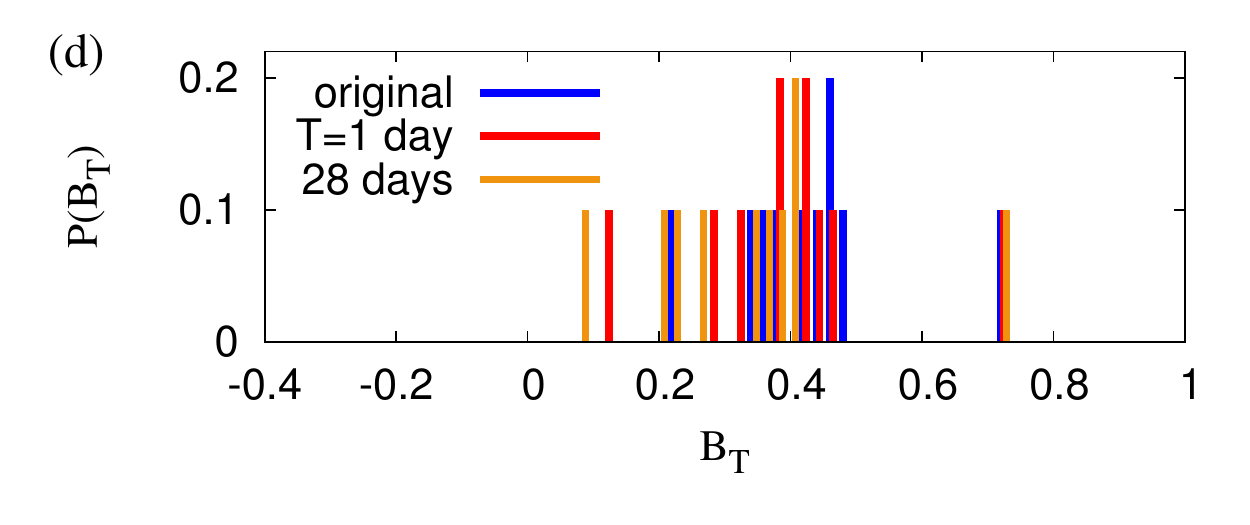}&
\includegraphics[width=.45\columnwidth]{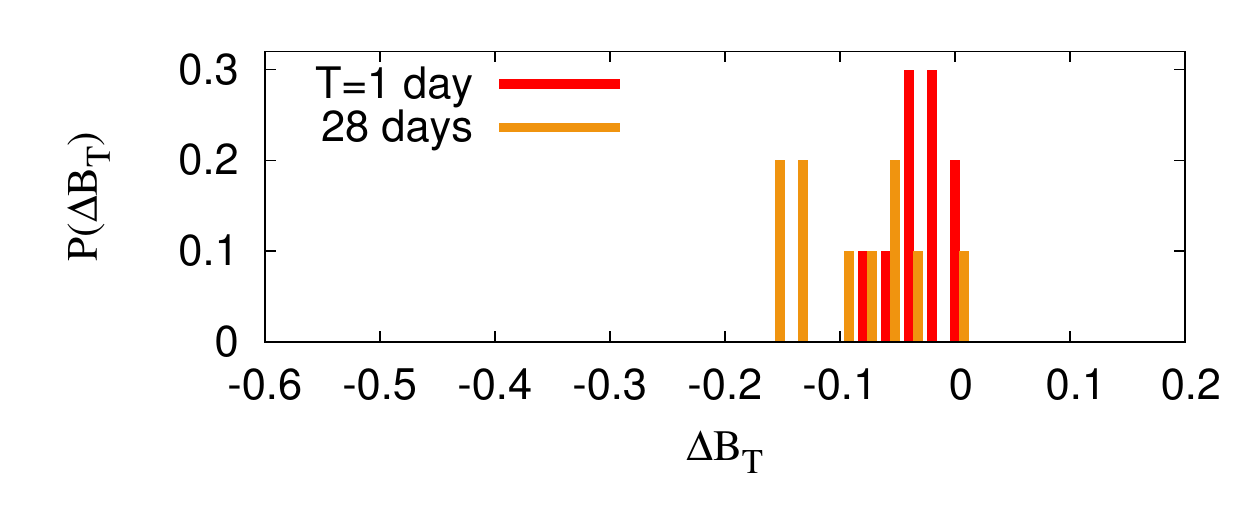}
\end{tabular}
\caption{Distributions $P(B_T)$ of the original and rescaled burstiness of invididual users with the same strength (left) and distributions $P(\Delta B_T)$ of the difference in burstiness, defined as $\Delta B_T=B_T-B_0$ (right). The individual users with the strengths $s_i=200$ (a), 400 (b), 800 (c), and 1600 (d) are analyzed. The numbers of users are correspondingly 1434, 344, 62, and 10.}
\label{fig:SMS_strength_sep}\end{figure*}

\begin{figure*}[!h]
\begin{tabular}{ccc}
\includegraphics[width=.3\columnwidth]{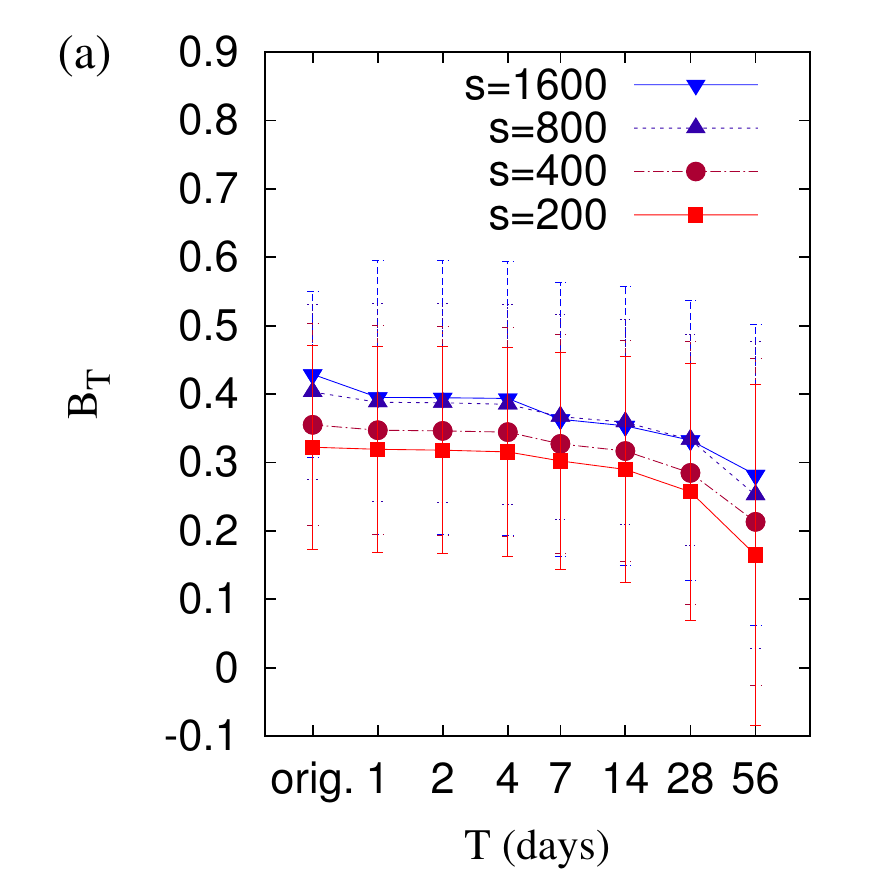}&
\includegraphics[width=.3\columnwidth]{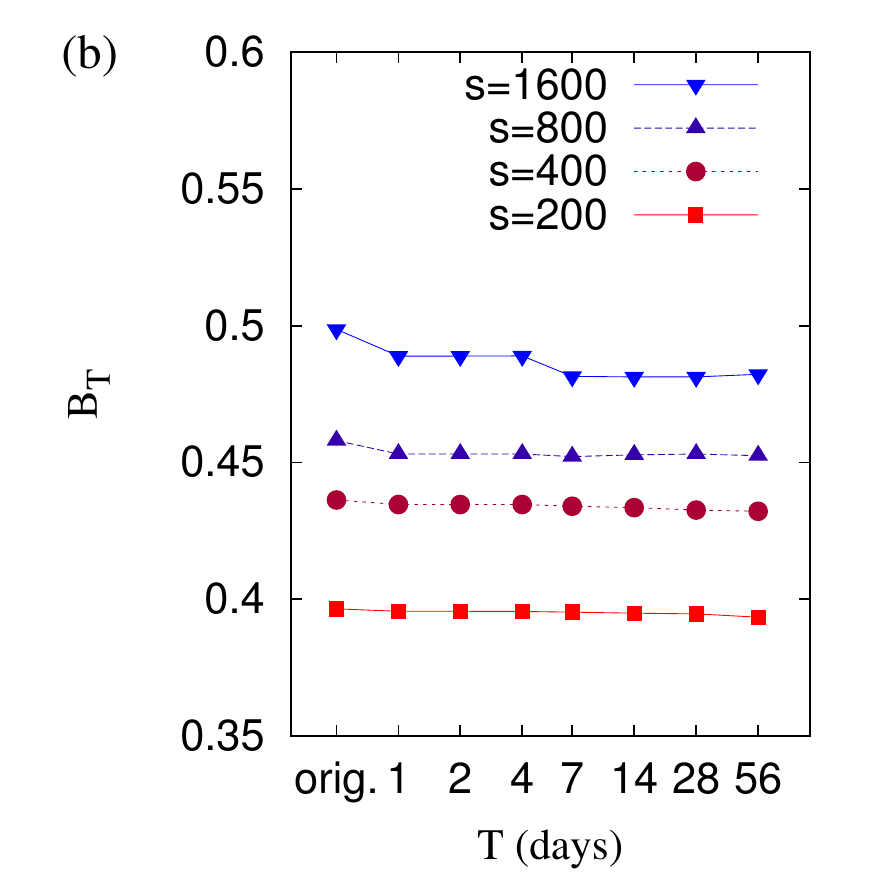}&
\includegraphics[width=.3\columnwidth]{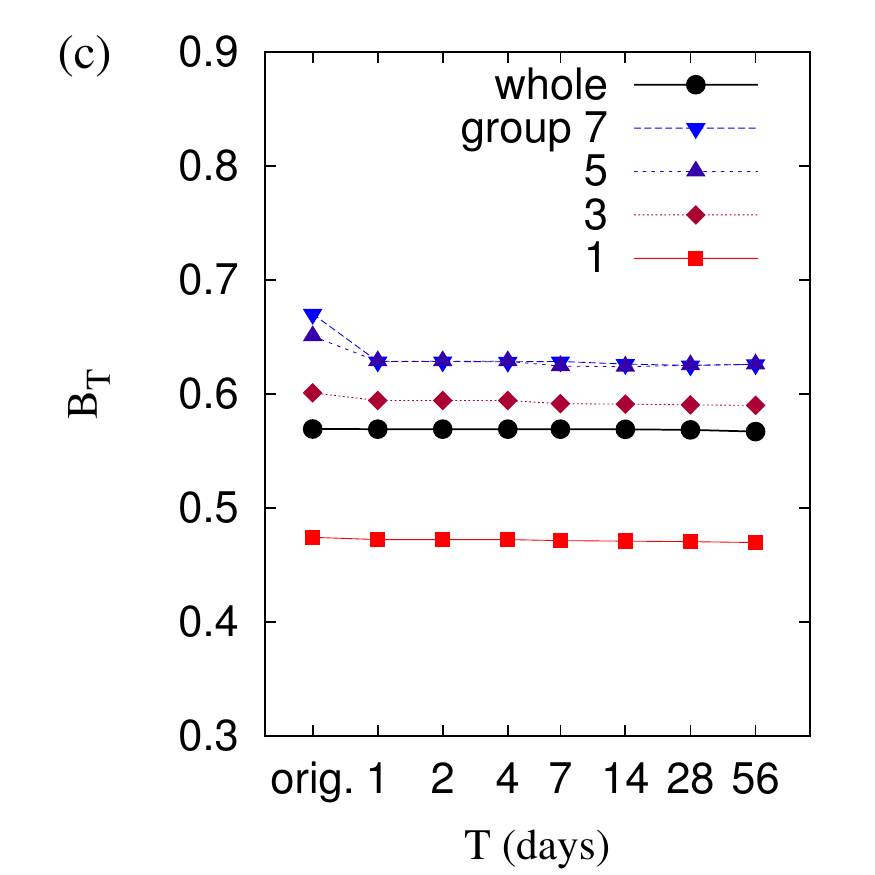}
\end{tabular}
\caption{Burstiness $B_T$ as a function of period of $T$: (a) the average and the standard deviation of $B_T$ obtained from the burstiness distribution in Fig.~\ref{fig:SMS_strength_sep}, (b) burstiness from groups with the same strength, and (c) burstiness from groups with broad ranges of strength.}
\label{fig:SMS_burstiness}\end{figure*}

\begin{figure*}[!h]
\begin{tabular}{cc}
\includegraphics[width=.45\columnwidth]{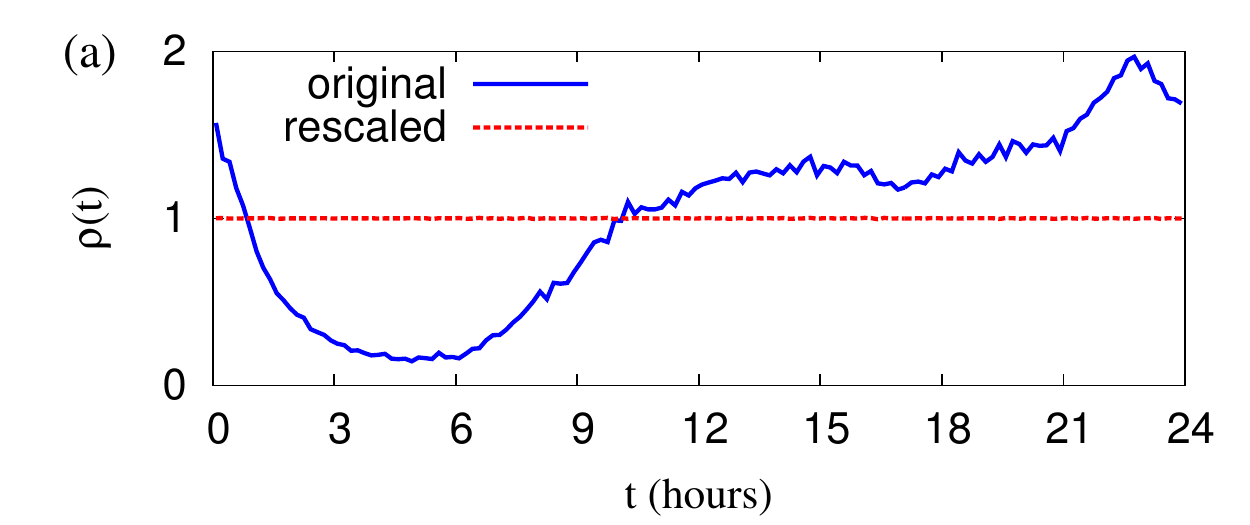}&
\includegraphics[width=.45\columnwidth]{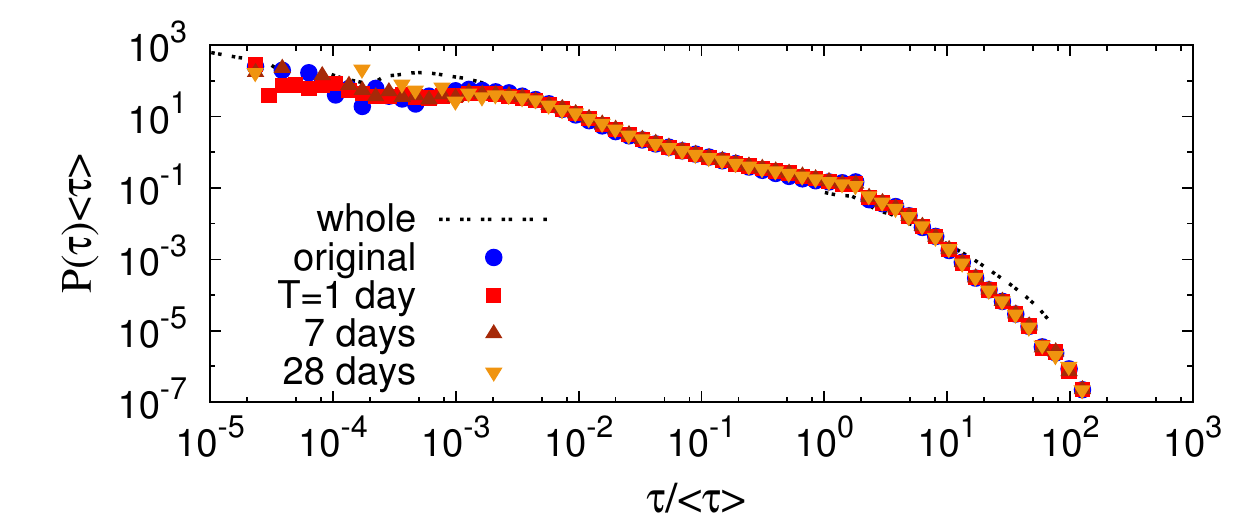}\\
\includegraphics[width=.45\columnwidth]{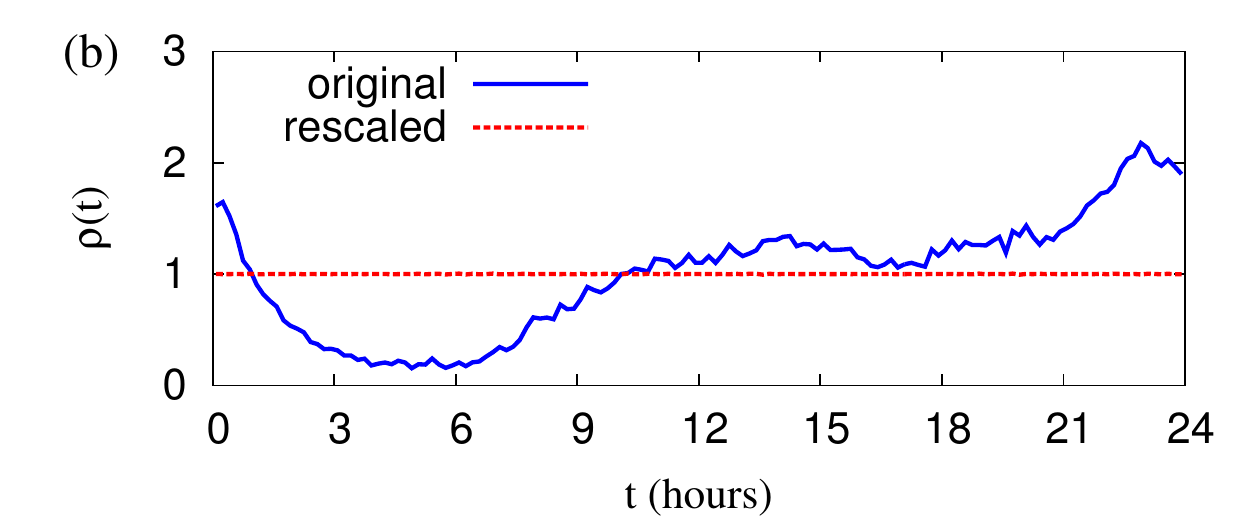}&
\includegraphics[width=.45\columnwidth]{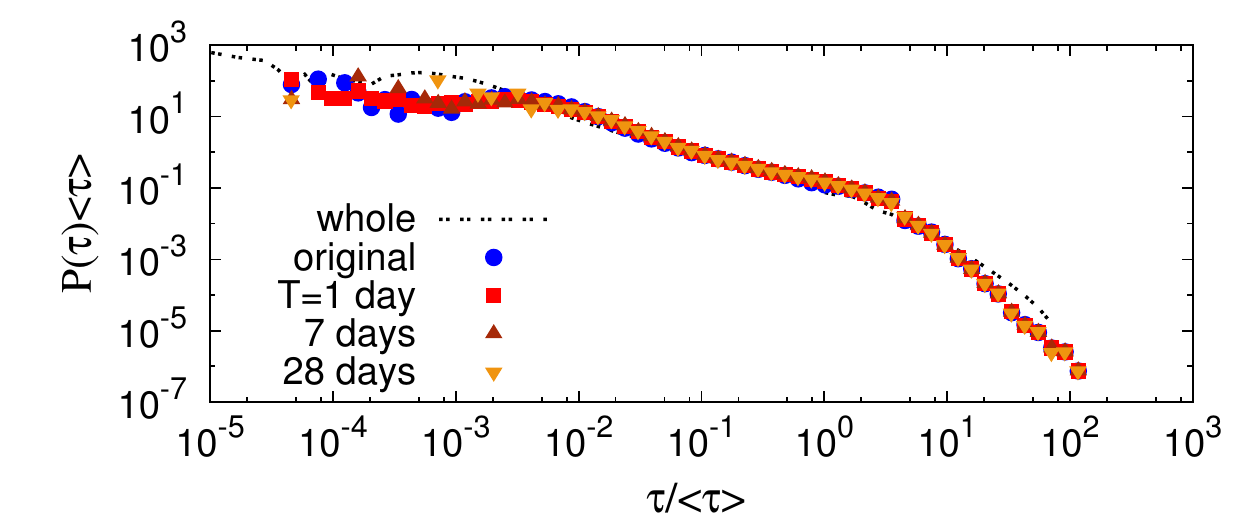}\\
\includegraphics[width=.45\columnwidth]{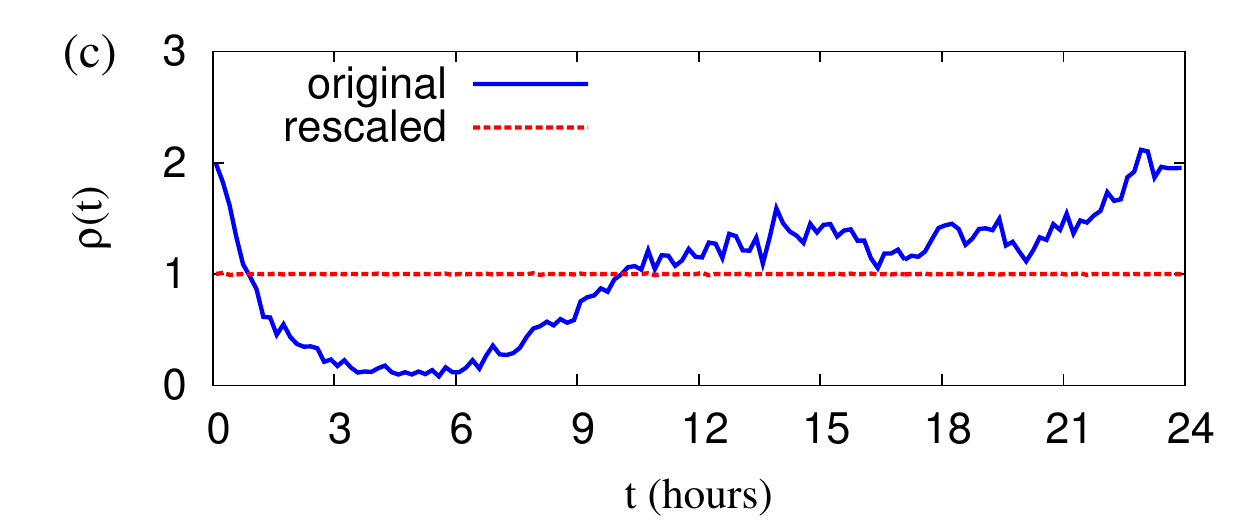}&
\includegraphics[width=.45\columnwidth]{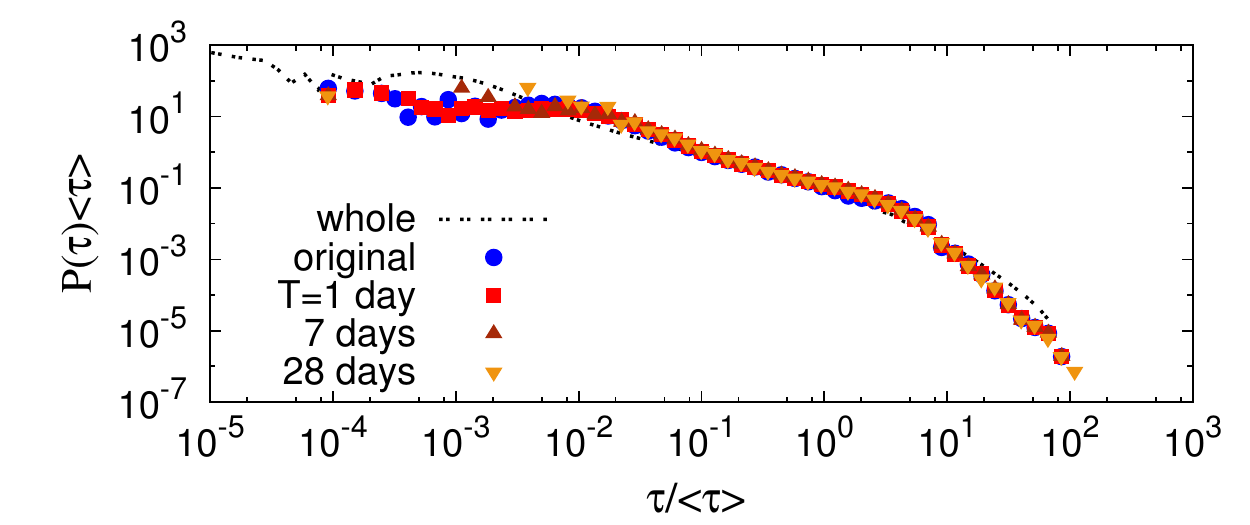}\\
\includegraphics[width=.45\columnwidth]{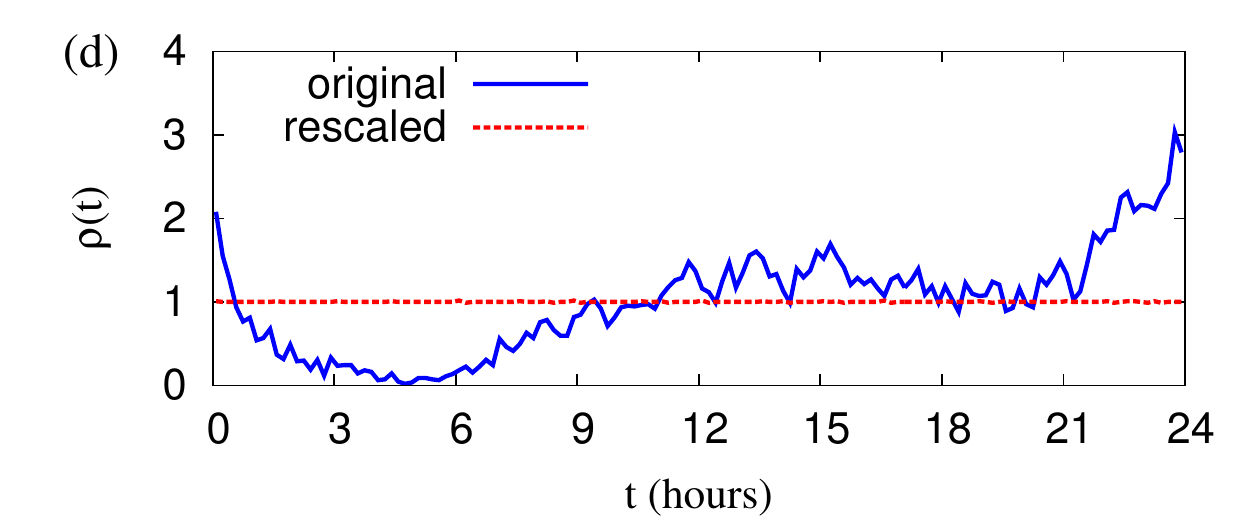}&
\includegraphics[width=.45\columnwidth]{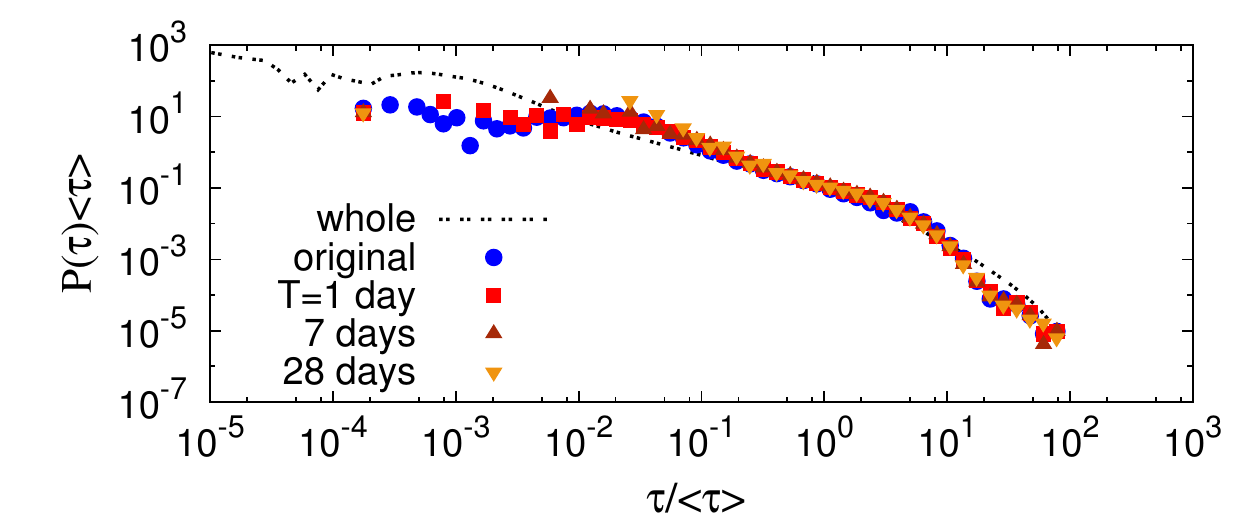}
\end{tabular}
\caption{De-seasoning of SM patterns for groups with the same strengths: $s=200$ (a), 400 (b), 800 (c), and 1600 (d). The original and rescaled distributions of burstiness are plotted in Fig.~\ref{fig:SMS_burstiness}(b).}
\label{fig:SMS_strength_merge}\end{figure*}

\begin{table}[!h]
    \caption{Strength dependent grouping in the SM dataset. For each group, the range of strength, also in terms of the ratio to the maximum strength $s_{\rm max}=37925$, the number of users, and its fraction to the whole population except for two most active users are summarized.}
\label{table:SMS_group}
\begin{indented}
\item[]\begin{tabular}{rrr}
\hline
group index & strength range ($\%$) & the number of users ($\%$)\\ 
\hline
0 & 0-94 (0-0.25) & 3685116 (87.2)\\
1 & 94-189 (0.25-0.5) & 294181 (6.9)\\
2 & 189-379 (0.5-1) & 152893 (3.6)\\
3 & 379-758 (1-2) & 63379 (1.5)\\
4 & 758-1517 (2-4) & 22507 (0.53)\\
5 & 1517-3034 (4-8) & 7505 (0.18)\\
6 & 3034-6068 (8-16) & 2393 (0.057)\\
7 & 6068-12136 (16-32) & 316 (0.007)\\
8 & 12136-24272 (32-64) & 30 (0.001)\\
whole & 0-24272 (0-64) & 4228320 (100)\\
\hline
\end{tabular}
\end{indented}
\end{table}

\begin{figure*}[!h]
\begin{tabular}{cc}
\includegraphics[width=.45\columnwidth]{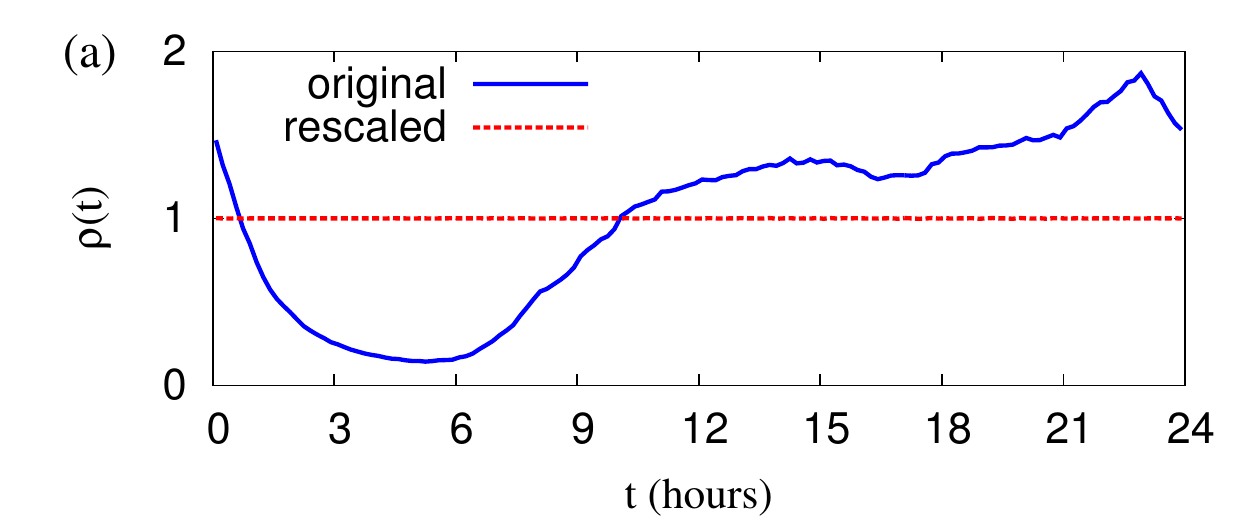}&
\includegraphics[width=.45\columnwidth]{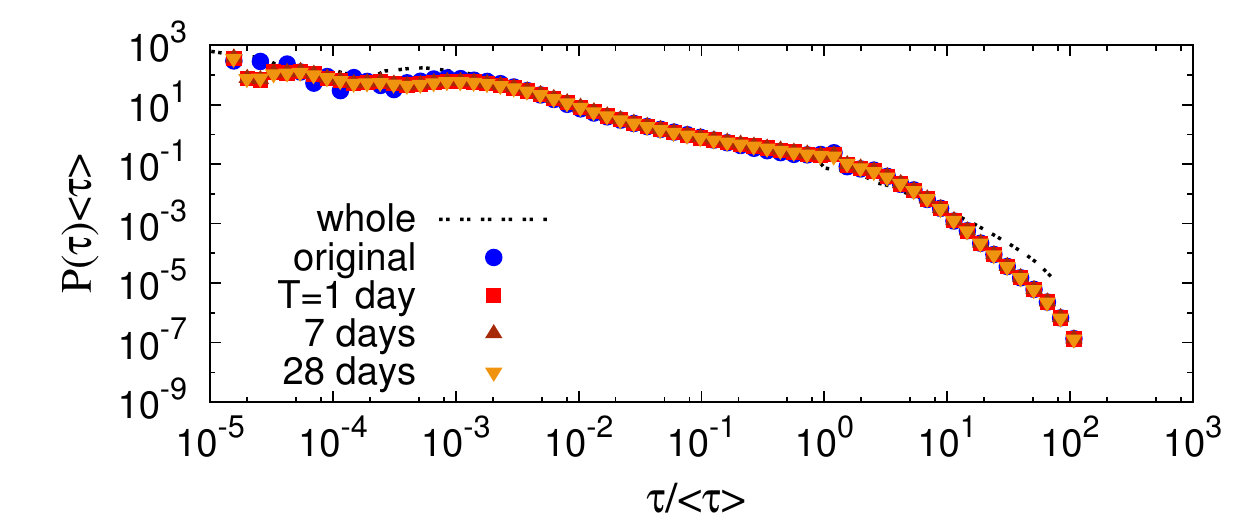}\\
\includegraphics[width=.45\columnwidth]{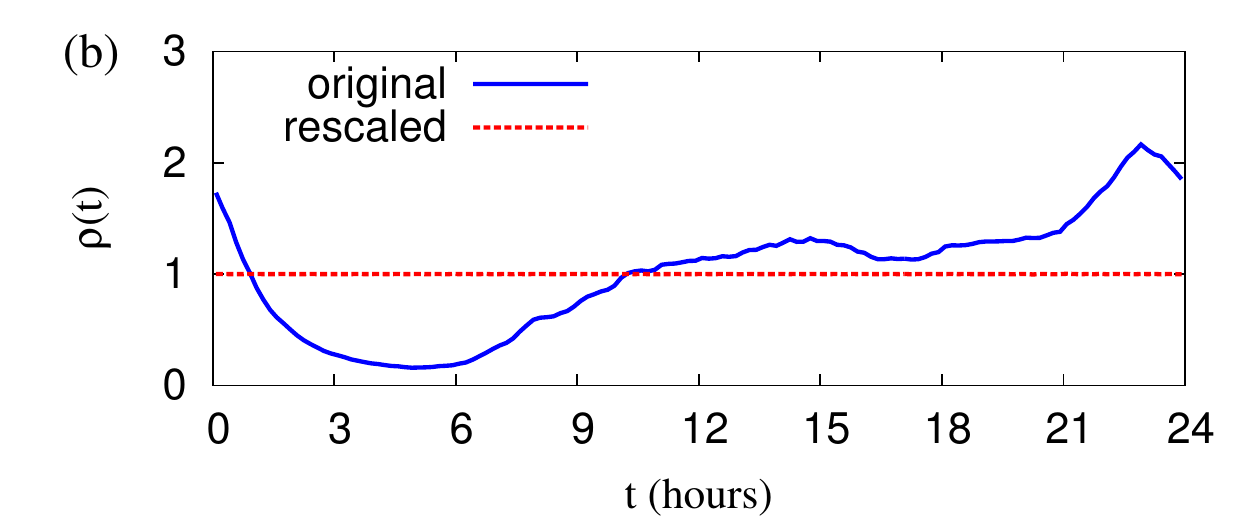}&
\includegraphics[width=.45\columnwidth]{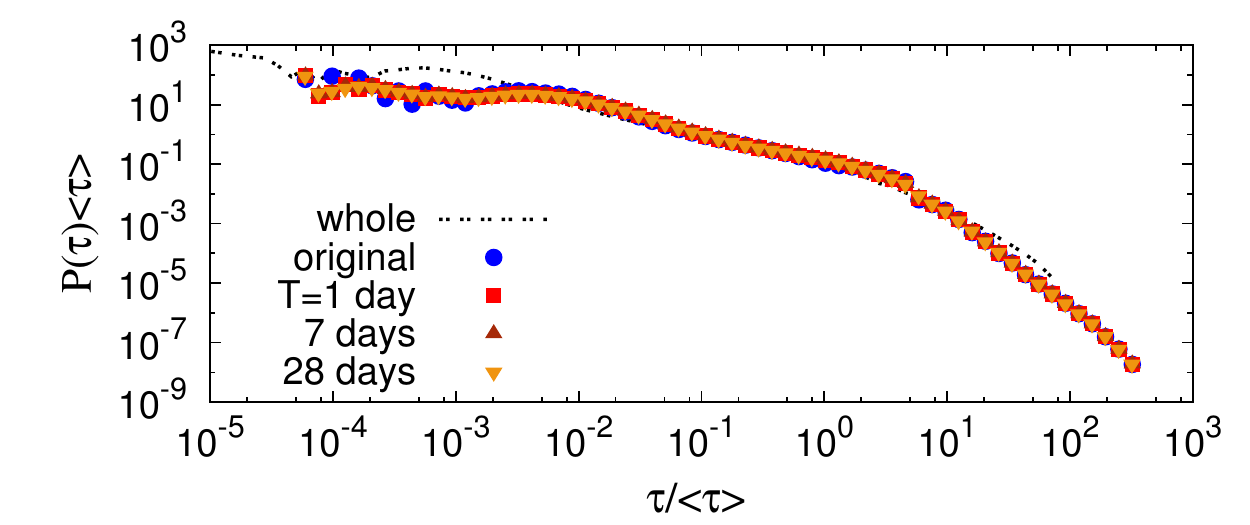}\\
\includegraphics[width=.45\columnwidth]{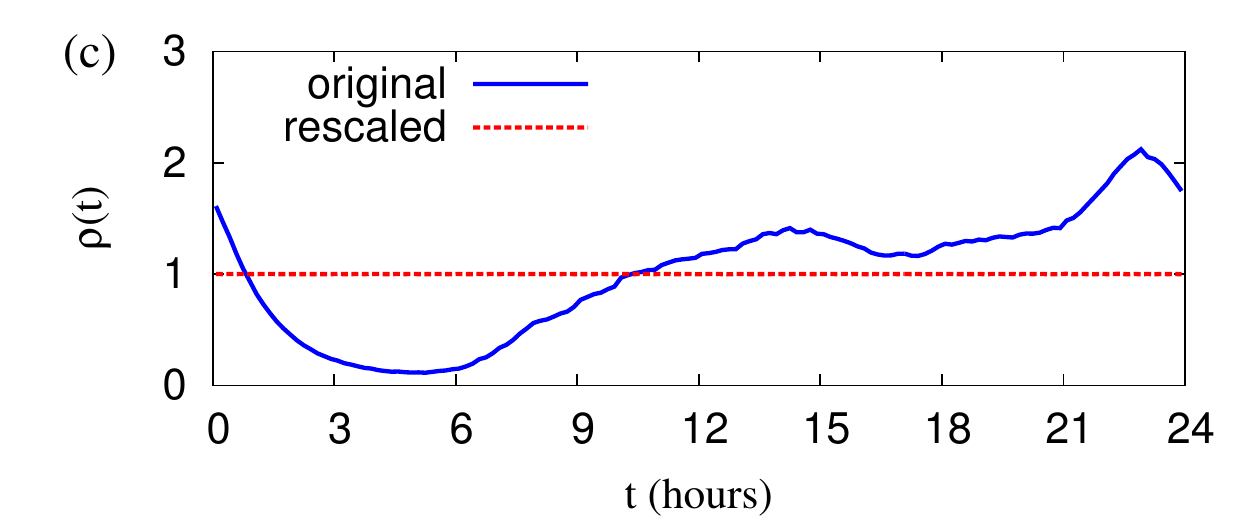}&
\includegraphics[width=.45\columnwidth]{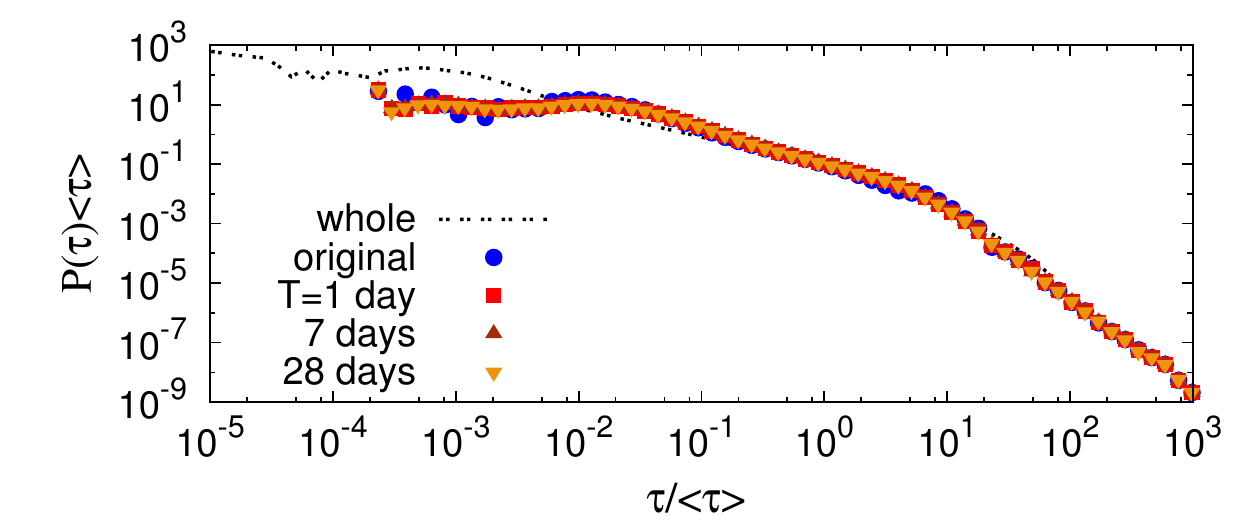}\\
\includegraphics[width=.45\columnwidth]{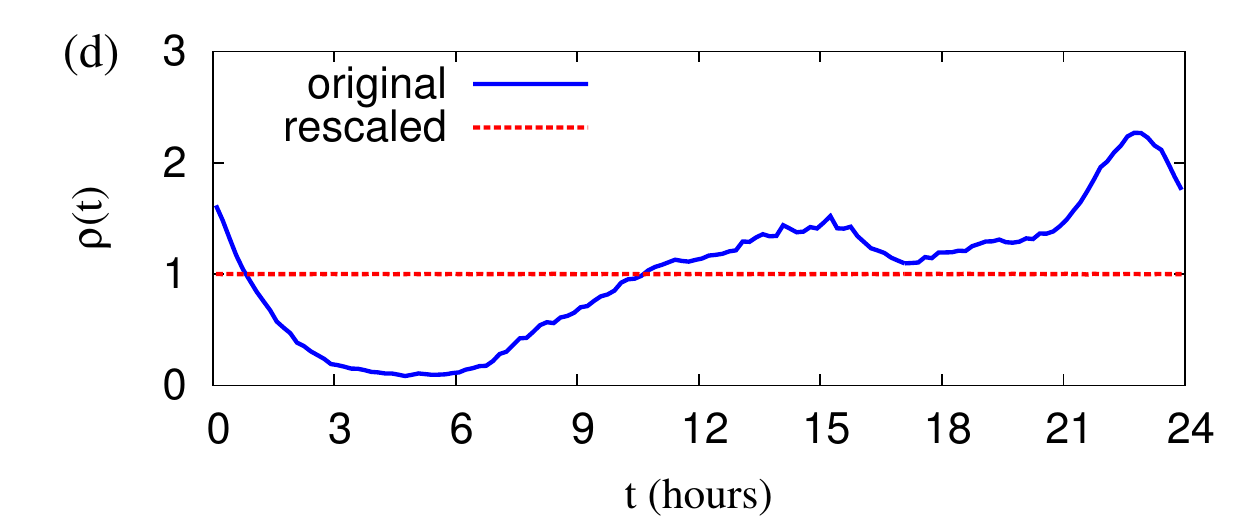}&
\includegraphics[width=.45\columnwidth]{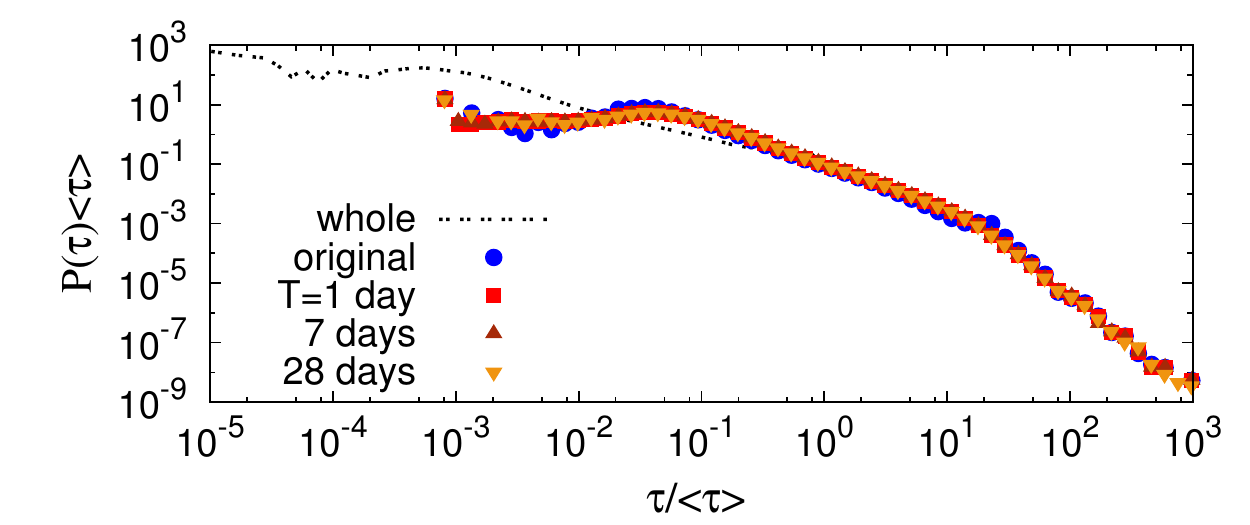}\\
\includegraphics[width=.45\columnwidth]{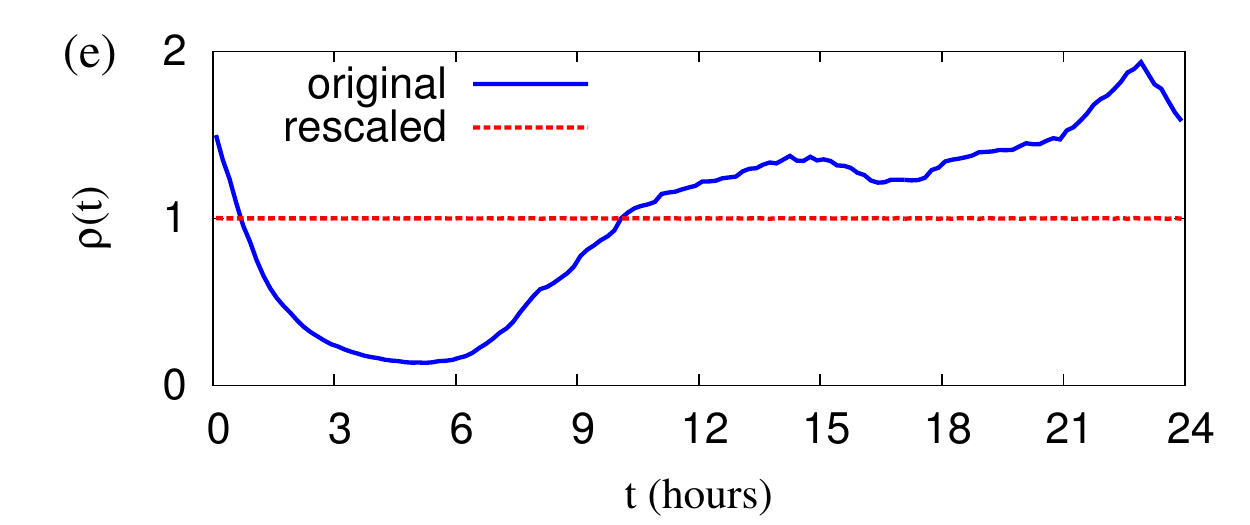}&
\includegraphics[width=.45\columnwidth]{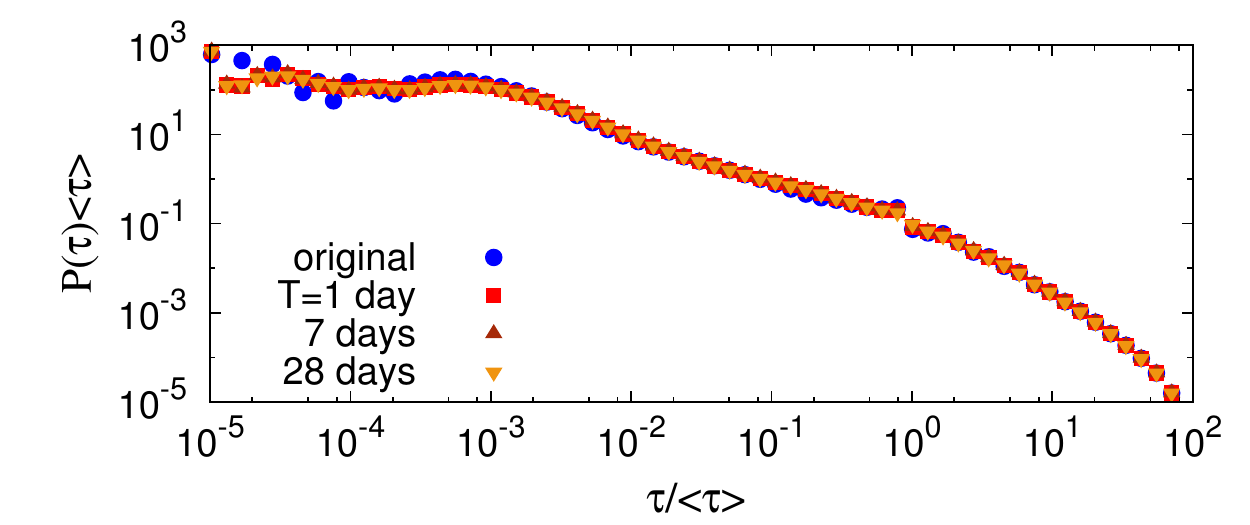}
\end{tabular}
\caption{De-seasoning of the SM patterns for groups of individuals with broad ranges of strength: groups 1 (a), 3 (b), 5 (c), 7 (d), and the whole population (e). For the details of groups, see Table~\ref{table:SMS_group}. The original and rescaled distributions of burstiness are plotted in Fig.~\ref{fig:SMS_burstiness}(c).}
\label{fig:SMS_group}\end{figure*}

\begin{figure*}[!h]
\begin{tabular}{cc}
\includegraphics[width=.45\columnwidth]{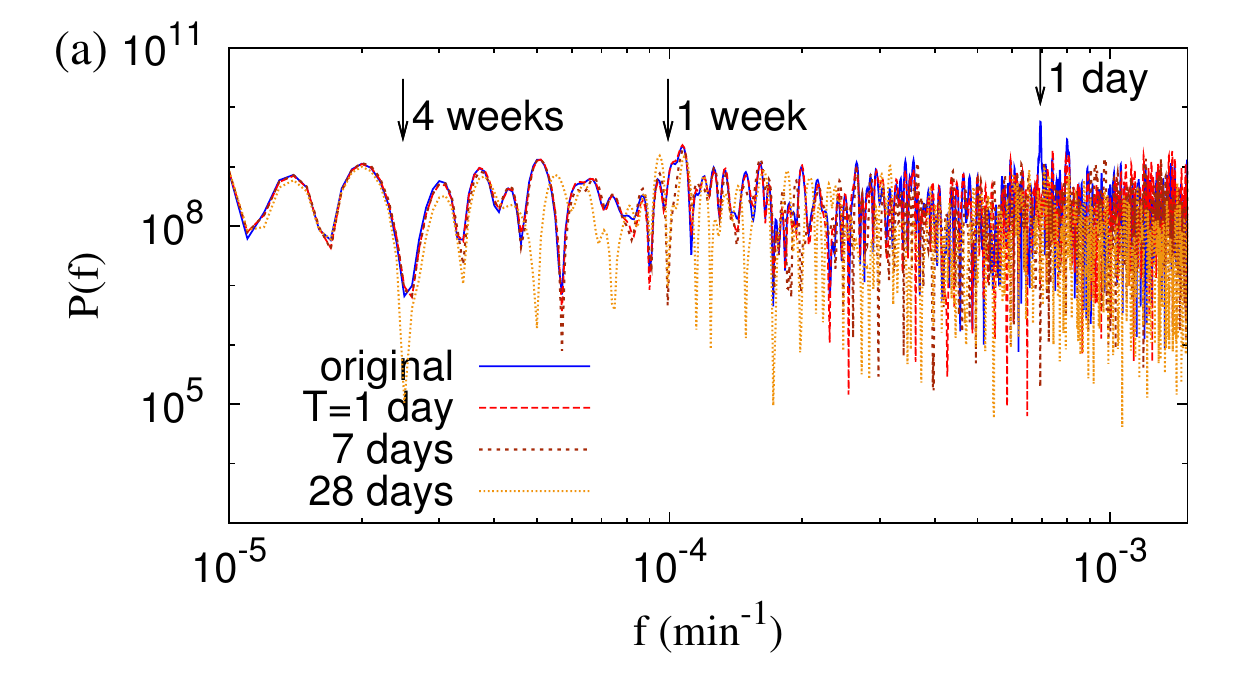}&
\includegraphics[width=.45\columnwidth]{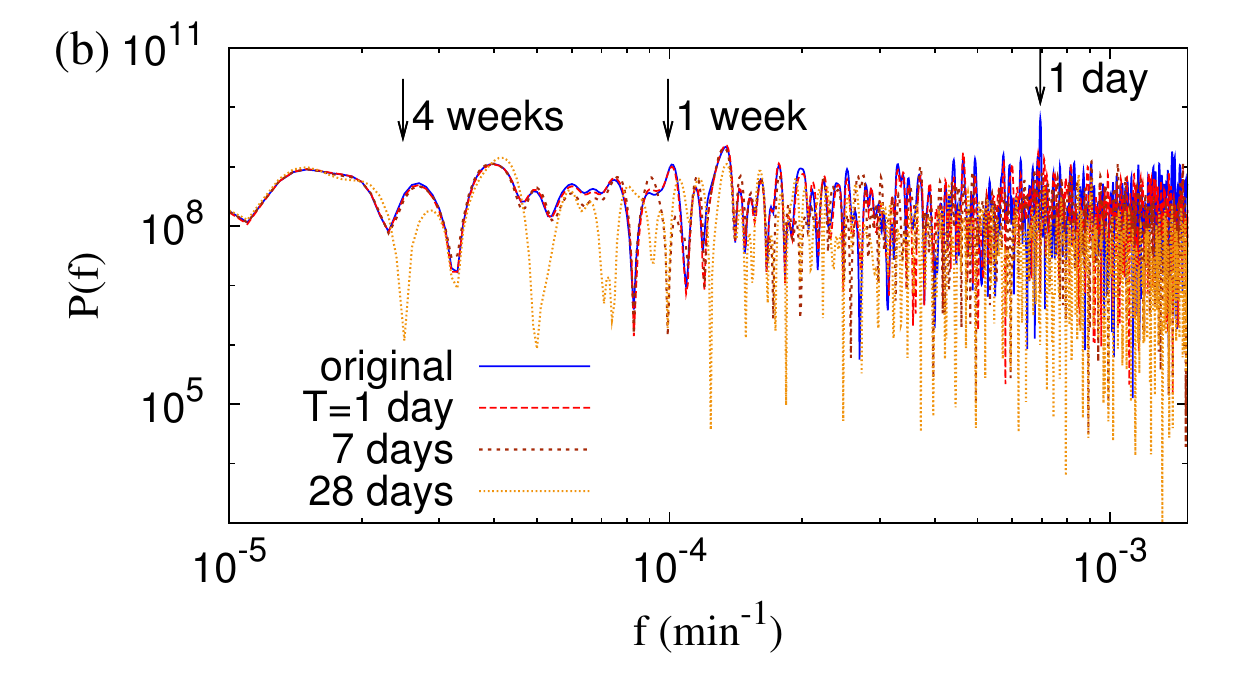}\\
\includegraphics[width=.45\columnwidth]{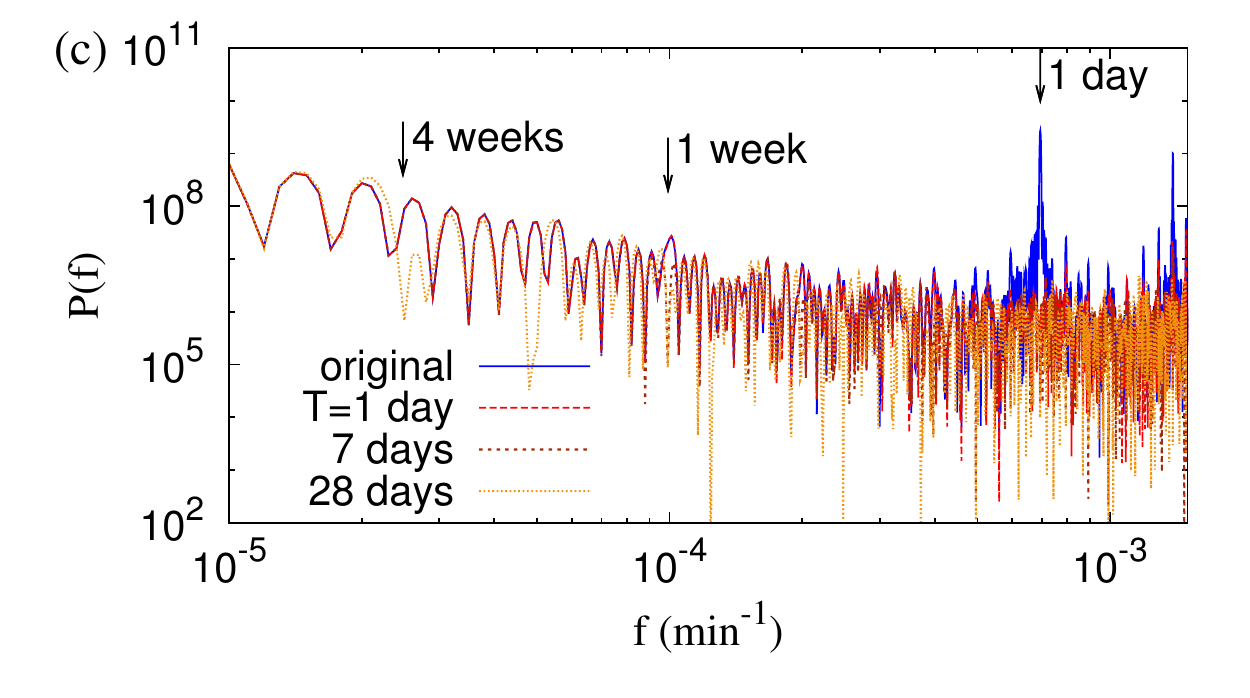}&
\includegraphics[width=.45\columnwidth]{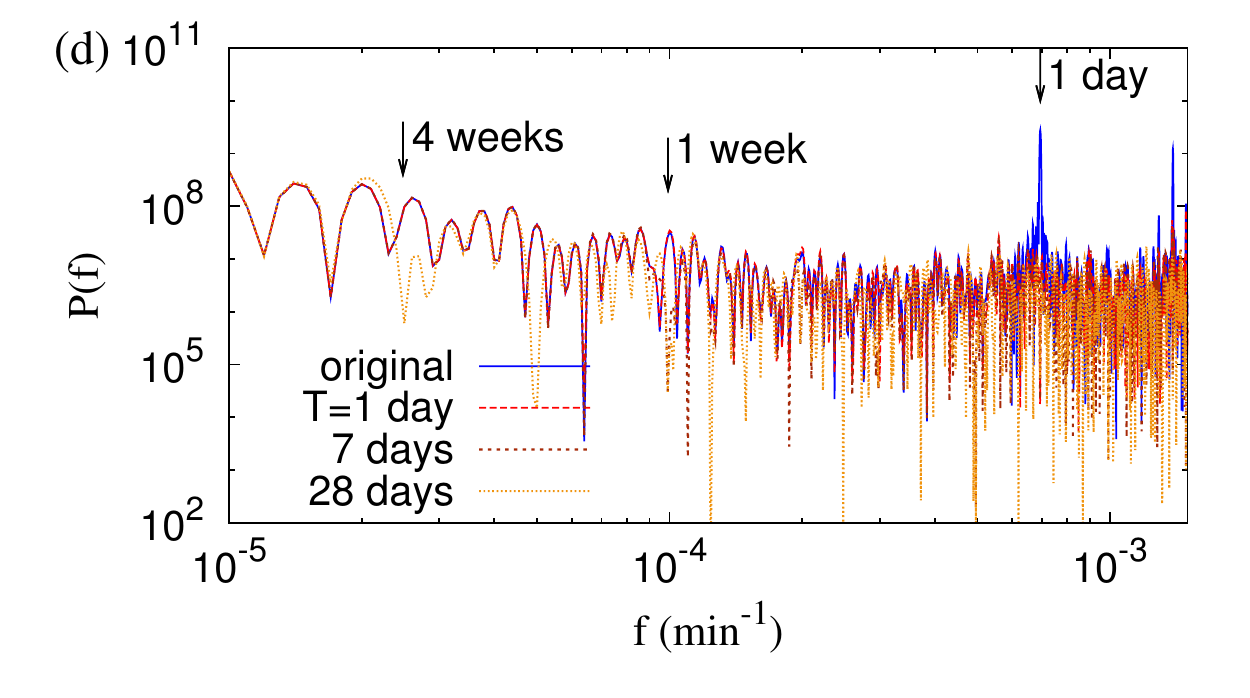}\\
\includegraphics[width=.45\columnwidth]{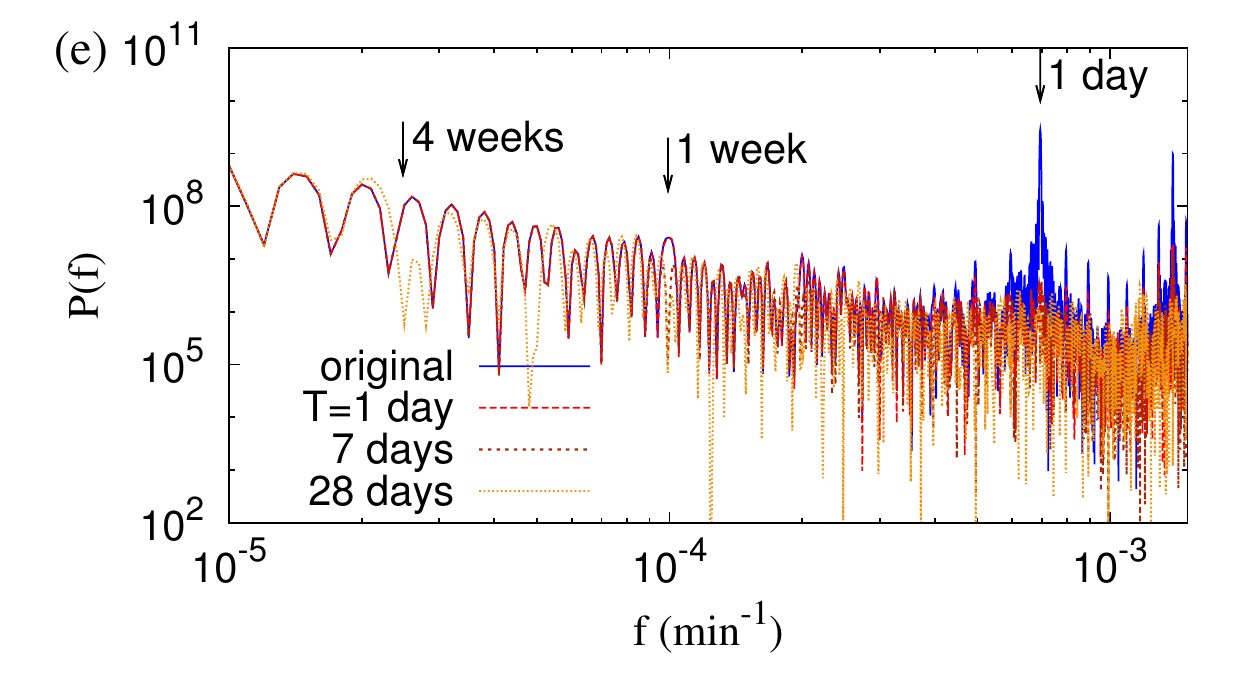}
\end{tabular}
\caption{Power spectra, $P(f)$, of the original and the rescaled event rates with various periods of $T$ for individual users with strengths 200 (a) and 800 (b), for groups with the same strengths of 200 (c) and 800 (d), and for the whole population (e). The circadian and weekly peaks in the original power spectra are successfully removed by the de-seasoning in various ways.}
\label{fig:SMS_powerSpec}\end{figure*}

We apply the same de-seasoning analysis described in the main text to the SM dataset. Figure~\ref{fig:SMS_individual} shows the various circadian activity patterns of individual users. The main difference from the MPC dataset is that the activity peak is around 11 PM. This feature becomes evident if the averaged event rates are obtained from the same strength groups or from the groups with broad ranges of strength, shown in the left columns of Figs~\ref{fig:SMS_strength_merge} and~\ref{fig:SMS_group}. For the details of the strength dependent grouping, see Table~\ref{table:SMS_group}.

The inter-event time distributions and their values of burstiness are also compared. At first, the distributions cannot be described by the simple power-law form but by the bimodal combination of power-law and Poisson distributions as suggested by~\cite{Wu2010}. As shown in Fig.~\ref{fig:SMS_burstiness}, the values of burstiness slowly decrease as the period $T$ increases up to 7 days, which implies that de-seasoning the circadian and weekly patterns does not considerably affect the bursty behavior.

Finally, we perform the power spectrum analysis for SM dataset in Fig.~\ref{fig:SMS_powerSpec} and find again that the cyclic patterns longer than $T$ cannot be removed by the de-seasoning with period $T$. All these results confirm our conclusion that the heavy tail and burstiness are not only the consequence of circadian and other longer cycle patterns but also due to other correlations, such as human task execution.
 
\section*{Acknowledgements}
Financial support by Aalto University postdoctoral program (HJ), from EU’s 7th Framework Program’s FET-Open to ICTeCollective project no. 238597 and by the Academy of Finland, the Finnish Center of Excellence program 2006-2011, project no. 129670 (MK, KK, JK), as well as by TEKES (FiDiPro) (JK) are gratefully acknowledged. The authors thank A.-L.~Barab\'asi for the mobile phone communication dataset used in this research.

\section*{References}
%\bibliography{h2jo-circadian}

\end{document}